\documentclass[sigconf,nonacm]{acmart}
\usepackage{xspace,balance,tabularx,multirow}
\usepackage{flushend}
\usepackage{tikz}
\usepackage{pgfplots}
\usetikzlibrary{patterns}
\pgfplotsset{compat=1.16}
\usepackage[ruled, vlined, linesnumbered]{algorithm2e}
\usepackage[noend]{algpseudocode}
\usepackage{xcolor}
\usepackage{float}
\usepackage{colortbl}
\usepackage{hyperref}
\usepackage{bbold}
\usepackage{textcomp}
\usepackage{pifont}
\SetKwComment{Comment}{$\triangleright$\ }{}
\usepackage{enumitem}
\usepackage{xcolor,framed}
\usepackage[aboveskip=-4pt]{subcaption}
\usepackage[normalem]{ulem}
\usepackage{tikz-3dplot}

\newcommand{\laks}[1]{{#1}}
\newcommand{\sq}[1]{{{#1}}}
\newcommand{\yiqiann}[1]{{{#1}}}
\newcommand{\edit}[1]{{{#1}}}

\newcommand{\LL}[1]{{#1}}
\newcommand{\shiqi}[1]{{{#1}}}
\newcommand{\yiqian}[1]{{{#1}}}
\newcommand{\W}[1]{{{#1}}}
\newcommand{\hl}[1]{{#1}}

\newcommand{\revise}[1]{{#1}}

\newcommand{\argmax}[1]{\underset{#1}{\operatorname{arg}\,\operatorname{max}}\;}

\definecolor{Red}{HTML}{E81123}
\definecolor{Orange}{HTML}{FF8C00}
\definecolor{Green}{HTML}{009E49}
\definecolor{LightBlue}{HTML}{00BCF2}
\definecolor{DeepBlue}{HTML}{001BA3}
\definecolor{Pink}{HTML}{8BC63E}
\definecolor{deepblue}{RGB}{2, 18, 129} 
\definecolor{plasma1}{RGB}{243, 247, 88}
\definecolor{plasma2}{RGB}{235, 152, 82}
\definecolor{plasma3}{RGB}{190, 81, 120}
\definecolor{plasma4}{RGB}{115, 34, 163}

\newcommand{\eat}[1]{} %comments out the arument. 

\makeatletter
\newcommand*\bigcdot{\mathpalette\bigcdot@{.5}}
\newcommand*\bigcdot@[2]{\mathbin{\vcenter{\hbox{\scalebox{#2}{$\m@th#1\bullet$}}}}}
\makeatother

\SetKw{Break}{break}

\newcommand{\ie}{{i.e.},\xspace}
\newcommand{\eg}{{e.g.},\xspace}

\newcommand{\stitle}[1]{\noindent{\bf #1.\/}}

\newtheorem{corollary}{Corollary}

\newtheorem{proposition}{proposition}

\newcommand{\Q}{\mathcal{Q}\xspace}
\newcommand{\G}{\mathsf{G}\xspace}
\newcommand{\Lset}{\mathsf{L}\xspace}
\newcommand{\x}{\mathbf{x}}
\newcommand{\y}{\mathbf{y}}

\newcommand{\R}{\mathcal{R}\xspace}
\newcommand{\I}{\mathcal{I}\xspace}

\newcommand{\Sset}{\mathcal{S}\xspace}
\newcommand{\Rset}{\mathcal{R}\xspace}

\newcommand{\M}{\mathsf{M}\xspace}

\newcommand{\BPoly}{\mathcal{B}\xspace}

\newcommand{\rim}{\texttt{RM}\xspace}
\newcommand{\mrim}{\texttt{MRIM}\xspace}
\newcommand{\advim}{\texttt{AdvIM}\xspace}
\newcommand{\prm}{\texttt{PRM}\xspace}
\newcommand{\im}{\texttt{IM}\xspace}
\newcommand{\ourproblem}{\texttt{IM-GM}\xspace}
\newcommand{\imm}{\textsf{IMM}\xspace}
\newcommand{\opimc}{\textsf{OPIM-C}\xspace}
\newcommand{\mgreedy}{\textsf{Greedy}\xspace}
\newcommand{\lgreedy}{\textsf{Local-Greedy}\xspace}
\newcommand{\tgreedy}{\textsf{Threshold-Greedy}\xspace}
\newcommand{\cgreedy}{\textsf{AMPSearch}\xspace}
\newcommand{\cseedselect}{\textsf{AMP}\xspace}
\newcommand{\cseedselectp}{\textsf{AMP-PM}\xspace}
\newcommand{\cgreedyp}{\textsf{AMPSearch-PM}\xspace}

\newcommand{\crounding}{\textsf{AMPRound}\xspace}
\newcommand{\ouralgo}{\textsf{RAMP}\xspace}
\newcommand{\mwgreedy}{$\mathsf{Maximum}$-$\mathsf{Weight}$-$\mathsf{Greedy}$\xspace}
\newcommand{\rma}{\textsf{RM-A}\xspace}
\newcommand{\naimm}{\textsf{CR-NAIMM}\xspace}
\newcommand{\advimm}{\textsf{AAIMM}\xspace}
\newcommand{\mclick}{\textsf{Most-Click}\xspace}

\newcommand{\ic}{\textsf{IC}\xspace}
\newcommand{\lt}{\textsf{LT}\xspace}

\newlength\lengtha 

\definecolor{forestgreen}{RGB}{34, 139, 34}

\definecolor{RYB1}{RGB}{192, 128, 255}
\definecolor{RYB2}{RGB}{255, 192, 32}
\definecolor{RYB3}{RGB}{139, 0, 0}
\definecolor{RYB4}{RGB}{0, 128, 255}

\newenvironment{customlegend}[1][]{%
    \begingroup
    \csname pgfplots@init@cleared@structures\endcsname
    \pgfplotsset{#1}%
}{%
    \csname pgfplots@createlegend\endcsname
    \endgroup
}%

\def\addlegendimage{\csname pgfplots@addlegendimage\endcsname}

\makeatletter
\newcommand\footnoteref[1]{\protected@xdef\@thefnmark{\ref{#1}}\@footnotemark}
\makeatother

\let\oldnl\nl% Store \nl in \oldnl
\newcommand{\nonl}{\renewcommand{\nl}{\let\nl\oldnl}}% Remove line number for one line

\newcommand{\trref}[1]{Appendix~{#1}}

\newcommand\vldbdoi{XX.XX/XXX.XX}
\newcommand\vldbpages{XXX-XXX}
\newcommand\vldbvolume{18}
\newcommand\vldbissue{2}
\newcommand\vldbyear{2025}
\newcommand\vldbauthors{\authors}
\newcommand\vldbtitle{\shorttitle} 
\newcommand\vldbavailabilityurl{https://github.com/waetr/IM_GM}
\newcommand\vldbpagestyle{empty}

\begin{document}
\title{Efficient and Effective Algorithms for A Family of Influence Maximization Problems with A Matroid Constraint}
\subtitle{[Technical Report]}

%%
%% The "author" command and its associated commands are used to define the authors and their affiliations.
\author{Yiqian Huang}
\orcid{0009-0001-5601-3439}
\affiliation{%
  \institution{National University of Singapore}
}
\email{yiqian@comp.nus.edu.sg}

\author{Shiqi Zhang}
\orcid{0000-0002-7155-9579}
\affiliation{%
  \institution{National University of Singapore}
}
\affiliation{%
  \institution{PyroWis AI}
}
\email{shiqi@pyrowis.ai}

\author{Laks V.S. Lakshmanan}
\orcid{0000-0002-9775-4241}
\affiliation{%
  \institution{University of British Columbia}
}
\email{laks@cs.ubc.ca}

\author{Wenqing Lin}
\orcid{0000-0003-4741-801X}
\affiliation{%
  \institution{Tencent}
}
\email{edwlin@tencent.com}

\author{Xiaokui Xiao}
\orcid{0000-0003-0914-4580}
\affiliation{%
  \institution{National University of Singapore}
}
\email{xkxiao@nus.edu.sg}

\author{Bo Tang}
\orcid{0000-0001-8424-0092}
\affiliation{%
  \institution{Southern University of Science and Technology}
}
\email{tangb3@sustech.edu.cn}

\begin{abstract}
Influence maximization (\im) is a classic problem that aims \edit{to identify} a small group of critical individuals, known as seeds, who can influence the largest number of users in a social network through word-of-mouth. This problem finds important applications including viral marketing, infection detection, and misinformation containment. The conventional \im problem is typically studied with the oversimplified goal of selecting a \textit{single} seed set. Many real-world scenarios call for \textit{multiple} sets of seeds, particularly on social media platforms where various viral marketing campaigns need different sets of seeds to propagate effectively. To this end, previous works have formulated various \im variants, central to which is the \laks{requirement} of multiple seed sets, \laks{naturally modeled as} a matroid constraint. However, the current best-known solutions for these variants either offer a \laks{weak} $(1/2-\epsilon)$-approximation, or offer a $(1-1/e-\epsilon)$-approximation \laks{algorithm that is very expensive}. We propose an efficient seed selection method called \cseedselect, an algorithm with a $(1-1/e-\epsilon)$-approximation guarantee for this family of \im variants. To further improve efficiency, we also devise a fast implementation, called \ouralgo. \revise{We extensively evaluate the performance of our proposal against 6 competitors across 4 \im variants and on 7 real-world networks, demonstrating that our proposal outperforms all competitors in terms of result quality, running time, and memory usage.}\eat{Additionally,} We have also deployed \ouralgo in a real industry strength application involving online gaming, where we show that our deployed solution significantly improves upon the baselines.
\end{abstract}

\maketitle

\pagestyle{\vldbpagestyle}
\begingroup\small\noindent\raggedright\textbf{PVLDB Reference Format:}\\
\vldbauthors. \vldbtitle. PVLDB, \vldbvolume(\vldbissue): \vldbpages, \vldbyear.\\
\href{https://doi.org/\vldbdoi}{doi:\vldbdoi}
\endgroup
\begingroup
\renewcommand\thefootnote{}\footnote{\noindent
This work is licensed under the Creative Commons BY-NC-ND 4.0 International License. Visit \url{https://creativecommons.org/licenses/by-nc-nd/4.0/} to view a copy of this license. For any use beyond those covered by this license, obtain permission by emailing \href{mailto:info@vldb.org}{info@vldb.org}. Copyright is held by the owner/author(s). Publication rights licensed to the VLDB Endowment. \\
\raggedright Proceedings of the VLDB Endowment, Vol. \vldbvolume, No. \vldbissue\ %
ISSN 2150-8097. \\
\href{https://doi.org/\vldbdoi}{doi:\vldbdoi} \\
}\addtocounter{footnote}{-1}\endgroup

%% VLDB block start %%%
\ifdefempty{\vldbavailabilityurl}{}{
\vspace{.3cm}
\begingroup\small\noindent\raggedright\textbf{PVLDB Artifact Availability:}\\
The source code, data, and/or other artifacts have been made available at \url{\vldbavailabilityurl}.
\endgroup
}

\section{Introduction}
Given a social network, the problem of influence maximization (\im) seeks \LL{to find} a small group of users who can (directly \laks{or} indirectly) influence as many users in the network as possible. It has been applied in various domains, including viral marketing~\cite{domingos2001mining,iyengar2011opinion}, efficient infection detection~\cite{leskovec2007cost}, and misinformation containment~\cite{budak2011limiting,simpson2022misinformation, tong2018misinformation}, among others. Viral marketing, for instance, typically involves a company setting up a campaign to promote a product by incentivizing a \eat{single}set of influential users (called ``seeds''), aiming to \edit{maximize the number of influenced users (referred to as ``spread'')} by \laks{triggering} a cascade of adoptions through word-of-mouth. Existing work on \im has mostly focused on this setting of \textit{single seed set}.

Real-world scenarios of viral marketing are often more complex, requiring \textit{multiple} sets of seeds, to optimize the \edit{overall spread produced by these sets}. This complexity is clearly evident in the rise of influencer marketing platforms, including the official creator marketplaces of social media platforms such as Instagram~\cite{instagram}, TikTok~\cite{tiktok}, and Douyin~\cite{douyin}, as well as over 80 third-party platforms~\cite{IMplatform}. These platforms act as intermediaries, aligning advertisers with key influencers for social advertising campaigns and aiding influencers in monetizing their online presence. \eat{For example}\yiqiann{E.g.,} the Douyin Xingtu platform reported a substantial user base of over 1.9 million advertisers and 3.2 million creators as of June 2022\edit{~\cite{oceanengine2022}}. A critical task for them\eat{these platforms} is coordinating diverse advertiser campaigns\eat{\edit{such as brand alliances of non-competitive businesses,}} and determining optimal seed sets \sq{such that the content created by the seeds can reach a broader audience through word-of-mouth.}
The above viral marketing scenarios that seek multiple seed sets can be modeled as an \im problem subject to a \textit{general matroid} constraint (\ourproblem). A general matroid \laks{(formal definition in \S~\ref{sec:imgm})} represents constraints that an \im solution must adhere to. It is defined as a pair $(U,\I)$, where $U$ is a ground set of elements from which seeds are selected, and $\I$ is a collection of subsets of $U$ (called independent sets), representing feasible solutions to the problem. For example, we show an instance of \ourproblem in the context of {\em revenue maximization}.
%\yiqiann{To further illustrate this, we introduce the following example of \ourproblem formulation in the application of coordinating campaigns.}
% \vspace{-0.2mm}
\begin{example}[Revenue Maximization\eat{ (\rim)}]
    Consider the scenario of incentivized social advertising~\cite{aslay2017revenue, han2021efficient} where we need to coordinate $T$ viral marketing campaigns, such that seed sets $S_i, S_j$ selected for different campaigns $i\neq j$ do not overlap, \edit{and the sum of spread produced by all campaigns is maximized. This scenario cannot be trivially treated as independently \laks{running} \im $T$ times, as a seed can \eat{coordinate}\laks{contribute to} more than one campaign, motivating the introduction of a matroid.} In this case, the ground set $U$ comprises all possible combinations of users and campaigns, while $\I$ corresponds to all feasible solutions \LL{$(S_1, ..., S_T)$ which satisfy $S_i\cap S_j = \emptyset$, $i\neq j$.} \qed 
\end{example}
%In addition, %Apart from viral marketing with multiple seed sets for different campaigns, 
%there exist numerous other real-world problems \cite{XXX} that can be modeled as
% \edit{The \ourproblem formulation exhibits some unique features that differentiate it from the vanilla \im. In the above example, if adopting the vanilla \im solution, all users will be selected as seeds for the first examined campaign. But for \ourproblem, it is desirable to distribute the budget across other campaigns to generate more spreads.}

The \ourproblem~\LL{formulation} also \LL{captures} numerous other real-world \laks{applications}~\cite{sun2018multi,chen2020scalable,sun2023scalable,zhang2023capacity,liao2023popularity}. %\laks{We illustrate two of them below.} 
%can also be applied in other real-world scenarios. 
We illustrate two such applications: {\em multi-round \im} and {\em adversarial attacks on \im}.

% \vspace{-0.2mm}
\begin{example}[Multi-Round \im\eat{ (\mrim)}] 
    In the recruitment-based campaign of the Xingtu \eat{platform}\laks{mobile app}~\cite{xingtu2023}, service providers select seeds from a candidate pool in multiple rounds, %These rounds occur each time an advertiser reviews the current pool before deciding to launch a single campaign. 
such that each round's seed set has a size $\leq k$ \edit{and the spread across all rounds is maximized}; this cardinality constraint can be captured using a\eat{general} matroid.  \qed  %which can also be represented by a general matroid. 
\end{example}

%Finally, we consider the application of limiting illegal activities.
\begin{example}[Adversarial Attacks on \im\eat{ (\advim)}]
In combating the spread of illicit activities such as pyramid scheme scams~\cite{feng2021case}, money laundering~\cite{isolauri2022money}, and phishing scams~\cite{vishwanath2015diffusion}, authorities identify not only \textit{malicious users} but also \textit{suspicious relationships} for blocking in order to control further spread \laks{of such activities.} %The rationale for blocking specific relationships, such as transactions and emails, is that innocent victims can be inadvertently involved in activities like making unsafe cross-country wire transfers or forwarding phishing emails. 
\laks{Owing to the sensitivity of blocking, in practice there are limits on the number of users and relationships that can be blocked.} 
This problem can be modeled as an instance of \ourproblem, with a\eat{general} matroid constraining the maximum number of \yiqiann{identified} users and relationships \laks{for blocking}. \qed 
\end{example}
% \citet{zhang2023capacity} also found the need for multiple seed sets in online gaming platforms, where \textit{different sets of close and influential friends} are chosen for different active players. This strategy is pivotal for effectively disseminating information and generating interest in specific gaming activities.

% To identify multiple pivotal seed sets for these scenarios, different \im variants can be adopted. For instance, multi-product influence maximization~\cite{du2017scalable} and revenue maximization~\cite{han2021efficient,aslay2017revenue} are suitable for viral marketing with different campaign seed sets. Non-adaptive multi-round influence maximization~\cite{sun2018multi} is apt for scenarios requiring multiple seed sets before a single campaign launch. In combating the spread of illicit activities, adversarial attacks on influence maximization~\cite{sun2023scalable} help identify suspicious users and relationships.

There exist a number of solutions to  \ourproblem~\cite{du2017scalable,sun2018multi,aslay2017revenue,tang2018social,el2020fairness,chen2020scalable,han2021efficient,sun2023scalable,zhang2023capacity,liao2023popularity,ChanN15}, most of which utilize the greedy approach introduced by~\citet{fisher1978analysis}. The greedy approach iteratively selects the element with the largest improvement of the objective function under the matroid constraint, resulting in a $(1/2-\epsilon)$-approximation for any given $\epsilon >0$. \LL{Subsequent work \cite{calinescu2011maximizing} improves upon this result and presents a $(1-1/e-\epsilon)$-approximation algorithm for submodular maximization under a matroid constraint}. Unlike the greedy approach, it first \edit{iteratively searches for} a fractional solution which contains a {\it fraction} for each element in $U$, with each fraction representing the probability that the element is selected. It then converts the fractional solution into a discrete one, by randomly rounding each fraction to either $0$ or $1$. However, this algorithm incurs significant computational cost and fails to handle even datasets with a few hundred users %~\cite{ozcan2021submodular} 
(see \S~\ref{sec:selection-overview}), because it requires sampling a large number of \edit{random} subsets of $U$ in the evaluation of candidate fractional solutions.

To address this issue, we propose \cseedselect, an algorithm that ensures $(1-1/e-\epsilon)$-approximation for \ourproblem while offering superior efficiency. 
\edit{\cseedselect shares a similar framework as mentioned above but performs the search and rounding steps \textit{deterministically}. The linchpin lies in the fact that \cseedselect avoids expensive random sampling in the assessment of fractional solutions by utilizing a new estimation method, which reduces computation costs by exploiting the characteristics of fractional solutions in \ourproblem.}
\eat{\cseedselect follows a two-step framework: the first step iteratively refines an approximate fractional solution, while the second step rounds this fractional solution into a feasible solution under the matroid constraint.}
We also present \ouralgo, an algorithm that further improves \cseedselect's efficiency by \yiqiann{adapting the doubling strategy from~\citet{tang2018online}, but with careful redesign to ensure the correctness for \ourproblem.} %, \ouralgo iteratively increases the number of RR sets, until achieving the desired approximation guarantee.
\revise{We experimentally evaluate the performance of \cseedselect and \ouralgo against 6 competitors across 4 \ourproblem instances, on \revise{7} real-world networks \eat{with up to 1.5 billion edges}ranging in size from thousands of nodes and edges to millions of nodes and billions of edges}. Notably, \cseedselect outperforms all competitors by up to 871\% in terms of result quality. \revise{Besides, \ouralgo surpasses the state-of-the-art by up to $1,000\times$ in running time and $10\times$ in memory usage, respectively.}\eat{In addition, \ouralgo has a superior memory footprint in all cases.}
In addition, we have deployed our solution \ouralgo on a Tencent online gaming platform, and observed that it attracted 160,000 more user engagements compared to the control group using the current state-of-the-art solution. 

To summarize, we make the following contributions in this work:
\begin{itemize}[topsep=2pt,itemsep=1pt,parsep=0pt,partopsep=0pt,leftmargin=11pt]
    \item We propose \laks{an efficient} $(1-1/e-\epsilon)$-approximation solution \cseedselect for \ourproblem (\S~\ref{sec:greedy}).  
    \item We provide a scalable implementation \ouralgo for \laks{further} improving the efficiency of \cseedselect (\S~\ref{sec:opimc}). 
    \item \revise{Our extensive experiments\eat{ across all three settings \rim, \mrim, and \advim} show that \cseedselect and \ouralgo can efficiently scale to datasets with billions of edges (\S~\ref{sec:exp})}. 
    \item We deploy our solution \sq{to a Tencent online game and significantly improve on the state of the art in user engagement}\eat{achieve \laks{a significant boost in user engagement} over existing practice} (\S~\ref{sec:deploy}). 
\end{itemize}

\section{Preliminaries}
We provide basic terminology and notations in \S~\ref{sec:notations}, followed by formulating influence maximization subject to a general matroid constraint (\ourproblem) in \S~\ref{sec:imgm}. We review typical instances of \ourproblem and their solutions in \S~\ref{sec:imgm-instance} and \S~\ref{sec:existing}, respectively.

\vspace{-0.5mm}
\subsection{Terminology and Notations}\label{sec:notations}
\stitle{Social networks}
A social network is a directed graph $\G=(V, E)$, where $V$ is a set of nodes representing users and $E$ is a set of edges representing relationships. Each directed edge $e_{i,j}\in E$ is associated with $p_{i,j}\in [0,1]$, and indicates that $v_j$\eat{is a follower of and} can be influenced by $v_i$ with probability $p_{i,j}$. We say $v_i$ (resp. $v_j$) the in-neighbor (resp. out-neighbor) of $v_j$ (resp. $v_i$), and denote the set of in-neighbors \edit{(resp. out-neighbors) of $v_i$} as $N^{in}_{i}$ (resp. $N^{out}_{i}$).
% For an undirected graph, we replace each undirected edge $(v_i,v_j)$ with two directed ones in opposing directions, \ie $(v_i,v_j)$ and $(v_j,v_i)$.

%In addition, certain types of matroids are particularly noteworthy. One example is the 

\stitle{Monotone and submodular functions}
Given a\eat{ground} set $U$, a non-negative set function $f$ is \textit{monotone} if $\forall{X}\subseteq Y\subseteq U, f(X) \leq f(Y)$. $f$ is \textit{submodular} if $\forall{X} \subseteq Y \subseteq U$ and $u_i\in U\backslash Y, f(u_i|X) \geq f(u_i|Y)$, where $f(u_i|X)=f(X\cup\{u_i\})-f(X)$ \yiqiann{denotes} the \textit{marginal gain} of $f$ after adding $u_i$ to $X$. As an example, given a collection $\R$ of subsets of $U$, the \textit{coverage} \yiqiann{function} of a set $S\subseteq U$, defined as
\begin{equation}\label{eq:coverage-function-raw}
\textstyle \Lambda_\R(S)=\sum_{R\in \R}\mathbb{I}(S\cap R \not=\emptyset), 
\end{equation}
where $\mathbb{I}(\cdot)$ is the indicator function, is \yiqiann{monotone and submodular.}\eat{Here, $\Lambda_\R(u_i|S)$ represents the marginal gain in the context of the coverage function.}

\stitle{Notations}
Throughout this paper, %we consistently utilize specific notation styles to represent different mathematical objects. An 
an upper-case letter $A$ (resp. a calligraphic upper-case letter $\mathcal{A}$) denotes a set (resp. a collection of sets). For a ground set $U$, a boldface lower-case letter $\mathbf{x} \in [0,1]^U$ represents a vector corresponding to $n$ elements of $U$, where $\mathbf{x}[i]$ denotes the value associated with the element $u_i\in U$. As a special case, $\mathbf{1}_i$ (resp. $\mathbf{1}_S$) signifies the indicator vector with value 1 for $u_i$ (resp. for each element $u_i\in S$) and value 0 otherwise. Table~\ref{tab:notations} lists the frequently used notations.

\begin{table}[!t]
\centering
\renewcommand{\arraystretch}{1.1}
\begin{footnotesize}
\caption{Frequently used notations.}\vspace{-2.4mm} \label{tab:notations}
\resizebox{\columnwidth}{!}{
\begin{tabular}{rp{2.7in}}	
    \toprule
    \bf Notation & \bf Description \\
    \midrule
    $\G=(V,E)$   &  A graph $\G$ with node set $V$ and edge set $E$.\\
    $\M=(U,\I)$   &  A matroid $\M$ with ground set $U$ and collection $\I$ of independent sets.\\
    $n$  &  The number of elements in $U$. \\ 
    $r, \BPoly$   &  The rank \edit{of $\M$} and the collection of all bases in $\M$.\\
    $h, k_i$   &  The number of partitions, and the capacity of the $i$-th partition in\eat{partition matroid} $\M$.\\
    $R,\R$   &  A random RR set and a collection of random RR sets.\\
    $\Lambda_\R,\sigma$   &  The coverage function \edit{in Eq.\eqref{eq:coverage-function-raw}} and the objective \edit{in Definition~\ref{def:imgm}}.\\
    % $\Lambda_\R(u_i|S)$   &  The marginal coverage $\Lambda_\R(S\cup\{u_i\})-\Lambda_\R(S)$.\\
    $\mathbf{x}$   &  The fractional solution of \edit{element} selection in\eat{ Eq.\eqref{eq:fractional}} \yiqiann{\S~\ref{sec:greedy}}.\\
    % \edit{$\Q_\R, q_R$}   & The collection of variables defined in Algorithm~\ref{alg:greedy-framework}, and the element in Eq.\eqref{eq:qr}.\\
    % $S^o, S^*$   &  The optimal result set for maximizing $\Lambda_\R(\cdot)$ and $\sigma(\cdot)$ subject to the matroid constraint.\\
    % $\epsilon_a$, $\epsilon_s$, $\epsilon_r$ & The error terms in the whole algorithm, node selection and RR set generations\\
    % $F$   &  The multilinear extension of a set function.\\
    % $\Omega(\mathbf{x})$   &  A random set containing each $u\in N$ independently with probability $x_u$.\\
    % $\mathbf{1}_S$   &  The indicator vector of set $S$.\\
    \bottomrule
\end{tabular}
}
\end{footnotesize}
% \vspace{-2.1mm}
\end{table}

% \vspace{-0.2mm}
\subsection{Problem Formulation}\label{sec:imgm}

% \edit{Prior to formulating Influence Maximization subject to the General Matroid Constraint (\ourproblem), this section first introduces necessary concepts about the diffusion model and matroid.}

% \stitle{Diffusion models}
Given a graph $\G=(V,E)$ and a set $S\subseteq V$ of chosen nodes (called seeds), a \textit{diffusion model} describes the process by which information spreads from $S$ to other nodes via social connections in $\G$. In this work, we consider the well-known Independent Cascade (\ic)~\cite{goldenberg2001talk} and Linear Threshold (\lt)~\cite{granovetter1978threshold} models, \edit{where a diffusion instance captures the stochastic diffusion process from $S$ and unfolds in discrete steps. At the beginning of a diffusion instance 
$\Lset$ of \ic or \lt, all seeds in $S$ are set to be active at step $0$, leaving other nodes inactive. In \yiqiann{\ic}}, at step $t>0$, each \edit{node} $v_i$ activated at step $t-1$ has one chance to independently activate its inactive out-neighbor $v_j$ with probability $p_{i,j}$. \eat{The \lt model}\yiqiann{\lt} requires that for each node $v_j$, (i) $\sum_{v_i\in N^{in}_j}p_{i,j}\leq 1$, and (ii) a threshold $\lambda_j\in [0, 1]$ is sampled uniformly at step $0$. At step $t>0$, an inactive node $v_j$ is activated \eat{if and only if}\yiqiann{iff} the sum of $p_{i,j}$ w.r.t.\ $v_j$'s activated in-neighbors $v_i$ exceeds $\lambda_j$. During a diffusion instance \edit{$\Lset$} of \ic or \lt, once a node is activated, it remains active in all subsequent steps. \edit{$\Lset$} terminates if no more nodes can be activated, and the set of eventually activated nodes is denoted as $\Gamma_\Lset(\G, S)$. Accordingly, the \edit{\textit{spread} of $S$ in $G$} is defined as 
\begin{equation}\label{eq:spread}
    \edit{\rho_{\G}(S)=\mathbb{E}[|\Gamma_\Lset(\G, S)|].}
\end{equation}

% These models capture the diffusion of a given item from $S$ in a stochastic manner and can be conceptualized through the concept of \textit{live-edge graph}~\cite{kempe2003maximizing}.
% Specifically, both models generate the live-edge graph $\Lset=(V, E_l)$, where $E_l$ is a subset randomly sampled from $E$. In the \ic model, $\Lset$ is sampled with the probability $\pr(\Lset)= \prod\limits_{e_{i,j}\in E_l}p_{i,j}\prod\limits_{e_{i,j}\in E\backslash E_l}(1-p_{i,j})$, whereas \lt samples $\Lset$ with the probability $\pr(\Lset)= \prod_{v_j\in V}\pr(N^{l}_j)$, where $N^{l}_j=E_l\cap N^{in}_j$ and
% $$\pr(N^{l}_j)=
% \begin{cases}
% p_{i,j}:e_{i,j}\in N^{l}_j & |N^{l}_j|= 1\\
% 1-\sum_{e_{i,j}\in N^{in}_j}p_{i,j} & N^{l}_j= \emptyset, \\
% 0& \text{otherwise}
% \end{cases}$$
% assuming $\sum_{e_{i,j}\in N^{in}_j}p_{i,j}\leq 1$. 

Based on these concepts, \citet{kempe2003maximizing} define the influence maximization (\im) problem as follows.

\vspace{-0.3mm}
\begin{definition}[\im]\label{def:im}
    Given a graph $\G$, a diffusion model and a cardinality $k$, the \im problem is to select a seed set $S\subseteq V$ with $|S|\leq k$ such that the spread $\rho_{\G}(S)$ defined in Eq.\eqref{eq:spread} is maximized. 
\end{definition}
\vspace{-0.3mm}

\edit{We next review concepts of matroids and then propose a generalized \im problem, viz., \im subject to a general matroid (\ourproblem).}

\stitle{Matroids} 
A \textit{general matroid} is a pair $\M=(U, \I)$, where $U$ is a finite set (called the ground set) with $|U|=n$ and $\I$ is a collection of subsets of $U$ (called independent sets) satisfying the following axioms: (i) $\I\not=\emptyset$; (ii) if $I\in \I$ and $J\subseteq I$, then $J \in \I$; (iii) if $I, J\in \I$ and $|I|<|J|$, then there exists $u_i\in J\backslash I$ such that $I\cup\{u_i\}\in \I$.
Of special interest is the \textit{partition matroid}, whose ground set $U$ is divided into $h\geq 1$ disjoint partitions $U_1, \dots, U_h$, each $U_i$ associated with a positive integer $k_i$. The set of independent sets $\I$ in the partition matroid is defined as $\I =\{S \subseteq U : |S \cap U_i| \leq k_i, \forall i \in[h]\}$. In the special case where $h=1$, the partition matroid reduces to a \textit{uniform matroid}, characterized by $\I=\{S \subseteq U: |S|\leq k\}$. 

Given a matroid $\mathsf{M}=(U, \I)$, any independent set $B \in \I$ with the largest cardinality, \ie $|B|=\max\{|I|:I\in \I\}$, is a \textit{base} and its cardinality $r$ is the \textit{rank} of $\M$. \yiqiann{E.g., $r=\sum_{i=1}^{h}k_i$ in the partition matroid and $r=k$ in the uniform matroid.} \LL{A matroid can have multiple bases.} We denote the collection of all bases of $\M$ as $\BPoly$.\eat{Notably, the rank $r$ represents the maximum number of elements that can be selected in the matroid, \eg $r=\sum_{i=1}^{h}k_i$ in the partition matroid and $r=k$ in the uniform matroid.}

\vspace{-0.3mm}
\begin{definition}[\ourproblem
]\label{def:imgm}
Given a graph $\G$, a diffusion model, and a matroid $\M=(U,\I)$ where $U$ is built on $\G$, \ourproblem aims to find a set $S\in\I$ that maximizes a given objective function $\sigma:2^U\rightarrow \mathbb{R}_+$, which is associated with the spread function in Eq.\eqref{eq:spread}. 
\end{definition}
\vspace{-0.3mm}

\edit{Observe in Definitions~\ref{def:im}-\ref{def:imgm} that \im is an instance of \ourproblem, where $\M$ is a uniform matroid and $\sigma(S)=\rho_{\G}(S)$. Besides \im, a plethora of \im variants~\cite{du2017scalable,sun2018multi,aslay2017revenue,chen2020scalable,han2021efficient,sun2023scalable,zhang2023capacity,liao2023popularity,ChanN15} can also be \laks{modeled using} \ourproblem.} \sq{The next section will discuss three instances, \LL{obtained from different instantiations of $\M$ and $\sigma$.}}

% \edit{Based on Definitions~\ref{def:im} - ~\ref{def:imgm}, the difference between \im and \ourproblem is twofold.
% First, in contrast to the cardinality constraint in \im, \ourproblem asks for a seed set $S$ satisfying a matroid constraint $\M$. 
%\eat{For example, one \ourproblem instance, called revenue maximization (\rim)~\cite{han2021efficient,aslay2017revenue, du2017scalable}, describes a scenario where $T$ advertisers individually request seeds for their campaigns, intending to host them collectively as an alliance. However, in this case, each seed $v_i$ is only able to coordinate $k_i\leq T$ campaigns. This results in the scenario not being trivially treated as independently repeating \im for $T$ times, as the number of campaigns seed $v_i$ coordinates may exceed $k_i$. This motivates the introduction of a partition matroid for \rim and highlights the importance of \ourproblem.}
% Second, the objective function in \ourproblem is not limited to $\rho_{\G}(S)$. 
%\eat{For example, the objective of \rim is the weighted sum of spreads from $T$ campaigns, reflecting the total revenue.}
% It is worth noting that \im is also an instance of \ourproblem, where $\M$ is a uniform matroid and $\sigma(S)=\rho_{\G}(S)$.
% % (\ie $U=V$, $\I=\{S\subseteq U:|S|\leq k\}$)
% }

%Formally, $S=\argmax{S\in\I}\sigma(S).$ 
% Here, $\sigma(\cdot)$ is defined as the form of the expected value over the distribution of the live edge graph(s).

\subsection{Representative \ourproblem Instances}\label{sec:imgm-instance}

% \begin{table*}[!t]
% \centering
% \renewcommand{\arraystretch}{1.2}
% \caption{Summarize the settings of \ourproblem instances.}\vspace{-2mm}
% \begin{small}
% \begin{tabular}{|c|c|c|c|c|}
% \hline
% \textbf{Instance}  & \textbf{Parameters}     & \textbf{Ground set}      & \textbf{Independent sets}      & \textbf{Objective function}  \\ \hline
% \textbf{Vanilla \im}  & $k$     & $V$      & $\{S:|S|\leq k\}$      & $\rho_{D,\G}(S)$  \\ \hline
% \textbf{Revenue maximization}  & $T,\{\alpha_t\}_{t\in[T]}, \{c_i\}_{i\in[n]}$     & $V\times [T]$      & $\{\bar{S}:\forall i\in [n], |\bar{S}\cap \bar{V}_i|\leq c_i\}$      & $\sum_{t=1}^T \alpha_t \cdot \rho_{D,\G}(S_t)$  \\ \hline
% \textbf{Multi-round \im}  & $k, T$     & $V\times [T]$      & $\{\bar{S}:\forall l\in [T], |\bar{S}\cap (V\times t)|\leq k\}$      & $\mathbb{E}[|\cup_{l=1}^T \Gamma(L_t, S_t)|]$  \\ \hline
% \textbf{Adversarial attack on \im}  & $k_N, k_E, S\subseteq V$     & $V\cup E$      & $\{A: |A\cap V|\leq k_N, |A\cap E|\leq k_E\}$      & $\rho_{D,\G}(S)-\rho_{D,\G'}(S)$  \\ \hline
% \end{tabular}
% \end{small}
% \vspace{-2mm}
% \end{table*}

\eat{\edit{This section further discusses three representative instances in the \ourproblem family, \LL{obtained from different instantiations of $\M$ and $\sigma$.}}}

\stitle{Revenue maximization (\rim)~\cite{han2021efficient,aslay2017revenue, du2017scalable}} 
Given a graph $G=(V,E)$, \yiqiann{a diffusion model, and} $T$ campaigns associated with $T$ unit revenues $\alpha_1,\dots,\alpha_T$,\eat{and a diffusion model describing how campaigns interact with each other, } the \rim problem aims to select seed sets $S_1, S_2, \dots, S_T\subseteq V$ for the $T$ campaigns such that the total revenue $\textstyle \sigma(S_1, S_2, \dots, S_T)=\sum_{t=1}^T \alpha_t \cdot \rho_G(S_t)$ is maximized.
\LL{This problem is subject to a user constraint that each node $v_i\in V$ can be included in at most $k_i$ seed sets, where $k_i \leq T$ is a number associated with node $v_i$.
This constraint corresponds to a partition matroid $\M = (U,\I)$, where the ground set $U=V\times [T]$ contains all pairs of nodes and campaigns. $U$ is partitioned into $|V|$ disjoint subsets $U_1, U_2,\dots,U_{|V|}$, where $U_i=\{(v_i,t): t\in[T]\}$. Accordingly, $\I=\{S\subseteq U:\forall v_i\in V, |S\cap U_i|\leq k_i\}$. As a special case, we have $k_i=1$ for all $v_i\in V$ in the scenario of incentivized social advertising~\cite{aslay2017revenue, han2021efficient}.} 
 
\stitle{Non-adaptive multi-round \im (\mrim)~\cite{sun2018multi}} 
Given a graph $G=(V,E)$, \edit{a diffusion model}, a cardinality $k$ and the number of campaign rounds $T$, the non-adaptive \mrim problem aims to select seed set $S_t\subseteq V$ in each round $t$, with $|S_t|\leq k$, such that the expected number of activated nodes over all rounds $\textstyle \sigma(S_1, S_2, \dots, S_T)$ $\textstyle=\mathbb{E}\left[\left|\bigcup_{t=1}^T \Gamma_{\Lset_t}(G, S_t)\right|\right]$ is maximized, where \edit{each diffusion instance $\Lset_t$ is obtained independently.}
% Akin to the revenue maximization, 
\LL{This constraint can also be viewed as a partition matroid with the ground set $U=V\times [T]$. However, in contrast with \rim, this ground set is partitioned w.r.t.\ rounds, \ie $U_t=\{(v_i,t):v_i \in V\}$, and $\I=\{S\subseteq U:\forall t\in [T], |S_t|\leq k\}$. } 

\stitle{Adversarial attacks on \im (\advim)~\cite{sun2023scalable}} 
Given a graph $G=(V,E)$, \edit{a diffusion model}, a set of contagious seeds $A \subset V$ and two cardinalities $k_v, k_e$, the \advim problem aims to select a blocking \LL{set $S$ consisting of a blocking} node set $S_v \subseteq V\backslash A$ with $|S_v|\leq k_v$ and a blocking edge set $S_e \subseteq E$ with $|S_e|\leq k_e$, \LL{\ie $S = S_v \cup S_e$,} such that the influence reduction $\sigma(S_v, S_e)=\rho_G(A)-\rho_{G\backslash S}(A)$ is maximized, where $G\backslash S=(V\backslash S_v, E\backslash S_e)$. \LL{The constraint in this problem corresponds to \yiqiann{a partition} matroid with ground set $U=(V\backslash A)\cup E$ and independent sets $\I=\{S\subseteq U: |S_v| \leq k_v, |S_e| \leq k_e\}$. } 

\vspace{-1mm}
\subsection{Approximation Solutions for \ourproblem{}s}\label{sec:existing}

Due to the NP-hardness of each \ourproblem problem and the inefficiency of Monte-Carlo simulations, most previous works borrow the idea of \textit{reverse reachable (RR) set} sampling~\cite{borgs2014maximizing}, originally designed for vanilla \im. \revise{Specifically, an RR set consists of nodes identified by simulating a reverse diffusion process of a given diffusion model from a node $v_i$ selected uniformly at random. For example, the reverse diffusion process of the \ic model is a stochastic breadth-first search starting from $v_i$. For each visited $v_a$, it explores each in-neighbor $v_b \in N^{in}_a$ with probability $p_{b,a}$.}
\citet{borgs2014maximizing} show that for any seed set $S\subseteq V$, the probability that an RR set overlaps $S$ equals $\frac{\rho_{\G}(S)}{|V|}$. \edit{That is,} if we have a \edit{collection} $\R$ of RR sets, we can use $\frac{|V|}{|\R|}\cdot \Lambda_{\R}(S)$ as an unbiased estimator of $\rho_{\G}(S)$. 
Accordingly, \citet{borgs2014maximizing} propose an approximation solution for \im, \yiqiann{which runs in two main steps:}
(i) \textit{RR set generation}: sample a set $\R$ of RR sets, with the size $|\R|$ sufficient to achieve the desired approximation; (ii) \textit{\edit{element} selection}: use a greedy algorithm for solving maximum coverage which returns a set $S$ that maximizes the coverage $\Lambda_\R(S)$.
% \begin{enumerate}
% [topsep=2pt,itemsep=1pt,parsep=0pt,partopsep=0pt,leftmargin=16pt]
%     \item \textit{RR set generation}: sample a set $\R$ of random RR sets, with the size $|\R|$ sufficient to achieve the desired approximation.
%     \item \textit{Node selection}: use a greedy algorithm for solving the maximum coverage problem that returns the seed set $S$ to maximize the coverage $\Lambda_\R(S)$.
% \end{enumerate}
\sq{Borgs et al.'s framework is also leveraged in various \ourproblem instances~\cite{aslay2017revenue,chen2020scalable, han2021efficient,sun2018multi,sun2023scalable,zhang2023capacity,liao2023popularity}. A central role in determining the approximation is played by various greedy element selection algorithms used in these works, briefly reviewed below.}

\stitle{\mgreedy~\cite{fisher1978analysis}}
This algorithm greedily selects elements with the maximum marginal gain subject to the matroid constraint. Specifically, it starts with a set $S=\emptyset$, and then iteratively adds $r$ elements to $S$. In each iteration, it adds $u_i$ to $S$ if $\Lambda_\R(u_i|S)$ is the largest among the \edit{elements} of $U\backslash S$ and the inclusion maintains the matroid constraint. This algorithm achieves a $1/2$-approximation subject to a general matroid constraint~\cite{fisher1978analysis} and is the most commonly adopted algorithm in prior works for \ourproblem problems~\cite{sun2018multi,sun2023scalable,aslay2017revenue,han2021efficient,zhang2023capacity,chen2020scalable}.

\stitle{\lgreedy~\cite{sun2018multi}} Designed for the partition matroid constraint, \lgreedy selects elements partition by partition. Specifically, it greedily selects elements in the first partition with \yiqiann{capacity} $k_1$. Once $k_1$ elements are selected, it proceeds to the second partition, and so on. This improves the efficiency by pruning the search space from $U$ to each partition. \citet{sun2018multi} prove that \lgreedy achieves a $1-e^{-(1-\frac{1}{e})}\approx 0.46$ approximation. 
% It does this by considering Local-Greedy as nested greedy processes: the inner process selects elements within a partition and achieves a $1-1/e$ approximation in this partition, while the outer process treats the selection in each partition as an item, achieving a $1-1/e$ approximation of the maximum marginal contribution.

\stitle{\tgreedy~\cite{buchbinder2017comparing}}
Unlike \mgreedy, \tgreedy adds {an element} $u_i$ to $S$ if $\Lambda_\R(u_i|S)$ exceeds a threshold, which starts at a predefined value and decreases by $(1-\xi)$ in each iteration. Compared to \mgreedy, this algorithm boosts efficiency by reducing the number of iterations from $r$ to $O\left(\xi^{-1}\cdot\ln (r/\xi)\right)$, \LL{while incurring a loss term} $\xi$, resulting in an approximation ratio of $1/2-\xi$~\cite{buchbinder2017comparing}.

\sq{Due to the variety of diffusion models and objectives\eat{functions}, \eat{different }\ourproblem solutions require two additional steps before  RR set generation\eat{phase}:} (i) select a strategy of RR set construction for an \ourproblem instance, \yiqian{where an RR set $R$ consists of \textit{elements} in $U$ instead of nodes in $V$}; (ii) rewire the coverage $\Lambda_{\R}$ and the objective $\sigma$ such that the scaled $\Lambda_\R(S)$ is an unbiased estimator of $\sigma(S)$, \ie $\sigma(S)=\frac{\kappa}{|\R|}\cdot\mathbb{E}[\Lambda_\R(S)]$, where the value of constant $\kappa$ depends on the given \ourproblem instance. \revise{
E.g., \mrim \cite{sun2018multi} first selects a random node $v_i\in V$ uniformly, then independently simulates $T$ reverse diffusion processes starting from $v_i$. Denoting the set of examined nodes in each process $i$ as $R'_i$, the RR set in \mrim is $R'_1\times\{1\}\cup\dots\cup R'_T\times\{T\}$, and we have $\kappa=|V|$. For the remaining instances, we refer interested readers to~\trref{\ref{sec:exp-detail}}.

}

\sq{To improve the efficiency of {Borgs et al.'s} framework for the vanilla \im}, subsequent solutions, \eg \textsf{TIM}~\cite{tang2014influence}, \imm~\cite{tang2015influence}, \textsf{DSSA}~\cite{nguyen2016stop}, and \opimc~\cite{tang2018online}, focus on reducing the number of sampled RR sets while ensuring the same approximation ratio.
As of now, \opimc is the state of the art and it has been applied to subsequent solutions~\cite{bian2020efficient, guo2020influence,guo2022influence,han2021efficient,zhang2023capacity}.
\yiqiann{Most scalable solutions of other \ourproblem instances also extend the aforementioned ones used in \im. E.g., in \rim, \cite{han2021efficient} utilizes a greedy variant by mixing \mgreedy and \tgreedy, and then follows \opimc to achieve an approximation not exceeding $1/2-\epsilon$.
% a $\lambda-\epsilon$ approximation in \rim, where
% \begin{equation}\label{equ:rm-approx}
% \lambda = \begin{cases}
%   1/3  & T=1 \\
%   \frac{1}{2(T+1)(\tau+1)} & T\in\{2,3\} \\ 
%   \frac{1}{(T+6)(\tau+1)} & T\geq 4
%   \end{cases}
% \end{equation}
% and depends on the number of campaigns $T$ and an accuracy-controlling parameter $\tau$.
Additionally, \cite{sun2018multi} and \cite{sun2023scalable} offer a $(1/2-\epsilon)$-approximation for \mrim and \advim, respectively, by employing \mgreedy and \imm.}
% The greedy algorithm begins with a solution set $S=\emptyset$. It iteratively adds a node $u$ that maximizes the marginal coverage $\Lambda_\R(u|S)$, while maintaining the constraint $S\cup\{u\}\in\I$. Since the coverage function is monotone and submodular (\ie $\Lambda_{\R}(S)\leq\Lambda_{\R}(T)$ and $\Lambda_\R(u|S)\geq\Lambda_\R(u|T)$ for any $S\subseteq T\subseteq V$ and $u\in V\backslash T$), the greedy algorithm yields a $(1-1/e)$-approximation under the cardinality constraint~\cite{nemhauser1978analysis}, but a $1/2$-approximation under the general matroid constraint~\cite{fisher1978analysis}. The following example shows that the approximation of $1/2$ is tight.

% \begin{example}
%     Consider the inputs of the greedy algorithm as follows:
%     \begin{itemize}
%         \item the node set $V=\{u,v,w\}$ and its partitions $V_1=\{u,v\},V_2=\{w\}$;
%         \item the RR sets containing $\theta_0$ numbers of $\{u,w\}$, $\theta_0$ numbers of $\{v\}$ and one $\{u\}$;
%         \item the matroid $\M=(V,\{S\subseteq V: |S\cap V_i|\leq 1,i=1,2\})$.
%     \end{itemize}
    
%     The greedy algorithm will first select $u$ which has the maximum coverage $\theta_0+1$, and then select the only feasible node $w$, such that the result is $\{u,w\}$ with the coverage $\theta_0+1$. However, the optimal set is $\{v,w\}$ with the coverage $2\theta_0$, hence the approximation ratio is $(\theta_0+1)/(2\theta_0)\rightarrow 1/2$ when $\theta_0$ grows.
% \end{example}

% \vspace{-1.9mm}
\section{Related Work}\label{sec:related}

\eat{\im was first formulated as a discrete optimization problem by \citet{kempe2003maximizing}, who proved its NP-Hardness and proposed a greedy approximation algorithm based on Monte-Carlo simulation. Motivated by its computational inefficiency, extensive research has been spurred on developing more scalable solutions for \im, as highlighted in the survey work~\cite{li2018influence}. }
\laks{Since its introduction in  \cite{kempe2003maximizing}, the \im problem has been extensively researched -- see~\cite{li2018influence} for a comprehensive survey. Here, we mainly focus on \im under a matroid constraint. Various kinds of matroids have been considered in previous \im studies.}
\yiqiann{Besides the three \ourproblem instances introduced in \S~\ref{sec:imgm-instance}, partition matroids are also employed in several \ourproblem instances, including capacity constrained \im~\cite{zhang2023capacity}, \im with fairness constraint~\cite{el2020fairness}, the lattice \im with partitioned budgets~\cite{chen2020scalable}, and so forth~\cite{tang2018social,ChanN15}. Furthermore, other general matroids are used in previous \ourproblem studies to accommodate more sophisticated scenarios. For instance, the popularity ratio maximization\eat{ (\prm)} problem~\cite {liao2023popularity} follows the multiple round setting in \mrim but aims to significantly surpass competitors' outreach in popularity by initiating influence cascades. This paper studies the intersection of a uniform matroid that restricts the overall cardinality with a partition matroid, resulting in a \textit{general matroid}\footnote{It is not a partition matroid, unlike what is claimed in \cite{liao2023popularity}. See \trref{\ref{sec:prm-note}} for details.} constraint. As another example, the general version of \im for multiple products~\cite{du2017scalable} requires a \textit{laminar matroid} constraint that models capacity limits within hierarchical community structures, such as the maximum number of selected customers in different states and the cities they contain.}
\eat{Within the scope of \ourproblem, previous work employs different types of matroids to accommodate their scenarios, including partition matroids~\cite{chen2020scalable, zhang2023capacity, el2020fairness, tang2018social}, laminar matroids~\cite{du2017scalable}, and general matroids that are not of specific types~\cite{liao2023popularity}. }While developing different approaches, the core ideas of these solutions lie in either \mgreedy, \lgreedy, or \tgreedy, all limited to at most a $(1/2-\epsilon)$-approximation. 
\eat{Besides instances of \ourproblem, several previous works explore \im under other real-world constraints. For example, the budgeted \im~\cite{bian2020efficient, guo2023efficient} employs a knapsack constraint, considering a non-uniform cost of selecting each seed. Some other \im variants~\cite{rajan2018multi, ijcai2019p831} formulate their problem under group constraints as instances of multi-objective optimization.}

\yiqiann{In theoretical computer science, another line of previous works aims to maximize a general submodular function under a matroid constraint. \cite{calinescu2011maximizing} proposed continuous greedy, the first polynomial-time algorithm with a $(1-1/e-\epsilon)$ approximation guarantee.} \cite{badanidiyuru2014fast} later refined it, achieving state-of-the-art results for general matroids. \yiqiann{However, adopting these algorithms in \ourproblem is highly inefficient, as detailed in \S~\ref{sec:selection-overview}.}
Subsequent research has focused on specific categories of matroids~\cite{buchbinder2017comparing, ene2019towards, henzinger2023faster}. \cite{buchbinder2017comparing} introduce some specific optimizations for partition matroids, resulting in an $O(n^{3/2})$-time algorithm. \sq{Prior works~\cite{ene2019towards,henzinger2023faster} further develop algorithms with $O(n\cdot \text{poly}(1/\epsilon, \log{n}))$ time complexity for partition, graphic, transversal, and laminar matroids.}
\eat{\cite{henzinger2023faster} extended this work, providing algorithms for transversal and laminar matroids with similar running time. The above approaches cannot be directly adopted for \ourproblem problems, as they assume the objective function is given by a value oracle, whereas obtaining the exact value of $\sigma(\cdot)$ in \ourproblem is \#P-Hard in general~\cite{chen2010scalable}.} \yiqiann{The above approaches cannot be directly adopted for \ourproblem instances, as they do not apply for general matroids outside the above matroid classes.}  

\LL{In the next two sections, we present our approach for solving \ourproblem. \eat{It provides a scalable $(1-1/e-\epsilon)$-approximation algorithm. }Given that our overall approach is based on generating a collection of RR sets and then selecting a set that maximizes coverage, we divide the presentation along the same theme.}

\pgfdeclarehorizontalshading{grad}{100bp}{
    color(0cm)=(white);
    color(30bp)=(white);
    color(38bp)=(plasma1);
    color(48bp)=(plasma2);
    color(56bp)=(plasma3);
    color(64bp)=(plasma4);
    color(74bp)=(deepblue);
    color(100bp)=(black)
}

\begin{figure*}[!t]
    \centering
    \hspace{-6.5mm}
    \subfloat[Polytopes]{
    \centering
    \tdplotsetmaincoords{70}{75}
    \begin{tikzpicture}
    		[tdplot_main_coords,
    		axis/.style={->},
                cube hidden/.style={dashed},
                cube hl/.style={fill opacity=.2,fill=gray},
                scale=0.5]    
    	\draw[axis] (0,0,0) -- (4,0,0) node[anchor=west]{\footnotesize $u_1$};
    	\draw[axis] (0,0,0) -- (0,4,0) node[right=-0.2em]{\footnotesize $u_2$};
    	\draw[axis] (0,0,0) -- (0,0,3.5) node[anchor=south]{\footnotesize $u_3$};

    	% draw the back-left of the cube
    	\draw[] (3,0,0) -- (3,0,3) -- (0,0,3);
    	%draw the front-right of the cube
            \draw (3,0,0) -- (3,3,0);
            \draw (0,0,3) -- (0,3,3);
    	% %draw the top of the cube
    	% \draw (0,0,3) -- (0,3,3);
     	%draw the triangle face of the cube
    	\draw[cube hl] (0,3,3) -- (3,3,0) -- (3,0,3) -- cycle;
 %     	%draw dashed lines to represent hidden edges
	% \draw[thin] (0,0,0) -- (0,3,0);
        \draw[thin] (0,3,0) -- (0,3,3);
        \draw[thin] (0,3,0) -- (3,3,0);
        \draw[fill=black] (0,0,0) node[below] {\footnotesize $\mathbf{0}$} circle(0.2em);
        \draw[fill=black] (0,3,3) node[above] {\footnotesize $[0,1,1]$} circle(0.2em) ;
        \draw[fill=black] (3,0,3) node[left] {\footnotesize $[1,0,1]$} circle(0.2em) ;
        \draw[fill=black] (3,3,0) node[below] {\footnotesize $[1,1,0]$} circle(0.2em) ;
        
    \end{tikzpicture}
    \vspace{2mm}\hspace{0mm}
    }
    \hspace{5mm}
    \subfloat[Previous solution]{
    \centering
    \tdplotsetmaincoords{70}{75}
    \begin{tikzpicture}
    		[tdplot_main_coords,
                grid/.style={very thin, gray},
    		axis/.style={->},
    		cube/.style={},
                cube hidden/.style={dashed},
                cube hl/.style={fill opacity=.3,fill=gray},
                scale=0.5]

            % draw the axes
    	\draw[axis] (0,0,0) -- (4,0,0) node[anchor=west]{\footnotesize $u_1$};
    	\draw[axis] (0,0,0) -- (0,4,0) node[right=-0.2em]{\footnotesize $u_2$};
    	\draw[axis] (0,0,0) -- (0,0,3.5) node[anchor=south]{\footnotesize $u_3$};
            \draw[grid] (1,0,0) -- (1,1,0) -- (0,1,0);
            \draw[grid] (2,0,0) -- (2,2,0) -- (0,2,0);
            \draw[grid] (3,0,0) -- (3,3,0) -- (0,3,0);
            \draw[grid] (0,1,0) -- (0,1,1) -- (0,0,1);
            \draw[grid] (0,2,0) -- (0,2,2) -- (0,0,2);
            \draw[grid] (0,3,0) -- (0,3,3) -- (0,0,3);
          %draw the triangle face of the cube
    	\draw[postaction={fill opacity=.3,fill=gray}] (0,1,1) -- (1,1,0) -- (1,0,1) -- cycle;
            \draw[blue,very thick] (0,0,0) -- (1,1,0);
          %draw the triangle face of the cube
    	\draw[postaction={fill opacity=.3,fill=gray}] (0,2,2) -- (2,2,0) -- (2,0,2) -- cycle;
            \draw[blue,very thick] (1,1,0) -- (2,1,1);
     	%draw the triangle face of the cube
    	\draw[cube hl] (0,3,3) -- (3,3,0) -- (3,0,3) -- cycle;
            \draw[blue,very thick] (2,1,1) -- (3,2,1);
            \draw[-stealth,blue,very thick, densely dashed] (3,2,1) -- (2,3,1) -- (3,3,0);
            % \draw[->] (0, -0.5, 1.5) -- (2/3, 2/3, 2/3);
            % \draw (0,-0.5,1.5) node[above=-2pt] {\footnotesize $\epsilon\mathbf{1}_{B_1}$};

            % \draw[->] (0, 0, 2.5) -- (4/3, 4/3, 4/3);
            % \draw (0, 0, 2.5) node[above=-2pt] {\footnotesize $\epsilon(\mathbf{1}_{B_1}+\mathbf{1}_{B_2})$};

            % \draw[gray] (0,0,0) -- (3,0,3);
            % \draw[gray] (0,0,0) -- (3,3,0);
            % \draw[gray] (0,0,0) -- (0,3,3);
            \draw[grid] (1,0,0) -- (1,0,1) -- (0,0,1);
            \draw[grid] (2,0,0) -- (2,0,2) -- (0,0,2);
            \draw[grid] (3,0,0) -- (3,0,3) -- (0,0,3);
            \draw[fill=black] (0,0,0) node[left] {\footnotesize $\mathbf{0}$} circle(0.2em);
            \draw[fill=white] (1,1,0) node[below=0.5em] {} circle(0.3em);
            \draw[fill=white] (2,1,1) node[below=0.5em] {} circle(0.3em) ;
            \draw[fill=white] (3,2,1) node[below=0.5em] {} circle(0.3em) ;
            \draw[fill=white] (3,3,0) node[below=0.5em] {} circle(0.3em);
            \draw (0,0,1) node[left] {\footnotesize $\frac{1}{3}$};
            \draw (0,0,2) node[left] {\footnotesize $\frac{2}{3}$};
            \draw (0,0,3) node[left] {\footnotesize $1$};
            \draw (1,0,0) node[left] {\footnotesize $\frac{1}{3}$};
            \draw (2,0,0) node[left] {\footnotesize $\frac{2}{3}$};
            \draw (3,0,0) node[left] {\footnotesize $1$};
    	
    \end{tikzpicture}
    \vspace{2mm}
    }
    \hspace{5mm}
    \subfloat[Searching]{
    \centering
    \tdplotsetmaincoords{70}{75}
    \begin{tikzpicture}
    		[tdplot_main_coords,
                grid/.style={very thin, gray},
    		axis/.style={->},
    		cube/.style={},
                cube hidden/.style={dashed},
                cube hl/.style={fill opacity=.3,fill=gray},
                scale=0.5]

    	% draw the axes
    	\draw[axis] (0,0,0) -- (4,0,0) node[anchor=west]{\footnotesize $u_1$};
    	\draw[axis] (0,0,0) -- (0,4,0) node[right=-0.2em]{\footnotesize $u_2$};
    	\draw[axis] (0,0,0) -- (0,0,3.5) node[anchor=south]{\footnotesize $u_3$};
    	% draw a grid in the x-y plane
    	\foreach \x in {1,2,3}
    	{
    		\draw[grid] (\x,0,0) -- (\x,3,0);
    		\draw[grid] (0,\x,0) -- (3,\x,0);
    	}
     %      % draw a grid in the y-z plane
    	\foreach \x in {1,2,3}
    	{
    		\draw[grid] (0,\x,0) -- (0,\x,3);
    		\draw[grid] (0,0,\x) -- (0,3,\x);
    	}

        \draw[red,very thick] (0,0,0) -- (0,1,0) -- (1,1,0);

     \draw[red,very thick] (1,1,0) -- (1,1,1) -- (1,2,1);

     \draw[-stealth,red,very thick] (1,2,1) -- (2,2,1) -- (2,2,2);
     	%draw the triangle face of the cube
    	\draw[cube hl] (0,3,3) -- (3,3,0) -- (3,0,3) -- cycle;
        
 	\draw[densely dashed, thin, red] (0,2,1) -- (1,2,1);
        \draw[densely dashed, thin, red] (1,2,0) -- (1,2,1);
        \draw[densely dashed, thin, red] (2,0,2) -- (2,2,2);
        \draw[densely dashed, thin, red] (0,2,2) -- (2,2,2);
        \draw[densely dashed, thin, red] (2,2,0) -- (2,2,1);

     %            % draw a grid in the x-z plane
    	\foreach \x in {1,...,3}
    	{
    		\draw[grid] (\x,0,0) -- (\x,0,3);
    		\draw[grid] (0,0,\x) -- (3,0,\x);
    	}

         \draw[fill=black] (0,0,0) node[below=0.5em] {} circle(0.2em);
         \draw[fill=white] (1,1,0) node[below=0.5em] {} circle(0.3em);
         \draw[fill=gray!30] (1,2,1) node[below=0.5em] {} circle(0.3em) ;
         \draw[fill=white] (2,2,2) node[below=0.5em] {} circle(0.3em) ;
    	
    \end{tikzpicture}
    \vspace{2mm}
    }
    \subfloat[Rounding]{
    \centering
    \hspace{5mm}
    \tdplotsetmaincoords{70}{75}
    \begin{tikzpicture}
    		[tdplot_main_coords,
                grid/.style={very thin,gray},
    		axis/.style={blue,->},
    		cube/.style={},
                cube hidden/.style={dashed},
                cube hl/.style={fill opacity=.3,fill=gray},
                scale=0.5]
    	% draw a grid in the x-y plane
    	\foreach \x in {0,3}
    	{
    		\draw[grid] (\x,0,0) -- (\x,3,0);
    		\draw[grid] (0,\x,0) -- (3,\x,0);
    	}
     %      % draw a grid in the y-z plane
    	\foreach \x in {0,3}
    	{
    		\draw[grid] (0,\x,0) -- (0,\x,3);
    		\draw[grid] (0,0,\x) -- (0,3,\x);
    	}
     %            % draw a grid in the x-z plane
    	\foreach \x in {0,3}
    	{
    		\draw[grid] (\x,0,0) -- (\x,0,3);
    		\draw[grid] (0,0,\x) -- (3,0,\x);
    	}
    			
        \filldraw[shading=grad,shading angle=-100] (0,3,3) -- (3,3,0) -- (3,0,3) -- cycle;

        %draw a vector from O to P
	\draw[-stealth,red,very thick] (2,2,2) -- (1,3,2) -- (0,3,3);

         \draw[draw=black,fill=white] (2,2,2) node[left] {} circle(0.3em) ;
        \draw[draw=black,fill=white] (0,3,3) node[below] {} circle(0.3em) ;
        \draw[draw=black,fill=white] (1,3,2) node[below] {} circle(0.3em) ;

        \draw[draw=black,shading=grad,shading angle=-90] (0,4.1,2.88) node[right] {\footnotesize $\max$} (0,4,3) rectangle (0,4.2,1) ;

        \draw[draw=black,shading=grad,shading angle=-90] (0,4.1,1.2) node[right] {\footnotesize $\min$};

        \draw (0,4.1,2.87) node[above] {\scriptsize $F(\mathbf{x})$};
    	
    \end{tikzpicture}
    \vspace{5.3mm}
    }
    \vspace{-2mm}
    \caption{Illustration of continuous greedy solutions and \cseedselect subject to the matroid with $U=\{u_1,u_2,u_3\}$ and $\I=\{I\subseteq U: |I|\leq 2\}$.}\label{fig:vis-polytope}
    \label{fig:enter-label}
\end{figure*}
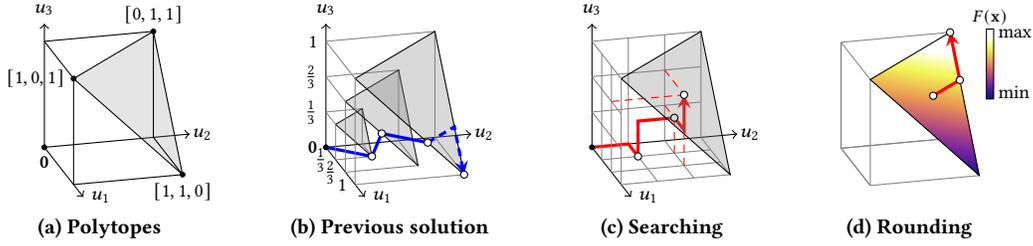

% \vspace{-0.4mm}
\section{Element Selection}\label{sec:greedy}
\eat{In this section, we suppose that a set of RR sets $\R$ is given and defer the discussion of how to determine the number of RR sets $|\R| = \theta$ to \S~\ref{sec:opimc}.} \sq{Given a set of RR sets $\R$, this section proposes} an element-wise Ascent algorithm over Matroid Polytopes for \edit{element} selection (\cseedselect), providing a $(1-1/e-\epsilon)$-approximation for maximizing the coverage $\Lambda_\R$ subject to the matroid constraint. 
\eat{\edit{In what follows},} We first discuss the challenges in adopting prior $(1-1/e-\epsilon)$-approximate solutions and the main idea of \cseedselect (\S~\ref{sec:selection-overview}), followed by detailed descriptions of two core subroutines \cgreedy/\cgreedyp (\S~\ref{sec:find-base}) and \crounding (\S~\ref{sec:det-rounding}). At last, we provide an analysis of correctness and time complexity (\S~\ref{sec:put-together}).

\subsection{Overview}\label{sec:selection-overview}

\stitle{Main idea of existing solutions}
\yiqiann{Several} continuous greedy algorithms~\cite{calinescu2011maximizing,badanidiyuru2014fast} have been proposed to find $(1-1/e-\epsilon)$-approximate solutions for maximizing \edit{any monotone submodular function $f$ (\eg $\Lambda_\R$)} subject to a general matroid constraint.\eat{\edit{As $\Lambda_\R$ is an instance of monotone and submodular functions, these solutions are applicable for maximizing it.}} \sq{First, these solutions expand} the original function $f$\eat{problem $\max_{S\in\I}\{f(S)\}$} to its corresponding multilinear extension, defined as follows.\eat{where the function $F$ is the \textit{multilinear extension} of $f$, defined as follows.}

\begin{definition}[Multilinear Extension]\label{def:multilinear-extension}
For a set function $f:2^U\rightarrow \mathbb{R}_+$, \eat{the multilinear extension of $f$ is the function $F:[0,1]^U\rightarrow \mathbb{R}_+$ defined as }its multilinear extension $F:[0,1]^U\rightarrow \mathbb{R}_+$ is defined as $F(\x)=\mathbb{E}\left[f(\Omega(\x))\right]$,
\eat{
\begin{equation}\label{eq:multilinear-expect-format}
F(\x)=\mathbb{E}\left[f(\Omega(\x))\right],
\end{equation}
}
where $\Omega(\x)$ is a random set that independently includes each element $u_i\in U$ from the fractional result $\x$ with probability $\x[i]$. 
\end{definition}

\sq{Second, these solutions aim to \eat{obtain a fractional solution $\x$ for}solve the multilinear optimization problem $\max_{\x\in P^\I}\{F(\x)\}$\eat{that maximizes the multilinear extension in Definition~\ref{def:multilinear-extension}}}, where $P^{\I}$ is the \textit{matroid polytope} defined as the convex hull (\ie all convex combinations of elements)\eat{(\ie the convex hull)} of $\left\{\mathbf{1}_I:I\in\I\right\}$. 
Notice that, due to the monotonicity of $f$, the fractional solution $\x$ that maximizes $F(\x)$ is in the \textit{base polytope} $\textstyle P^{\BPoly} = \big\{ \x\in P^\I: \sum_{i\in [n]} \x[i] =r \big\}$~\cite{edmonds1971matroids}.
\eat{Based on the definition of \textit{base polytope} $\textstyle P^{\BPoly} = \big\{ \x\in P^\I: \sum_{i\in [n]} \x[i] =r \big\}$~\cite{edmonds1971matroids} and the monotonicity of $f$, the fractional solution $\x$ that maximizes $F(\x)$ is in $P^\BPoly$.}\eat{Therefore, previous approaches aim to obtain a fractional solution $\x \in P^{\BPoly}$ to solve the multilinear optimization problem.}
\sq{To find such a solution $\x \in P^{\BPoly}$,} given a parameter $\epsilon$, these approaches start from $\x=\mathbf{0}$ and \LL{update} $\x$ by an iterative \textit{hill-climbing search} process: at each step $t$, this process examines \textit{basic solution} $B_t \in \BPoly$ and the corresponding fractional solution $\epsilon\cdot\mathbf{1}_{B_t}$ such that $F(\x)$ can be improved after adding this fractional solution to $\x$. After $1/\epsilon$ steps, this process returns a fractional solution $\textstyle 
\x=\sum_{t=1}^{1/\epsilon}\epsilon\cdot\mathbf{1}_{B_t},$\eat{\begin{equation}\label{eq:fractional}
\textstyle 
\x=\sum_{t=1}^{1/\epsilon}\epsilon\cdot\mathbf{1}_{B_t},    
\end{equation}}
which is in $P^{\BPoly}$ by definition. \sq{Finally}, upon obtaining $\x \in P^{\BPoly}$, these solutions employ a \textit{randomized rounding} process that eventually converts $\x$ to a discrete solution $S\in \BPoly$, thus solving the original problem $\max_{S\in\I}\{f(S)\}$.
% These faces scale $P^\BPoly$, with $\mathbf{0}$ as the origin, by $1/3\times$, $2/3\times$, and $1\times$, respectively.

 % and the vertices of this face are the indicators of bases.

% \begin{definition}[Matroid Polytope \& Base Polytope]\label{def:matroid-polytope}
% For a matroid $\M$, the matroid polytope $P^{\I}$ and the base polytope $P^{\BPoly}$ are defined as
% \begin{align*}
%     P^{\I}&=conv\left\{\mathbf{1}_I:I\in\I\right\},\\
%     P^{\BPoly}&=conv\left\{\mathbf{1}_B:B\in\BPoly\right\},
% \end{align*}
% where $conv\{\cdot\}$ denotes the convex hull. 
% % Equivalently, $P^{\BPoly}$ is the face $\{\x:\sum\limits_{i\in [U]} \x[i] = r\}$ of $P^{\I}$~\cite{edmonds1971matroids}.
% \end{definition}

\vspace{-1mm}
\begin{example}[Illustrative Example] \label{matroid-ex} 
\hl{Consider a matroid with $U=\{u_1,u_2,u_3\}$ and $\I=\{I\subseteq U: |I|\leq 2\}$. As depicted in Figure~\ref{fig:vis-polytope}(a), the origin represents $\x=\mathbf{0}$, and the vertices $[1,1,0]$, $[1,0,1]$, $[0,1,1]$ indicate all bases. Additionally, the polyhedron and the shaded face correspond to the matroid polytope and base polytope, respectively. Assuming $\epsilon=1/3$, the three shaded faces shown in Figure~\ref{fig:vis-polytope}(b) represent the feasible spaces for each search step. The solid and dashed blue lines illustrate the search and rounding process of previous solutions, respectively. Starting from $\x=\mathbf{0}$, the search process of~\citet{calinescu2011maximizing} selects the direction of $B_1$ (\ie $\mathbf{1}_{B_1}=[1,1,0]$) and moves $\x$ to $\x+\epsilon\cdot\mathbf{1}_{B_1}=\left[\frac{1}{3},\frac{1}{3},0\right]$ at the first step, reaching the smallest shaded face. Subsequently, it selects $\mathbf{1}_{B_2}=[1,0,1]$ and moves $\x$ to $\x+\epsilon\cdot\mathbf{1}_{B_2}=\left[\frac{2}{3},\frac{1}{3},\frac{1}{3}\right]$, located in the medium shaded face. At last, it updates $\x$ along the direction of $B_3$, bringing $\x$ to $\left[1,\frac{2}{3},\frac{1}{3}\right]$ in the base polytope. Upon completing the search, to round the fractional solution $\x$, the algorithm~\cite{chekuri2010dependent} moves $\x$ on the base polytope in random directions until it reaches a vertex, \eg $[1,1,0]$, indicating the final solution $S=\{u_1,u_2\}$.}
\qed \end{example} 
\vspace{-1mm}

\stitle{Limitations of existing solutions}
Existing algorithms suffer from severe efficiency issues \hl{in the hill-climbing search stage}, since computing the multilinear extension $F$ of a general monotone submodular function $f$ is expensive.
Specifically, the multilinear extension of a set function $f$ is
\vspace{-1mm}
\begin{equation}\label{eq:multilinear-full-format}
F(\x)=\sum_{S\subseteq U}\Bigg(\prod_{u_i\in S}\x[i]\prod_{u_i\in U\backslash S}(1-\x[i])\Bigg)\cdot f(S). 
\end{equation}
Computing $F(\x)$ exactly involves enumerating all $2^{n}$ subsets of $U$. Previous works leverage random subset sampling to estimate $F(\x)$, however they still have a prohibitively expensive time complexity. 
E.g., the approach of \cite{calinescu2011maximizing} which selects the $t$-th basic solution ${B_t}$ that yields an improvement in $F$ along the direction of the base $B_t$, requires $O\left({n}^7\log({n})\right)$ samples. 
% across all $O\left(n^6\right)$ search iterations. 
\eat{To mitigate the efficiency issue, } \eat{ propose an improved algorithm by selecting}\yiqiann{The improved algorithm}~\cite{badanidiyuru2014fast} selects each basic solution ${B_t}$ using \tgreedy, which includes an element $u_i$ into $B_t$ if the partial derivative $\partial F(\x)/\partial \x[i]$ exceeds a pre-defined threshold. This compromises the result quality (see \S~\ref{sec:existing}) yet  generates $O({n}r\epsilon^{-4}\cdot\log^2{\left({n}/{\epsilon}\right)})$ samples in total.
% For instance, \citet{calinescu2011maximizing} select the $t$-th partial fraction $\epsilon\mathbf{1}_{B_t}$ that yields an improvement in $F$ w.r.t.\ the base $B\in \BPoly$. This requires a total of $O\left({n}^7\log({n})\right)$ samples. 
% % across all $O\left(n^6\right)$ search iterations. 
% To mitigate the efficiency issue, \citet{badanidiyuru2014fast} select each partial fraction $\epsilon\mathbf{1}_{B_t}$ following the idea of \tgreedy, which includes the element $u_i$ into $B_t$ if the partial derivative $\partial F(\x)/\partial \x[i]$ exceeds the pre-defined threshold. This results in a compromised result quality as described in \S~\ref{sec:existing} and still generates $O\left({n}r\epsilon^{-4}\cdot\log^2{\left(\frac{{n}}{\epsilon}\right)}\right)$ samples in total.

\hl{Additionally, in the rounding stage, existing randomized techniques~\cite{ageev2004pipage,chekuri2010dependent} focus on the efficiency of maintaining $\x\in P^{\BPoly}$ rather than the quality of the rounded result. To elaborate, the state-of-the-art swap rounding~\cite{chekuri2010dependent} iteratively eliminates each distinct item $u_i$ from the difference between the current basic solution ${B_{t}}$ and the previous merged one ${B_{t-1}}$ until all basic solutions are identical. This is done by accepting $u_i$ w.p.\ $1/t$ if $u_i \in B_{t} \backslash B_{t-1}$ and with the remaining probability if $u_i \in B_{t-1} \backslash B_{t}$. As proved by \cite{chekuri2010dependent}, swap rounding returns a discrete set $S\in\BPoly$ satisfying $f(S)\geq (1-\psi)F(\x)$ w.p.\ at least $1-e^{-\frac{F(\x)\psi^2}{8 F^*}}$, where $F^*=\max_{\x}F(\x)$. Due to the additional error term $\psi$, more iterations in the search step or more RR sets are required to ensure the desired total error $\epsilon$, resulting in significant computational overhead. Furthermore, the failure probability is $e^{-\frac{F(\x)\psi^2}{8 F^*}}\geq e^{-\frac{\psi^2}{8}}> 0.88$, which can severely compromise the quality of the rounded result.}
% Despite that the swap rounding returns a discrete set $S\in\BPoly$,
% the randomized process only offers a poor guarantee related to the result quality $f(S)$. To elaborate, the following lemma provides a concentration bound associated with the solution quality and the failure probability of swap rounding.

% \begin{lemma}[\cite{chekuri2010dependent}]
%     Denote $F^*=\max\limits_{\x\in[0,1]^U}F(\x)$. Given the input $\x \in P(\BPoly)$, for any $\psi > 0$, the random set $S$ returned by swap rounding satisfies $$\Pr[f(S)\leq (1-\psi)F(\x)]\leq e^{-\frac{F(\x)\psi^2}{8 F^*}}.$$
% \end{lemma}

% Even when $\psi=1$, we have $e^{-\frac{F(\x)\psi^2}{8 F^*}}\geq e^{-\frac{\psi^2}{8}}> 0.88$, rendering this bound meaningless in practice.

\stitle{Our proposal} To address the above issues, we consider the multilinear extension of the coverage function $\Lambda_\R(\cdot)$ and \textit{show that it can be evaluated efficiently and deterministically}. Specifically, by plugging $\Lambda_\R(\cdot)$ defined in Eq.\eqref{eq:coverage-function-raw} into Eq.\eqref{eq:multilinear-full-format}, we have
\vspace{-1mm}
\begin{equation*}
F(\x) = \sum_{R\in \R}\sum_{S \subseteq U}\left(\prod_{u_i\in S}\x[i]\prod_{u_j\in U\backslash S}(1-\x[j])\right)\cdot\mathbb{I}(R\cap S\neq\emptyset),  
\end{equation*}
% 
% F(\x)&=\sum_{S \subseteq U}\left(\prod_{u_i\in S}\x[i]\prod_{u_j\in U\backslash S}(1-\x[j])\right)\cdot \sum_{R\in \R}\mathbb{I}(R\cap S\neq\emptyset)\\
% &=\sum_{R\in \R}\sum_{S \subseteq U}\left(\prod_{u_i\in S}\x[i]\prod_{u_j\in U\backslash S}(1-\x[j])\right)\cdot\mathbb{I}(R\cap S\neq\emptyset).
% 
which can be converted to $\textstyle F(\x)=\sum_{R\in \R}\mathbb{E}\left[\mathbb{I}(R\cap \Omega(\x)\neq\emptyset)\right]$.
Since $\mathbb{E}\left[\mathbb{I}(R\cap \Omega(\x)\neq\emptyset)\right]=1-\prod_{u_i\in R}(1-\x[i])$ for a given $R\subseteq U$, $F(\x)$ %of $\Lambda_\R(\cdot)$ 
can be further rewritten as
\begin{equation}\label{eq:multilinear-coverage-format}
F(\x)=\sum_{R\in \R}\left(1-\prod_{u_i\in R}(1-\x[i])\right).
\end{equation}
%In accordance with Eq.\eqref{eq:multilinear-coverage-format}, 
Based on this, we can efficiently calculate the exact value of $F(\x)$ by maintaining \edit{a collection $\Q_\R$, consisting of} the variable 
\begin{equation}\label{eq:qr}
    \textstyle q_R=\prod_{u_i\in R}(1-\x[i])
\end{equation}
for each $R\in \R$ from the first iteration. This enables us to develop new search and rounding algorithms capable of achieving the desired $(1-1/e-\epsilon)$-approximation efficiently. 
\revise{
It is noteworthy that, besides instances of \ourproblem, Eq.\eqref{eq:multilinear-coverage-format} has the potential to improve \im scenarios that are solved by leveraging the multilinear extension in Eq.\eqref{eq:multilinear-full-format}. This includes \im variants constrained by user groups~\cite{rajan2018multi} and group fairness~\cite{ijcai2019p831}.
}

\begin{small}
\begin{algorithm}[!t]
\edit{
\KwIn{The matroid $\M$, an RR collection $\R$, $0<\epsilon\leq1$}
\KwOut{A discrete set $S\in \BPoly$ maximizing the coverage}
}
\edit{$\Q_\R\gets\{q_R:R\in\R\}$;}~$\forall R\in\R,q_R\gets 1$;\\
$\forall t=1,2,\dots,1/\epsilon, B_t \gets \emptyset$;~$\x\gets \mathbf{0}$;\\
\For{$t=1,2,\dots,1/\epsilon$}{
    \If{$\M$ is a partition matroid}{
        $B_t,\edit{\Q_\R}\gets\cgreedyp(\R, \x, \epsilon,\edit{\Q_\R})$;
    }
    \Else{
        $B_t,\edit{\Q_\R} \gets \cgreedy(\R, \x, \epsilon,\edit{\Q_\R})$
    }
    $\x\gets \x+\epsilon\cdot\mathbf{1}_{B_t}$;
}
$S\gets\crounding(\R, \epsilon, \edit{\Q_\R}, B_1, B_2, \dots, B_{1/\epsilon})$;\\
\Return{$S$;}
\caption{\cseedselect$(\M, \R, \epsilon)$}
\label{alg:greedy-framework}
\end{algorithm}
\end{small}

As outlined in Algorithm~\ref{alg:greedy-framework}, our proposed \cseedselect shares a similar framework with existing continuous greedy methods but introduces two new search strategies, \cgreedy \edit{ (Algorithm \ref{alg:cgreedy})} for the general matroid and \cgreedyp \edit{ (Algorithm \ref{alg:cgreedy-partition})} for the partition matroid, and a deterministic rounding algorithm, \crounding\edit{ (Algorithm \ref{alg:crounding})}. For ease of presentation, we assume w.l.o.g.\ that $1/\epsilon$ is an integer, since \LL{for any given $\epsilon'$, we can use $\frac{1}{\lceil 1/\epsilon' \rceil}$ as \eat{the value for }$\epsilon$.}
\hl{During each search iteration $t$, \cgreedy and \cgreedyp adopt a more refined approach compared to the coarse-grained search of previous works~\cite{calinescu2011maximizing,badanidiyuru2014fast}. Specifically, they aim for the \textit{steepest} improvement in $F$ and include the element $u_i$ with the largest $\partial F(\x)/\partial \x[i]$ into $B_t$, resulting in quicker convergence and fewer search iterations. \cgreedyp improves the efficiency of \cgreedy further by pruning the search space $U$ to \sq{one partition}\eat{one of the partitions $U_l$}, reducing the number of evaluations of partial derivatives by a factor of $r/k^*$, where $k^*=\max\{k_1,k_2,\dots k_h\}$.}
As for the rounding phase, unlike the randomized methods, our proposed \crounding \textit{deterministically} decides the elimination of a distinct element $u_i$ based on the partial derivative w.r.t. $\x_i$. This ensures that $F(\x)$ does not decrease throughout rounding until $\x$ becomes integral, yielding the rounded result $S$ satisfying $\Lambda_\R(S)\geq F(\x)$ \textit{without failure probability}.

%\begin{example}[Running example (contd.)]
\hl{For the illustrative example in Figure~\ref{fig:vis-polytope}, the red arrowed lines in Figures~\ref{fig:vis-polytope}(c) and (d) trace the movement of $\x$ in the search and rounding process of \cseedselect, respectively. Initially, when $\x=\mathbf{0}$ and \cseedselect starts to search for $B_1$, $u_2$ is first included as $\partial F(\x)/\partial \x[2]$ is the largest, updating $\x$ to $\x=\left[0,\frac{1}{3},0\right]$. Next, as $\partial F(\x)/\partial \x[1]$ becomes the largest w.r.t.\ the current $\x$, $u_1$ is included in $B_1$, leading to an update of $\x$ to $\left[\frac{1}{3},\frac{1}{3},0\right]$. Analogously, $u_3$ and $u_2$ (resp.\ $u_1$ and $u_3$) are selected and the corresponding basic solution is $B_2=\{u_2,u_3\}$ (resp.\ $B_3=\{u_1, u_3\}$), eventually yielding a fractional result $\x=\left[\frac{2}{3},\frac{2}{3},\frac{2}{3}\right]$ in the base polytope, as shown in Figure~\ref{fig:vis-polytope}(c). After completing the search, as illustrated in Figure~\ref{fig:vis-polytope}(d), \cseedselect moves $\x$ to a vertex by ensuring the directions are towards maximizing $F(\x)$ (\ie the brighter area), thereby guaranteeing $F(\x)$ is non-decreasing.} \qed 

%\end{example} 

% To summarize, the proposed \cseedselect achieves the $(1-1/e-\epsilon)$ approximation for maximizing $\Lambda_\R$ subject to the matroid constraint. In the partition matroid case, \cseedselect runs with a time complexity of $O\left(k^*\cdot\epsilon^{-1}\cdot \sum_{R\in\R}|R|\right)$, where $k^*=\max\{k_1,k_2,\dots k_h\}$. For general matroids, it requires $O\left(r\epsilon^{-1}\cdot\sum_{R\in\R}|R|\right)$, \hl{assuming the cost of checking the feasibility of the matroid constraint is neglected.}

\subsection{Element-wise Ascent Search}\label{sec:find-base}
We first present our \edit{element} selection algorithm \cgreedy for the general matroid constraint (pseudocode in Algorithm~\ref{alg:cgreedy}). Specifically, Line 1 initializes by setting $B= \emptyset$ and the vector $\y= \x$. It then proceeds to iteratively populate $B$ through $r$ iterations, each time adding an element to construct a basic solution. 
In Line 3, during each iteration $i$, \cgreedy examines all elements in the remaining set $U\backslash B$ and selects the element $u_{j}$ that maximizes the value of $\sum_{R\in\R:{u_j}\in R} q_R/(1-\y[j])$, while adhering to the matroid constraint. {This value corresponds to the partial derivative $\partial F(\y)/\partial \y[j]$ as shown below. \textit{All proofs are available in \trref{\ref{sec:proof-all}}.}}

\begin{lemma}\label{lem:marginal-reflect}
    $\textstyle \forall \y\in[0,1]^U \text{ and }u_i\in U, \frac{\partial F(\y)}{\partial \y[i]}=\sum\limits_{R\in\R:u_i\in R}\frac{q_R}{1-\y[i]}.$
\end{lemma}
In Lines 4 - 5, \cgreedy updates $\y[j]$ to $\y[j]+\epsilon$ and includes $u_j$ in $B$, followed by maintaining the correctness of each $q_R$ value as per Eq.\eqref{eq:qr}. In particular, for each RR set $R$ containing $u_j$, \cgreedy updates the corresponding $q_R$ by scaling it with $\frac{1-\y[j]}{1-\y[j]+\epsilon}$. This process removes the outdated factor $1-\y[j]+\epsilon$ and includes the latest value of $1-\y[j]$. 

\begin{small}
\begin{algorithm}[t]
\edit{\KwIn{An RR collection $\R$, a fractional solution $\x\in [0,1]^U$, $0<\epsilon\leq1$, $\Q_\R$ defined in Algorithm 1}
\KwOut{A basic solution $B\in \BPoly$, an updated collection $\Q_\R$}
}
$B\gets \emptyset$;~$\y\gets \x$;\\
\For{$i=1,2,\dots,r$}{
$u_j\gets \argmax{u_j:~u_j\in U\backslash B,~B\cup\{u_j\}\in\I}\left\{\sum\limits_{R\in\R:~u_j\in R}\frac{q_R}{1-\y[j]}\right\}$;\\
$\y[j]\gets \y[j]+\epsilon$;~$B\gets B\cup\{u_j\}$;\\
$\forall R\in\R:u_j\in R,~q_R\gets q_R\cdot\frac{1-\y[j]}{1-\y[j]+\epsilon};$\\
}
\Return{$B,\Q_\R$;}\\
\caption{\cgreedy~$(\R, \x, \epsilon, \Q_\R)$}
\label{alg:cgreedy}
\end{algorithm}
\end{small}

We next introduce the algorithm \cgreedyp tailored for the partition matroid constraint. In contrast to \cgreedy which selects the next element from $U\backslash B$, \hl{Algorithm~\ref{alg:cgreedy-partition} selects an arbitrary partition $U_l$ of $U$ satisfying $|B\cap U_l|<k_l$, and chooses the element $u_j$ from $U_l\backslash B$ during each iteration, thus reducing the search space and running time of \cgreedy by a factor of $r/k^*$, where $k^*=\max\{k_1,k_2,\dots k_h\}$.} The following lemma offers a guarantee of both algorithms, bounding how much $F(\x)$ increases as it approaches the optimal result $S^o$ for maximizing $\Lambda_\R(\cdot)$ s.t.\ the matroid constraint.

\begin{lemma}\label{lem:greedy-approx}
For any input fractional result $\x$, the \LL{output} $B$ of both \cgreedy and \cgreedyp satisfies
$\textstyle \Lambda_\R(S^o)-F(\x+\epsilon\cdot\mathbf{1}_B)\leq \frac{\Lambda_\R(S^o)-F(\x)}{1+\epsilon}$, where $S^o=\argmax{S\in\I}\Lambda_\R(S)$.
\end{lemma}

\begin{small}
\begin{algorithm}[!t]
\edit{\KwIn{An RR collection $\R$, a fractional solution $\x\in [0,1]^U$, $0<\epsilon\leq1$, $\Q_\R$ defined in Algorithm 1}
\KwOut{A basic solution $B\in \BPoly$, an updated collection $\Q_\R$}
}
$B\gets \emptyset$;~$\y\gets \x$;\\
\For{$i=1,2,\dots,r$}{
$U_l\gets$ an arbitrary partition such that $|B\cap U_l|< k_l$;\\
$u_j\gets \argmax{u_j:u_j\in U_l\backslash B}\left\{\sum\limits_{R\in\R:~u_j\in R}\frac{q_R}{1-\y[j]}\right\}$;\\
\hl{$\y[j]\gets \y[j]+\epsilon$;~$B\gets B\cup\{u_j\}$;}\\
\hl{$\forall R\in\R:u_j\in R,~q_R\gets q_R\cdot\frac{1-\y[j]}{1-\y[j]+\epsilon}$;}\\
}
\Return{$B,\Q_\R$;}\\
\caption{\cgreedyp~$(\R, \x, \epsilon, \edit{\Q_\R})$}
\label{alg:cgreedy-partition}
\end{algorithm}
\end{small}

\subsection{Element-wise Ascent Rounding}\label{sec:det-rounding}
We introduce the rounding algorithm \crounding, outlined in Algorithm~\ref{alg:crounding}. \hl{Given the basic solutions $B_1, B_2, \dots, B_{1/\epsilon}$,} \crounding maintains a vector 
\begin{equation}\label{eq:y-format}
    \textstyle \y=\sum_{l=t+1}^{1/\epsilon}\epsilon\cdot\mathbf{1}_{B_{l}}+t\cdot \epsilon \cdot \mathbf{1}_{B_t},
\end{equation}
which represents the value of the fractional solution \hl{(\ie $\x$)} after merging the first $t$ fractional solutions $\epsilon\mathbf{1}_{B_1}, \epsilon\mathbf{1}_{B_2}, \dots, \epsilon\mathbf{1}_{B_{t}}$ corresponding to the \edit{basic solutions} $B_1, B_2, ..., B_t$. It is initialized to $\y=\sum_{l=1}^{1/\epsilon}\epsilon\mathbf{1}_{B_{l}}$.
\crounding then scans each pair of successive basic solutions from $(B_1, B_2)$ to $(B_{1/\epsilon-1}, B_{1/\epsilon})$. For each pair $(B_t,B_{t+1})$, it iteratively eliminates distinct elements between the two basic solutions until $B_{t+1}$ is identical to $B_{t}$, and thereby merged with all preceding basic solutions $B_{t-1}, \dots, B_1$.
More precisely, \crounding picks an arbitrary element $u_i$ from $B_t\backslash B_{t+1}$, then identifies an element $u_j \in B_{t+1} \backslash B_t$ such that swapping $u_i$ and $u_j$ leads to $B_t\backslash\{u_i\}\cup\{u_j\}$ and $B_{t+1}\backslash\{u_j\}\cup\{u_i\}$ being bases of the matroid, as outlined in Lines 4-5. Notably, the existence of such an element $u_j$ is assured by the following matroid exchange property.

\vspace{-0.3mm}
\begin{lemma}[Matroid Exchange property~\cite{chekuri2010dependent}]\label{lem:matroid-exchange1}
Let $B_1, B_2$ be two bases of $\M$. For any $u_i\in B_1\backslash B_2$, there exists $u_j\in B_2\backslash B_1$ such that $B_1\backslash\{u_i\}\cup\{u_j\}$ and $B_{2}\backslash\{u_j\}\cup\{u_i\}$ are also bases.
\end{lemma}
\vspace{-0.3mm}

Upon identifying a conflicting element pair $(u_i,u_j)$, Lines 6-13 \eat{focus on retaining }\yiqiann{retain} the one with the larger partial derivative. If the partial derivative w.r.t. $u_i$ is not less than that w.r.t. $u_j$, \crounding includes $u_i\in B_{t}\backslash B_{t+1}$ into $B_{t+1}$ while removing $u_j$ from it. 
Correspondingly, it updates the fractional result $\y$ by adding $\epsilon(\mathbf{1}_i-\mathbf{1}_j)$ as per Eq.\eqref{eq:y-format}, and modifies the values of $q_R$ for all $R$ that include $u_i$ or $u_j$ based on Eq.\eqref{eq:qr}. 
Conversely, if $u_j$ is to be preserved, \crounding updates $B_{t}$ by replacing $u_i$ with $u_j$, and adds $t\epsilon\cdot\left(\mathbf{1}_{j} - \mathbf{1}_{i}\right)$ to $\y$ since the $t\epsilon\mathbf{1}_{B_{t}}$ term in Eq.\eqref{eq:y-format} is outdated. Finally, it updates the corresponding $q_R$ values. 
In each iteration, $F(\y)$ remains non-decreasing and the quality of the rounded result \eat{upon completion }is guaranteed by the following lemma.

\begin{small}
\begin{algorithm}[!t]
\edit{\KwIn{An RR collection $\R$, $0<\epsilon\leq1$, $\Q_\R$ defined in Algorithm 1, basic solutions $B_1, B_2, \dots, B_{1/\epsilon}$}
\KwOut{A discrete set $S\in\BPoly$}
}
$\y\gets \epsilon\cdot\sum_{l=1}^{1/\epsilon}\mathbf{1}_{B_l}$;\\
\For{$t=1,2,\dots,1/\epsilon-1$}{
\While{$B_t\not=B_{t+1}$}{
    $u_i\gets$ an arbitrary element in $B_t\backslash B_{t+1}$;\\
    $u_j\gets$ an element in $B_{t+1} \backslash B_t$ s.t.\ $B_t\backslash\{u_i\}\cup\{u_j\}\in\BPoly$ and $B_{t+1}\backslash\{u_j\}\cup\{u_i\}\in\BPoly$;\\
    \If{$\sum\limits_{R\in\R:~u_i\in R}\frac{q_R}{1-\y[i]}\geq\sum\limits_{R\in\R:~u_j\in R}\frac{q_R}{1-\y[j]}$}{
        $B_{t+1}\gets B_{t+1}\backslash\{u_{j}\}\cup\{u_{i}\}$;~$\y\gets \y+\epsilon\left(\mathbf{1}_{i} - \mathbf{1}_{j}\right)$;\\
        $\forall R\in\R:u_{i}\in R,~q_R\gets q_R\cdot\frac{1-\y[i]}{1-\y[i]+\epsilon};$\\
        $\forall R\in\R:u_{j}\in R,~q_R\gets q_R\cdot\frac{1-\y[j]}{1-\y[j]-\epsilon};$\\
    }
    \Else{
        $B_{t}\gets B_{t}\backslash\{u_{i}\}\cup\{u_{j}\}$;~$\y\gets \y+t\epsilon\left(\mathbf{1}_{j} - \mathbf{1}_{i}\right)$;\\
        $\forall R\in\R:u_{j}\in R,~q_R\gets q_R\cdot\frac{1-\y[j]}{1-\y[j]+t\epsilon};$\\
        $\forall R\in\R:u_{i}\in R,~q_R\gets q_R\cdot\frac{1-\y[i]}{1-\y[i]-t\epsilon};$\\
    }
    }
    }
    \Return{$B_{1/\epsilon}$;}
\caption{\crounding~$(\R, \epsilon, \Q_\R, B_1, B_2, \dots, B_{1/\epsilon})$}
\label{alg:crounding}
\end{algorithm}
\end{small}

\begin{lemma}\label{the:cround-approx}
For any $1/\epsilon$ basic solutions $B_1,\dots,B_{1/\epsilon}$ with $\x=\epsilon\cdot\sum_{t=1}^{1/\epsilon}\mathbf{1}_{B_t}$,
% The value of $F(\y)$ maintains non-decreasing in each unit step. This implies that upon completion, 
\crounding returns an $S\in \BPoly$ satisfying $\Lambda_\R(S)\geq F(\x)$.
\end{lemma}

In the example shown in Figure~\ref{fig:vis-polytope}(d), \crounding takes the basic solutions $B_1=\{u_1, u_2\}$, $B_2=\{u_2,u_3\}$, and $B_3=\{u_1, u_3\}$ as inputs. Initially, \crounding sets $\y$ to $\frac{1}{3}\cdot\left(\mathbf{1}_{B_1}+\mathbf{1}_{B_2}+\mathbf{1}_{B_3}\right)=\left[\frac{2}{3},\frac{2}{3},\frac{2}{3}\right]$. In the first iteration, $(u_1,u_3)$ is the only conflicting pair for merging $B_1$ and $B_2$. \crounding accepts $u_3$ but removes $u_1$ from $B_1$ based on their partial derivatives, yielding \LL{the merged} $B_1=B_2=\{u_2,u_3\}$. \crounding then updates $\y=\y+\frac{1}{3}\cdot\left(\mathbf{1}_{3} - \mathbf{1}_{1}\right)=\left[\frac{1}{3},\frac{2}{3},1\right]$. In the second iteration, \yiqian{$(u_2,u_1)$} is the only conflicting pair for merging (the updated) $B_2$ and $B_3$. Following similar operations as in the previous iteration, \LL{the merged} $B_2=B_3=\{u_2,u_3\}$ and the fractional result becomes  $\y=\y+\frac{1}{3}\cdot\left(\mathbf{1}_{2} - \mathbf{1}_{1}\right)=\left[0,1,1\right]$. Hence, the final discrete result after rounding is $S = \{u_2,u_3\}$. \qed

\subsection{Putting It Together}~\label{sec:put-together}

\stitle{Correctness} 
Recall in Algorithm~\ref{alg:greedy-framework} that \cseedselect iteratively invokes \cgreedy or \cgreedyp in each iteration $t$. We denote the fractional result at the end of iteration $t$  as $\x_t=\epsilon\cdot\sum_{i=1}^t \mathbf{1}_{B_i}$,  with $\x_0 \gets \mathbf{0}$. We can then rewrite the improvement of $F(\x_t)$ \LL{compared to $F(\x_{t-1})$} (see Lemma~\ref{lem:greedy-approx})  as
$\Lambda_\R(S^o)-F(\x_t)\leq \frac{\Lambda_\R(S^o)-F(\x_{t-1})}{1+\epsilon}$.
By induction, this leads to
$\Lambda_\R(S^o)-F(\x_t)\leq \frac{\Lambda_\R(S^o)}{(1+\epsilon)^{t}}$.
Upon completion of $1/\epsilon$ search iterations, this inequation becomes 
$F(\x_{1/\epsilon})\geq \big(1-{(1+\epsilon)^{-{1}/{\epsilon}}}\big)\cdot \Lambda_\R(S^o)$. Given this inequation and Lemma~\ref{the:cround-approx}, the result $S$ returned by \cseedselect satisfies the following theorem.
\begin{theorem}\label{thm:gans-correct}
The result $S$ of the \cseedselect algorithm satisfies
\vspace{-0.5mm}
$$\textstyle\Lambda_\R(S)\geq \left(1-\frac{1}{(1+\epsilon)^{\frac{1}{\epsilon}}}\right)\Lambda_\R(S^o).$$
\end{theorem}
\vspace{-0.5mm}
\shiqi{
As per Theorem~\ref{thm:gans-correct}, by taking as input $\textstyle \epsilon_s=\max\big\{\epsilon_s:\frac{1}{\epsilon_s}\in\mathbb{N},~1-{(1+\epsilon_s)^{-\frac{1}{\epsilon_s}}}\geq 1-\frac{1}{e}-\epsilon\big\}$ where \yiqiann{$\mathbb{N}=\{0,1,2,\dots\}$}\eat{ is the set of non-negative integers}, the \cseedselect algorithm achieves a $(1-1/e-\epsilon)$-approximation for maximizing the coverage $\Lambda_\R(S)$. This input $\epsilon_s$ can be \eat{efficiently }obtained by binary search.
% We set the $\epsilon_s$ to the largest value satisfying Eq.\eqref{eq:epsr-epss} with $\frac{1}{\epsilon_s}$ being integral.
}
% By the fact that $1-\frac{1}{(1+\epsilon)^{\frac{1}{\epsilon}}}\geq 1-\frac{1}{e}-\frac{\epsilon}{2e}$, \cseedselect achieves an $1-1/e-\epsilon$ approximation for maximizing the coverage $\Lambda_\R(S)$.

\stitle{Running time for the partition matroid}
In each iteration, \cgreedyp computes the partial derivative w.r.t. each $u_j$ from a given partition $U_l$, incurring a cost of $O\left(\sum_{u_j\in U_l}\sum_{R \in \R}\mathbb{I}(u_j \in R)\right)$. Upon completion, \cgreedyp selects $k_l$ elements from each $U_l$, leading to a total cost of $\textstyle O\left(\sum_{l=1}^{h}\left(k_l\cdot\sum_{u_j\in U_l}\sum_{R \in \R}\mathbb{I}(u_j \in R)\right)\right)=O\left(k^*\cdot\sum_{R\in\R}|R|\right),$ where $k^*=max\{k_1,k_2,\dots k_h\}$.
\crounding operates over \yiqian{$1/\epsilon-1$} iterations. In each iteration $t$, it computes partial derivatives for every conflicting pair $(u_i, u_j)$, with $u_i \in B_t\backslash B_{t+1}$ and $u_j \in B_{t+1} \backslash B_t$. Hence, the total running time of \crounding is $ O\left(\sum_{t=1}^{1/\epsilon-1}\sum_{R \in \R}\sum_{u_i\in (B_{t}\backslash B_{t+1})\cup(B_{t+1}\backslash B_{t})}\mathbb{I}(u_i \in R)\right),$
% \begin{align*}
%     &O\left(\sum_{t=1}^{1/\epsilon-1}\left(\sum_{u_i\in B_{t}\backslash B_{t+1}}\sum_{R \in \R}\mathbb{I}(u_i \in R)+\sum_{u_j\in B_{t+1}\backslash B_{t}}\sum_{R \in \R}\mathbb{I}(u_j \in R)\right)\right)\\
%     &=O\left(\textstyle \epsilon^{-1}\sum_{R\in\R}|R|\right).
% \end{align*}
which equals $O\textstyle\left(\epsilon^{-1}\sum_{R\in\R}|R|\right)$. As shown in Algorithm~\ref{alg:greedy-framework}, \cseedselect runs \cgreedyp for $1/\epsilon$ iterations and invokes \crounding once. Therefore, the running time of \cseedselect subject to the partition matroid is $$O\left(k^*\cdot\epsilon^{-1}\cdot \sum_{R\in\R}|R|\right).$$
\edit{As discussed in \S~\ref{sec:opim-time}, $\sum_{R\in\R}|R|$ is nearly linear to the size of the graph and seed sets. For instance, $\sum_{R\in\R}|R|=O(k\epsilon^{-2}|V|\ln|V|)$ in vanilla \im, and this running time becomes $O\left(k^2\epsilon^{-3}|V|\ln|V|\right)$.}
% \begin{equation*}\label{eq:greedy-partition-time}
%     \textstyle O\left(k^*\cdot\epsilon^{-1}\cdot \sum_{R\in\R}|R|\right). 
% \end{equation*}
%where $k^*=\max\{k_1,k_2,\dots k_h\}$.

% \begin{theorem}
%      The \cseedselect algorithm in Algorithm~\ref{alg:greedy-framework} runs in $$O(k^*\epsilon^{-1}\sum_{R\in\R}|R|)$$ time subject to partition matroid constraint, where $k^*=max\{k_1,k_2,\dots k_h\}$.
% \end{theorem}

\stitle{Running time for general matroid}
Unlike the partition matroid case, the general matroid scenario incurs additional cost to confirm if $B\in \I$. Specifically, Line 3 of Algorithm~\ref{alg:cgreedy} checks whether $B\cup \{u_i\}\in \I$ given that $B\in \I$, and we assume that this operation can be finished by an \textit{incremental independence call} in $\phi_{i}$ time. Additionally, \yiqiann{Line 5 of Algorithm~\ref{alg:crounding} identifies} an element $u_j\in B_{t+1}\backslash B_{t}$ satisfying the matroid exchange property for given $B_t,B_{t+1}\in\BPoly$. We assume that each identification can be completed by an \textit{exchange call} in $\phi_{e}$ time.\eat{\LL{A naive implementation of} \cgreedy (see Algorithm~\ref{alg:cgreedy}) typically examines all elements in $U$ and requires  $O(n)$ incremental independence calls for $n$ elements in each iteration.} \LL{\yiqiann{Following \cite{henzinger2023faster}, we implement \cgreedy (see Algorithm~\ref{alg:cgreedy}) by examining the elements in $U$} in decreasing order of their partial derivatives and selecting the first $u_j$ with $B \cup \{u_j\} \in \I$. Whenever an element is found to be incompatible, \ie adding it breaks independence, we can prune it from the candidate list for further iterations.}
\yiqiann{With the above process}, \cgreedy incurs\eat{just} $O(n)$ independence \edit{calls} across {all} iterations. Notice that \cgreedy runs in $r$ iterations, and each examines the derivatives of all $n$ elements. Hence, the time complexity of \cgreedy is $\textstyle O\left(n\hl{\phi_i}+r\left(\sum_{u_i\in U}\sum_{R \in \R}\mathbb{I}(u_i \in R)\right)\right)=O\left(n\hl{\phi_i}+r\cdot\sum_{R\in\R}|R|\right).$
Regarding \crounding, as $|B_{t+1}\backslash B_t| \leq r$, $O(r)$ exchange calls are required for finding a feasible $u_j$ for each of the elements in $B_{t+1}\backslash B_t$. Hence, the entire algorithm requires $O(r/\epsilon)$ exchange calls. \hl{Combining this with the cost analyzed above, \crounding costs a time of $O\left(\epsilon^{-1}\left(r\phi_e+\sum_{R\in\R}|R|\right)\right)$. Therefore, the running time of \cseedselect subject to the general matroid is $O\left(\epsilon^{-1}\left(r\cdot\sum_{R\in\R}|R|+n\phi_i+r\phi_e\right)\right).$}
% \begin{equation*}\label{eq:greedy-general-time}
% \textstyle O\left(\epsilon^{-1}\left(r\cdot\sum_{R\in\R}|R|+n\phi_i+r\phi_e\right)\right).
% \end{equation*}
% } 

As an example, \yiqiann{in the general version of \im for multiple products~\cite{du2017scalable}} subject to \LL{the laminar matroid constraint, the cost of each incremental independence \edit{call} and exchange call is $\phi_{i}=O(\log{n})$ and $\phi_{e}=O(\log{n})$\eat{by maintaining top trees}~\cite{henzinger2023faster}. Here, the running time of \cseedselect is $O\left(\epsilon^{-1}\left(r\sum_{R\in\R}|R|+(n+r)\log{n}\right)\right)$.
}

% which includes $m$ subsets $U_1, U_2, \dots U_m$ of $U$ and corresponding weights $w_1, w_2, \dots w_m$. In the laminar matroid, any two distinct subsets $U_i, U_j$ satisfy either $U_i\subset U_j$, or $U_j\subset U_i$, or $U_i\cap U_j = \emptyset$, and the definition of $\I$ is $\{S\subseteq U: \forall U_i\in [m],|S\cap U_i|\leq w_i\}$. The laminar matroid is used to model the hierarchical community structures~\cite{du2017scalable}. For example, each $U_i$ represents one of the cities in a state and consists of all users underneath.

% \begin{theorem}
%      The \cseedselect algorithm in Algorithm~\ref{alg:greedy-framework} runs in $$O(r\epsilon^{-1}\sum_{R\in\R}|R|+\epsilon^{-1}(n+r^2)M)$$ time under general matroid constraint, and $$O(k^*\epsilon^{-1}\sum_{R\in\R}|R|)$$ time under partition matroid constraint, where $k^*=max\{k_1,k_2,\dots k_h\}$.
% \end{theorem}

\section{RR Set Generation Scheme}\label{sec:opimc}
Recall in Theorem~\ref{thm:gans-correct} that the proposed \cseedselect offers the desired approximation guarantee w.r.t.\ the coverage function $\Lambda_\R(\cdot)$. \LL{In this section we present a Rapid version of \cseedselect, called \ouralgo, which helps \cseedselect to rapidly determine the required number of RR sets, ensuring the approximation guarantee for the original objective function $\sigma(\cdot)$.} \ouralgo follows the state-of-the-art \opimc approach~\cite{tang2018online} but is generalized to \ourproblem.\eat{ Given an \LL{error tolerance $\epsilon$, \ouralgo ensures that the output $S$ of \cseedselect satisfies $\sigma(S)\geq(1-1/e-\epsilon)\cdot\sigma(S^*)$ w.h.p.\ where $S^*$ is the optimal solution.}} \shiqi{In what follows, we briefly introduce the \ouralgo algorithm in \S~\ref{sec:opim-main-idea} and its time complexity in \S~\ref{sec:opim-time}.}
% returns a set $S$ satisfying $\Lambda_\R(S)\geq \left(1-(1+\epsilon)^{-\frac{1}{\epsilon}}\right)\cdot\Lambda_\R(S^o)$, where $S^o=\argmax{S\in\I}\{\Lambda_\R(S)\}$.

\subsection{\ouralgo}\label{sec:opim-main-idea}
\stitle{Main idea}
\shiqi{\ouralgo runs \yiqiann{iteratively} to mitigate the generation of excessive RR sets. Initially, it generates two distinct collections of RR sets, denoted as $\R_1$ and $\R_2$, each with a size of $\theta_1$. At each iteration $i$, the \edit{element} selection algorithm \cseedselect is performed on $\R_1$. Upon selecting a set $S$, $\R_2$ is leveraged to verify whether $S$ satisfies $\sigma(S)\geq(1-1/e-\epsilon)\cdot\sigma(S^*)$, \yiqian{where $S^{*}=\text{arg~max}_{S\in\I}\sigma(S)$.} If it does, $S$ is returned as the solution. If not, \ouralgo first increases the sizes of $\R_1$ and $\R_2$ for the next iteration $i+1$ by including new RR sets, such that each size is equal to $\theta_{i+1} = 2\cdot\theta_i$, and then repeats the above process. \ouralgo runs in at most $i_{max}$ iterations, and the result $S$ in iteration $i_{max}$ is obtained based on $\R_1$ with $|\R_1|=\theta_{max}$, where $\theta_{max}$ is a carefully-designed constant and ensures that $S$ still provide the desired approximation in this worst case.}
% Notably, $\sigma^l(S)$ is independent of $S$ as it is derived on $\R_2$ but not $\R_1$. Hence, $\sigma^l(S)$ is greatly tightened as per the associated concentration bounds.

\begin{small}
\begin{algorithm}[!t]
\edit{\KwIn{The graph $\R$, the constant 
$\kappa$, the matroid $\M$, an error tolerance $\epsilon$, a failure probability $\delta$}
\KwOut{An \ourproblem solution $S\in\I$ that provides a $1-1/e-\epsilon$ approximation w.p.\ at least $1-\delta$}
}
$\epsilon_s\gets$ Eq.\eqref{eq:eps_s};\\
$\theta_{max}\gets$ Eq.\eqref{equ:theta_max};~$i_{max}\gets \ln{{\kappa}}$; $\theta_{1}\gets\left\lceil\frac{\theta_{max}}{2^{i_{\max}}}\right\rceil$;\\
Generate $\R_1$ and $\R_2$ with $|\R_1|=|\R_2|=\theta_{1}$;\\
\For{$i=1,2,\dots,i_{max}$}{
$S\gets$~\cseedselect$(\R_1, \M, \epsilon_s)$;\\
Compute $\sigma^u(\Sset^*)$ by Eq.\eqref{eq:ub-opim1} on $\Rset_1$ with $p_f=\frac{\delta}{3\cdot i_{max}}$\;
Compute $\sigma^l(\Sset)$ by Eq.\eqref{eq:lb-opim1} on $\Rset_2$ with $p_f=\frac{\delta}{3\cdot i_{max}}$\;
\lIf{$\frac{\sigma^l(S)}{\sigma^u(S^*)}\geq 1-1/e-\epsilon$~\textbf{or}~$i=i_{max}$}{\Return{$S$}}
$\theta_{i+1} \gets 2\cdot\theta_i$;\\
Increase the sizes of $\R_1, \R_2$ such that $|\R_1|=|\R_2|=\theta_{i+1}$;\\
}
\caption{\ouralgo$(\G, \kappa, \M, \epsilon, \delta)$}
\label{alg:ouralgo}
\end{algorithm}
\end{small}

\stitle{Detailed implementation}
\shiqi{As illustrated in Algorithm~\ref{alg:ouralgo}, \ouralgo takes as input a graph $\G$, an \ourproblem instance associated with the matroid $\M$ and the constant $\kappa$, \yiqian{an error tolerance} $\epsilon$, and \yiqian{a failure probability} $\delta$. Recall in \S~\ref{sec:existing} that the constant $\kappa$ is the factor to ensure that $\Lambda_\R(\cdot)$ is an unbiased estimator of $\sigma(\cdot)$ for \LL{given} \ourproblem instance. As shown in Table~\ref{tab:time-rrset}, $\kappa=|V|$ in vanilla \im and \mrim, $\kappa=\sum_{t=1}^T \alpha_t\cdot|V|$ in \rim, and $\kappa=\rho_G(A)-|A|$ in \advim.
Akin to the \yiqian{input} $\epsilon_s$ discussed in \S~\ref{sec:put-together}, Line 1 of Algorithm~\ref{alg:ouralgo} introduces
\begin{equation}\label{eq:eps_s}
\textstyle\epsilon_s=\max\left\{\epsilon_s:\frac{1}{\epsilon_s}\in\mathbb{N},~1-{(1+\epsilon_s)^{-\frac{1}{\epsilon_s}}}\geq 1-\frac{1}{e}-\frac{\epsilon}{2}\right\}
\end{equation}
for the \cseedselect algorithm, \revise{such that its result $S$ satisfies {$\textstyle
    \Lambda_{\R_1}(S)\geq \left(1-1/e-\epsilon/2\right)\cdot\Lambda_{\R_1}(S^o_1),$} where {$S^{o}_1=\text{arg~max}_{S\in\I}\Lambda_{\R_1}(S)$}}.
In Line 2, \ouralgo initializes three constants: (i) the maximum number $\theta_{max}$ of required RR sets, (ii) the maximum number of iterations $i_{max}$, and (iii) the number of RR sets $\theta_{1}$ in the first iteration. \eat{Specifically, }To ensure that the result $S$ returned in iteration $i_{max}$ achieves $\sigma(S)\geq \left(1-1/e-\epsilon\right)\cdot\sigma(S^*)$ w.p.\ at least $1-\delta/3$, \yiqiann{where $S^{*}=\text{arg~max}_{S\in\I}\sigma(S)$, }we set $\theta_{max}$ to    
\begin{equation}\label{equ:theta_max}
  \theta_{max}={\frac{8\kappa\Big((1-\frac{1}{e}-\frac{\epsilon}{2})\sqrt{\ln\frac{6}{\delta}}+\sqrt{(1-\frac{1}{e}-\frac{\epsilon}{2})(\ln{|\mathcal{B}|}+\ln\frac{6}{\delta})}\Big)^2}{\epsilon^2\cdot \sigma^l(S^*)}}.
\end{equation}
In Eq.\eqref{equ:theta_max}, $\sigma^l(S^*)$ is a lower bound of $\sigma(S^*)$.
For \im, \rim, and \mrim, $\sigma^l(S^*)$ can be set as $k$, $\max\limits_{i\leq i\leq n}\{\alpha_i\}\cdot|V|$, and $Tk$, respectively. In \advim, $\sigma(S^{*})$ can be arbitrarily close to $0$ on degenerate instances. Since we are only interested in problem instances where the influence reduction is at least 1, \ie $\sigma(S^{*})\geq 1$, w.l.o.g.\ we can set $\sigma^l(S^*) \gets 1$. % \yiqian{In addition, we can let $\kappa\gets |V|-|A|$ when computing Eq.\eqref{equ:theta_max} in \advim, given that $|V|-|A|\geq \rho_G(A)-|A|$.}

At each iteration $i$, upon obtaining \eat{two collections of RR sets }$\R_1$ and $\R_2$ with $|\R_1|=|\R_2|=\theta_i$, \ouralgo employs \cseedselect to select a set $S$ \eat{that maximizes}to maximize the coverage on $\R_1$ (Lines 4-5). To verify whether current $S$ achieves the desired approximation ratio (Lines 6-8), \ouralgo next computes an upper bound of the optimal solution $\sigma^u(S^*)$, and a lower bound of the current solution $\sigma^l(S)$ by leveraging $\R_1$ and $\R_2$, respectively. The intuition behind this verification is that if $\frac{\sigma^l(S)}{\sigma^u(S^*)}$ reaches $1-1/e-\epsilon$, then $\frac{\sigma(S)}{\sigma(S^*)}\geq\frac{\sigma^l(S)}{\sigma^u(S^*)}$ also satisfies this approximation. By setting
\begin{align}
    \sigma^{u}(S^*) & = \textstyle\left( \sqrt{\Lambda^u_{\R_1}(S^*)-\frac{\ln{p_f}}{2}}+\sqrt{\frac{-\ln{p_f}}{2}}\right)^2\cdot \frac{\kappa}{|\Rset_1|}, \textrm{ and} \label{eq:ub-opim1}
\end{align}
\begin{align}
    \sigma^{l}(S) & = \left(\textstyle\left( \sqrt{\Lambda_{\Rset_2}(S)-\frac{2\ln{p_f}}{9}}-\sqrt{\frac{-\ln{p_f}}{2}}\right)^2+ \textstyle \frac{\ln{p_f}}{18}\right)\cdot \frac{\kappa}{|\Rset_2|}\label{eq:lb-opim1}
\end{align}
% \begin{equation}\label{eq:ub-opim1}
%     \sigma^{u}(S^*) = \left(\textstyle \sqrt{\Lambda^u_{\R_1}(S^*)-\frac{\ln{p_f}}{2}}+\sqrt{\frac{-\ln{p_f}}{2}}\right)^2\cdot \frac{\kappa}{|\Rset_1|}, \textrm{ and}
% \end{equation}
% \begin{equation}\label{eq:lb-opim1}
% \sigma^{l}(S) = \left(\left(\textstyle \sqrt{\Lambda_{\Rset_2}(S)-\frac{2\ln{p_f}}{9}}-\sqrt{\frac{-\ln{p_f}}{2}}\right)^2+ \textstyle \frac{\ln{p_f}}{18}\right)\cdot \frac{\kappa}{|\Rset_2|}
% \end{equation}
with $p_f=\frac{\delta}{3\cdot i_{max}}$, we can obtain that $\sigma^{u}(S^*)$ is an upper bound of $\sigma(S^*)$ w.p.\ at least $1-\frac{\delta}{3\cdot i_{max}}$ and $\sigma^{l}(S)$ is a lower bound of $\sigma(S)$ w.p.\ at least $1-\frac{\delta}{3\cdot i_{max}}$.
In Eq.\eqref{eq:ub-opim1}, \yiqian{$\Lambda^u_{\R_1}(S^*)$ is an upper bound of $\Lambda_{\R_1}(S^*)$}, which can be set to $\frac{\Lambda_{\R_1}(S)}{1-\frac{1}{e}-\frac{\epsilon}{2}}$\eat{ as per Eq.\eqref{eq:epsr-epss}}. To tighten \eat{this upper bound}$\Lambda^u_{\R_1}(S^*)$, we follow the idea of~\cite{tang2018online} that computes it based on the current fractional solution $\x$ during the iterations of \cseedselect.
See \trref{\ref{sec:tight-bound}} for a detailed explanation.

To summarize, based on the above settings of $\theta_{max}$, $\sigma^{u}(S^*)$, and $\sigma^{l}(S)$, by the union bound, the correctness of \ouralgo follows. 
\begin{theorem}\label{the:correctness-opim}
    Given a graph $\G$, an \ourproblem instance associated with the matroid $\M$ and the constant $\kappa$, an error tolerance $\epsilon$, and a failure probability $\delta$, \ouralgo returns a set $S$ satisfying $\sigma(S)\geq (1-1/e-\epsilon)\cdot \sigma(S^*)$ with probability at least $1-\delta$, where $S^{*}=\argmax{S\in\I}\sigma(S)$.
\end{theorem}
}

\subsection{Time Complexity Analysis}\label{sec:opim-time}
Denote the expected time of generating an RR set as $EPT$, and the expected size of an RR set as $EPS$. In each iteration $i$, the computational overhead of \ouralgo comes from (i) generating RR sets\eat{, whose cost is} in $O(\mathbb{E}[\theta_{i}\cdot EPT])$, and (ii) \LL{an invocation of} \cseedselect\eat{, whose cost is} in $O\left(k^*\cdot\epsilon_s^{-1}\cdot \sum_{R\in\R_1}|R|\right)=O\left(k^*\cdot\epsilon_s^{-1}\cdot\mathbb{E}[\theta_{i}\cdot EPS]\right)$ for the partition matroid, as analyzed in \S~\ref{sec:put-together}. Let $i'$ be the iteration where \ouralgo stops. \yiqiann{As proved in~\cite{tang2015influence}, $\mathbb{E}[{\theta_{i'}}\cdot EPT]=\mathbb{E}[\theta_{i'}]\cdot EPT$ and $\mathbb{E}[{\theta_{i'}}\cdot EPS]=\mathbb{E}[\theta_{i'}]\cdot EPS$. Thus,} the time complexity of \ouralgo for the partition matroid is
$\textstyle O\left(\mathbb{E}\left[\sum_{i=1}^{i'}\theta_{i}\right]\left(EPT+\frac{k^*}{\epsilon_s}\cdot EPS\right)\right)$,
where $k^*=\max\{k_1,\dots k_h\}$, and $\mathbb{E}\left[\sum_{i=1}^{i'}\theta_{i}\right]$ is bounded as follows.
\begin{lemma}\label{lem:rr-set-num}
    When $\delta\leq 1/2$, \ouralgo totally generates $O((\ln{|\mathcal{B}|}+\ln{(1/\delta)})\kappa\epsilon^{-2}/\sigma(S^*))$ RR sets in expectation.
\end{lemma}
Given this lemma and $\epsilon_s=O(\epsilon)$, when $\delta\leq 1/2$, the expected time complexity of \ouralgo subject to the partition matroid is
$$\textstyle O\left(\frac{(\ln{|\mathcal{B}|}+\ln{(1/\delta)})\kappa\epsilon^{-2}}{\sigma(S^*)} \cdot \left(EPT+\epsilon^{-1}\cdot k^*\cdot EPS\right)\right).$$
Akin to the above analysis, \edit{when $\delta\leq 1/2$}, the time complexity of \ouralgo for the general matroid is
\begin{equation*}
    \textstyle O\bigg(\frac{(\ln{|\mathcal{B}|}+\ln{(1/\delta)})\kappa\epsilon^{-2}}{\sigma(S^*)}\cdot\left(EPT+\epsilon^{-1}r\cdot EPS\right)+\frac{\left(r\cdot\phi_e+n\cdot\phi_i\right)\cdot \ln \kappa}{\epsilon} \bigg).
\end{equation*}
% \begin{align*}
%     \textstyle O&\bigg(\frac{(\ln{|\mathcal{B}|}+\ln{(1/\delta)})\kappa\epsilon^{-2}}{\sigma(S^*)}\cdot\left(EPT+\epsilon^{-1}r\cdot EPS\right)\\
%     &+\epsilon^{-1}\cdot\left(r\cdot\phi_e+n\cdot\phi_i\right)\cdot \ln \kappa \bigg).
% \end{align*}
For \im, \cite{tang2015influence} shows that $\textstyle EPT=O\left(\frac{|E|}{|V|}\rho^*\right)$ and $EPS=O(\rho^*)$ under \ic and \lt models, where $\rho^*=\max_{v_i\in V}\rho_{\G}(\{v_i\})$. \edit{Besides \im, the original works of \rim, \mrim, and \advim~\cite{han2021efficient, sun2018multi, sun2023scalable} also provide analyses for $EPT$ and $EPS$, which are summarized in Table~\ref{tab:time-rrset}.}
\shiqi{Since \eat{each $EPT$ and $EPS$ in Table~\ref{tab:time-rrset} is }they are related to $\rho^*$ and $\rho^*\leq \sigma(S^*)$,} the above complexities can be further simplified\eat{ for each instance}. \yiqiann{E.g., for \mrim with $\kappa=|V|$ and $|\mathcal{B}|=\binom{|V|}{k}^T$, the time complexity is $\textstyle O\left(T^2k\epsilon^{-2}_t(\ln{|V|}+\ln{(1/\delta)})\cdot(|E|+k\epsilon^{-1}|V|)\right)$.}

\begin{table}[!t]
\centering
\renewcommand{\arraystretch}{1.2}
\caption{$\kappa$, $EPT$, and $ EPS$ in instances ($\rho^*=\max\limits_{v_i\in V}\rho_{\G}(\{v_i\})$).}\vspace{-4mm}
\begin{footnotesize}
% \resizebox{\columnwidth}{!}{
\begin{tabular}{lrrr}
\toprule
\textbf{Name}   & $\kappa$     & $EPT$     & $EPS$  \\
\midrule
\textbf{Vanilla \im}   & $|V|$ & $O\left(\frac{|E|}{|V|}\rho^*\right)$     & $O(\rho^*)$  \\ 
\textbf{\rim}  & $\sum_{t=1}^T \alpha_t\cdot|V|$ &  $O\left(\frac{|E|}{|V|}\rho^*\right)$     & $O(\rho^*)$  \\
\textbf{\mrim}  & $|V|$ & $O\left(\frac{T\cdot|E|}{|V|}\rho^*\right)$     & $O\left(T\cdot\rho^*\right)$  \\
\textbf{\advim}  & $\rho_G(A)-|A|$ & $O\left(\frac{|V\backslash A|}{\rho_{\G}(A)-|A|}\cdot\frac{|E|}{|V|}\rho^*\right)$     & $O\left(\rho^*\right)$  \\
\bottomrule
\end{tabular}
% }
\end{footnotesize}\label{tab:time-rrset}
\end{table}

\input{fig/fig-rm-spread}
\input{fig/fig-mrim-spread}
\input{fig/fig-advim-spread}

\section{Experiments}\label{sec:exp}
We first introduce the experimental settings in \S~\ref{sec:exp-setting}, and then evaluate the performance of our proposal in \S~\ref{sec:exp-performance}. All experiments are conducted on a Linux machine with Intel Xeon(R) Gold 6240@2.60GHz CPU and 377GB RAM in single-thread mode. \eat{As a preview, in our experiments, the case consuming the most memory (Figure~\ref{fig:scale-rm-time}(f)) occupies no more than 30\% of the total memory.}

\subsection{Experimental Setup}\label{sec:exp-setting}

\stitle{Datasets}
\revise{\edit{We experiment with \revise{7} real-world datasets that are widely adopted in previous works and publicly available at SNAP~\cite{snapnets}: Facebook~\cite{zhu2020pricing,li2018automated,becker2021influence}, Wiki~\cite{li2018automated}, Twitter~\cite{becker2021influence}, Google+~\cite{zhu2020pricing}, Pokec~\cite{tang2018online}, LiveJournal~\cite{tang2018online,zhu2020pricing}, and Twitter-large~\cite{tang2018online}. The Twitter-large dataset is the largest dataset ever used in instances \rim, \mrim, and \advim.}
The dataset statistics are listed in Table~\ref{tab:dataset}.
% In addition, previous instances of \ourproblem were evaluated on the graph with at most xxx edges~\cite{xxx}. In contrast, this work includes the LiveJournal graph containing tens of millions of edges. 
}

\begin{table}[!t]
\centering
\caption{\revise{Dataset statistics ($K=10^3, M=10^6, B=10^9$).}}\label{tab:dataset}
\vspace{-3mm}
\renewcommand{\arraystretch}{1.2}
\begin{footnotesize}
\begin{tabular}{ccccc}
\toprule
\textbf{Name}     & \textbf{\#nodes} & \textbf{\#edges} & \revise{\textbf{Type}} & \revise{\textbf{Avg. degree}} \\ 
% \textbf{Name}     & Facebook & Wiki & Twitter & Google+ & Pokec & LiveJournal \\ 
\midrule 
Facebook & $4.0K$ & $176.4K$ & \revise{friendship} & \revise{$43.6$}\\
Wiki & $7.1K$ & $103.6K$ & \revise{who-votes-on-whom} & \revise{$14.5$}\\
Twitter & $81.3K$ & $1.7M$ & \revise{who-follows-whom} & \revise{$21.7$}\\
Google+ & $107.6K$ & $13.7M$ & \revise{who-follows-whom} & \revise{$283.3$}\\
Pokec & $1.6M$ & $30.6M$ & \revise{friendship} & \revise{$18.7$}\\
LiveJournal & $4.8M$ & $69.0M$ & \revise{friendship} & \revise{$14.2$}\\
Twitter-large & $41.7M$ & $1.5B$ & \revise{who-follows-whom} & \revise{$35.2$}\\
\bottomrule
\end{tabular}
\end{footnotesize}
\end{table}

\stitle{\ourproblem instances and configurations}
We focus on the instances \im, \rim, \mrim, and \advim, as illustrated in \edit{\S~\ref{sec:imgm}-\ref{sec:imgm-instance}.} Following the default configurations in \yiqiann{related works}~\cite{han2021efficient, sun2018multi, sun2023scalable, guo2020influence}, we use the \ic model in \rim, \mrim, \im and the \lt model in \advim. We set $p_{i,j}=1/|N^{in}_j|$ for all $e_{i,j}\in E$. \eat{In addition, }We configure $T=10,\alpha_i = 1,k_j=1$ for all $i\in [T],v_j\in V$ for \rim, set $T=20,k=100$ for \mrim, and $k=2000$ for \im. In the absence of a ground-truth set $A$ for \advim, where $|A|$ represents \shiqi{the number of contagious seeds} \eat{the existing spread }as detected by the authority, we generate a synthetic set $A$. This is accomplished by initially adding the node $v_i$\eat{, which has the largest out-degree,} with the largest out-degree to set $A$. Following this, each out-neighbor $v_j$ of $v_i$ is independently added to $A$ with a probability of $p_{i,j}$. Furthermore, we set $k_v=500$ and $k_e=1000$, which are comparable in magnitude to $|A|$.

\stitle{Algorithms and constant settings} 
We evaluate the performance of 9 algorithms in total, which are divided into two parts: 5 algorithms for \edit{element} selection, and 4 algorithms for their scalable implementation. These algorithms are as follows.
\begin{itemize}[topsep=2pt,itemsep=1pt,parsep=0pt,partopsep=0pt,leftmargin=11pt]
    \item \textit{\edit{Element} selection}: \shiqi{\mgreedy, \lgreedy, and \tgreedy (baselines in \S~\ref{sec:existing}); \cseedselect and \cseedselectp for the general matroid and partition matroid, respectively (Algorithm~\ref{alg:greedy-framework}).}
    \item \textit{Scalable solutions}: \shiqi{\rma for \rim~\cite{han2021efficient}, \naimm for \mrim~\cite{sun2018multi}, and \advimm for \advim~\cite{sun2023scalable} (baselines); \ouralgo (Algorithm~\ref{alg:ouralgo}).}
\end{itemize}
For all \edit{element} selection algorithms, we apply the lazy evaluation technique in \cite{minoux1978accelerated,leskovec2007cost} to improve their empirical efficiency.
For \cseedselect and \cseedselectp, $\epsilon$ is set to $1/2$ and $1/8$. Based on Theorem~\ref{thm:gans-correct}, these settings provide an approximation of 0.56 and 0.61 for \eat{the RR set coverage function}\yiqiann{maximizing} $\Lambda_{\R}$, respectively, \shiqi{indicating a focus on efficiency but at the cost of result quality, and vice versa.} For \tgreedy, $\xi$ is set to {the default value} $0.05$~\cite{bian2020efficient}. 
For all scalable implementations, we set $\delta=1/|V|$ by default and vary $\epsilon$ from 0.5 to 0.1~\cite{tang2018online,tang2015influence}. For a fair comparison, we have also made other necessary modifications to scalable competitors and refer interested readers to \trref{\ref{sec:exp-detail}} for details. In addition, for each algorithm, we estimate the objective function value $\sigma$ by averaging over $10,000$ Monte Carlo simulations. We repeat each algorithm 5 times and report the average when evaluating the running time and the objective function value. All algorithms are implemented in C++ with \texttt{-O3} optimization\eat{, available at: {\color{blue}\url{https://anonymous.4open.science/r/IMGM-920B/}}}.

\input{fig/rm-time}

\input{fig/fig-im-spread}

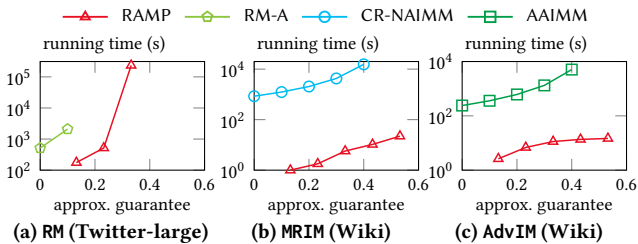
\begin{figure}[!t]
\centering

% \vspace{-1.5mm}
\begin{tikzpicture}
    \begin{customlegend}[legend columns=4,
        legend entries={\ouralgo, \rma, \naimm, \advimm
        }
        ,
        legend style={at={(0.45,1.05)},anchor=north,draw=none,font=\footnotesize,column sep=0.1cm}]
    \addlegendimage{line width=0.25mm,mark size=2pt,mark=triangle,color=Red}
    \addlegendimage{line width=0.25mm,mark size=2pt,mark=pentagon,color=Pink}
    \addlegendimage{line width=0.25mm,mark size=2pt,mark=o, color=LightBlue}
    \addlegendimage{line width=0.25mm,mark size=2pt,mark=square, color=Green}
    \end{customlegend}
    %  \begin{customlegend}[legend columns=6,
    %     legend entries={},
    %     legend style={at={(0.45,0.7)},anchor=north,draw=none,font=\footnotesize,column sep=0.1cm}]
 
    % \end{customlegend}
\end{tikzpicture}
\vspace{-1mm}
\\[-\lineskip]
\hspace{-3.1mm}\subfloat[\revise{\rim (Twitter-large)}]{
\begin{tikzpicture}[scale=1]
    \begin{axis}[
        height=\columnwidth/2.8,
        width=\columnwidth/2.25,
        ylabel={running time (s)},
        xlabel={approx. guarantee},
        xmin=0, xmax=0.6,
        ymin=100, ymax=312000,
        xtick={0,0.2, 0.4,0.6},
        xticklabel style = {font=\footnotesize},
        xticklabels={0,0.2, 0.4,0.6},
        ytick={100, 1000, 10000, 100000},
        yticklabels={$10^2$, $10^3$, $10^4$, $10^5$},
        ymode=log,
        log basis y={10},
        every axis y label/.style={at={(current axis.north west)},right=8.5mm,above=0mm},
        label style={font=\footnotesize},
        tick label style={font=\footnotesize},
        every axis x label/.style={at={(current axis.south)},right=0mm,above=-7mm},
        label style={font=\footnotesize},
        tick label style={font=\footnotesize},
    ]

    \addplot[line width=0.2mm,mark size=2.0pt,mark=pentagon,color=Pink]
        plot coordinates { 
        (0,513.603)
        (0.1, 2087.931)
        % (0.2,1003543.871)
        % (0.3,13.252)
        % (0.4,101.423)
    };

    \addplot[line width=0.2mm,mark size=2.0pt,mark=triangle,color=Red]
        plot coordinates { % ours
        (0.13212+0, 175.318)
        (0.13212+0.1, 517.097)
        (0.13212+0.2, 234613.649)
        % (0.13212+0.3, 2127025.820)
        % (0.13212+0.4, 35.460)
    };
    
    \end{axis}
\end{tikzpicture}\vspace{1mm}\hspace{1mm}
}
\hspace{-4.2mm}\subfloat[\mrim (Wiki)]{
\begin{tikzpicture}[scale=1]
    \begin{axis}[
        height=\columnwidth/2.8,
        width=\columnwidth/2.25,
        ylabel={running time (s)},
        xlabel={approx. guarantee},
        xmin=0, xmax=0.6,
        ymin=1, ymax=20000,
        xtick={0,0.2, 0.4,0.6},
        xticklabel style = {font=\footnotesize},
        xticklabels={0,0.2, 0.4,0.6},
        ytick={1, 100, 10000},
        yticklabels={$10^0$, $10^2$, $10^4$},
        ymode=log,
        log basis y={10},
        every axis y label/.style={at={(current axis.north west)},right=8.5mm,above=0mm},
        label style={font=\footnotesize},
        tick label style={font=\footnotesize},
        every axis x label/.style={at={(current axis.south)},right=0mm,above=-7mm},
        label style={font=\footnotesize},
        tick label style={font=\footnotesize},
    ]

    \addplot[line width=0.2mm,mark size=2.0pt,mark=o, color=LightBlue]
        plot coordinates { % CR-MRIMM
        (0, 847.220)
        (0.1, 1235.159)
        (0.2, 2048.609)
        (0.3, 4266.819)
        (0.4, 15437.723)
    };

    \addplot[line width=0.2mm,mark size=2.0pt,mark=triangle,color=Red]
        plot coordinates { % ours
        (0.13212+0, 1)
        (0.13212+0.1, 1.784)
        (0.13212+0.2, 5.708)
        (0.13212+0.3, 10.442)
        (0.13212+0.4, 22.186)
    };
    \end{axis}
\end{tikzpicture}\vspace{1mm}\hspace{1mm}%
}%
\hspace{-4.1mm}\subfloat[\advim (Wiki)]{
\begin{tikzpicture}[scale=1]
    \begin{axis}[
        height=\columnwidth/2.8,
        width=\columnwidth/2.25,
        ylabel={running time (s)},
        xlabel={approx. guarantee},
        xmin=0, xmax=0.6,
        ymin=1, ymax=10000,
        xtick={0,0.2, 0.4,0.6},
        xticklabel style = {font=\footnotesize},
        xticklabels={0,0.2, 0.4,0.6},
        ytick={1, 100, 10000},
        yticklabels={$10^0$, $10^2$, $10^4$},
        ymode=log,
        log basis y={10},
        every axis y label/.style={at={(current axis.north west)},right=8.5mm,above=0mm},
        label style={font=\footnotesize},
        tick label style={font=\footnotesize},
        every axis x label/.style={at={(current axis.south)},right=0mm,above=-7mm},
        label style={font=\footnotesize},
        tick label style={font=\footnotesize},
    ]

    \addplot[line width=0.2mm,mark size=2.0pt,mark=square, color=Green]
        plot coordinates { % AAIMM
        (0, 239.339)
        (0.1, 358.641)
        (0.2, 611.916)
        (0.3, 1320.682)
        (0.4, 5054.434)
    };

    \addplot[line width=0.2mm,mark size=2.0pt,mark=triangle,color=Red]
        plot coordinates { % ours
        (0.13212+0, 2.625)
        (0.13212+0.1, 6.846)
        (0.13212+0.2, 11.308)
        (0.13212+0.3, 13.567)
        (0.13212+0.4, 14.711)
    };
    \end{axis}
\end{tikzpicture}\vspace{1mm}\hspace{1mm}%
}%
\vspace{-3mm}
\caption{\revise{Total running time of scalable algorithms.}} \label{fig:scale-all-time}
% \vspace{-4mm}
\end{figure}
% \input{fig/scale-rm-time}
% \input{fig/scale-other-time}

% \vspace{-2mm}
\subsection{Performance Evaluations}\label{sec:exp-performance}

\stitle{Performance of \edit{element} selection in \rim, \mrim, and \advim}
In the first set of experiments, we evaluate the result quality and running time of \edit{element} selection algorithms on the \ourproblem instances \rim, \mrim, and \advim. For a fair comparison, we provide all algorithms with the same \LL{collection of RR sets $\R$}. Furthermore, we vary $|\R|$ from $2^5$ to $2^{21}$ in \rim; and from $2^1$ to $2^{17}$ in \mrim and \advim. \shiqi{The choices of the smallest and largest values for $|\R|$ roughly correspond to $\theta_{1}$ and $\theta_{max}$ in Algorithm~\ref{alg:ouralgo} when $\epsilon=0.1$.}

\noindent
{\it Result quality.} As shown in Figures~\ref{fig:rm-spread}-\ref{fig:advim-spread}, both \cseedselect and \cseedselectp return better solutions compared to their competitors in terms of related objective functions across all datasets. \revise{For instance,  \cseedselect or \cseedselectp outperforms all competitors by up to $86\%$ on Pokec in \rim, $708\%$ on Wiki in \mrim, and $205.8\%$ on Twitter-large in \advim}.
To explain, \cgreedy and \cgreedyp select each element based on the largest partial derivative and subsequently update the value of $q_R$ for each $R$ containing the selected element. In contrast to the competitors \mgreedy, \lgreedy, and \tgreedy, \shiqi{our} update strategy circumvents the direct exclusion of each related $R$ and offers a more fine-grained marginal coverage for each element. As a result, it allows for a more refined and potentially more effective selection process.
Furthermore, we observe that the result quality of \tgreedy is much worse than other approaches when the number of RR sets is small. This is because it disregards all elements whose marginal coverage is below the minimum threshold, rendering the number of selected elements fewer than expected.

% \stitle{Running time}
\noindent 
\textit{Running time.} In the second set of experiments, we use \rim as the instance and report the running times of \edit{element} selection algorithms in Figure~\ref{fig:rm-time}. Similar trends were observed in the running time results for other instances. Specifically, \lgreedy and \cseedselectp (with $\epsilon=1/2$) are the fastest and second-fastest algorithms respectively, across all datasets, each being $4-10\times$ faster than \mgreedy. Recall from the first set of experiments that \lgreedy, despite being the fastest, is the second least effective algorithm, quality wise. Furthermore, \shiqi{we compare \cseedselect and \cseedselectp with the same $\epsilon$ values across these problems using partition matroids. In particular, }\cseedselectp improves the running time by an order of magnitude. These observations demonstrate the efficiency of \cseedselectp, as proved in \S~\ref{sec:put-together}. Note that, in Figure~\ref{fig:rm-time}(e), the running time of \cseedselect decreases when $|\R|\geq 2^{15}$, due to the lazy evaluation technique. 
We also report the time taken to generate $\R$. RR set generation becomes more expensive than all \eat{\edit{element} selection }methods as the size of $\R$ increases. For instance, when $|\R|> 2^9$ on Google+, the time taken to generate $\R$ exceeds that of \cseedselectp by more than an order of magnitude. %Specifically, the cost of RR set generation becomes overwhelming if a large number of redundant RR sets are generated. 
This motivates the design of an efficient scheme for RR set generation, as proposed in \S~\ref{sec:opimc}.

\noindent 
\textbf{Performance of \edit{element} selection in \im.}
In the third set of experiments, we evaluate the performance of the proposed \edit{element} selection methods on the conventional \im problem. Here, we only compare \cseedselect with \mgreedy and \tgreedy as \cseedselect and \cseedselectp (resp. \mgreedy and \lgreedy) are \LL{essentially equivalent.} Figure~\ref{fig:im-spread} reports the spread of each method on different datasets \edit{as} $|\R|$ is varied from $2^1$ to $2^{17}$. Notably, \shiqi{\cseedselect outperforms competitors by up to $871\%$ on LiveJournal. This is attributed to the more fine-grained marginal coverage, as discussed earlier.}

\noindent 
\textbf{Effect of $\epsilon$ in \cseedselect.}
Next, we examine the impact of the $\epsilon$ parameter in the proposed \cseedselect algorithm. Here, we focus on the result qualities of \cseedselect and \mgreedy, as shown in Figures~\ref{fig:rm-spread}-\ref{fig:im-spread}. Note that \cseedselect is equivalent to \mgreedy when $\epsilon=1$. Based on observations in the first and third sets of experiments, the result quality of \cseedselect improves significantly by decreasing $\epsilon$ from $1$ to $1/2$ (\ie increasing the number of iterations in Line 4 of Algorithm~\ref{alg:greedy-framework} from 1 to 2). This improvement becomes less pronounced with further increases in the number of iterations. For example, in \mrim, the solution quality of \cseedselect improves by $84\%$ on Pokec when $|\R|=2^9$ as $\epsilon$ decreases from $1$ to $1/2$. In contrast, the improvement is only $7\%$ when comparing $\epsilon=1/2$ with $\epsilon=1/8$.
% the result qualities of 4 variants of our proposal, \cseedselect and \cseedselectp with $\epsilon=1/2$ and $\epsilon=1/8$, differ by less than $3\%$ in most cases;

% \subsection{Performance of \ouralgo}\label{sec:exp-opim}
\stitle{Performance of scalable implementations}
\revise{Finally, we evaluate the running times of \ouralgo and all scalable competitors by varying $\epsilon$ from $0.5$ to $0.1$. In Figure \ref{fig:scale-all-time}, the approximation \yiqiann{ratio} of each solution is used as the x-axis, with \ouralgo having $1-1/e-\epsilon$ and all competitors having $1/2-\epsilon$;} in Figure~\ref{fig:scale-all-time}(a), we exclude approaches that run more than 72 hours. \revise{Due to space constraints, we only report results for the largest dataset on which the baseline algorithm with $\epsilon=0.5$ can terminate within 72 hours. These datasets are Twitter-large in \rim and Wiki in \mrim and \advim. Interested readers can find more experimental results, including the memory usage of scalable algorithms, in \trref{\ref{sec:more-exp}}.}

\eat{Figure \ref{fig:scale-all-time} reports the running time for each scalable implementation for \rim, \mrim, or \advim on the largest available dataset; \LL{in Figure~\ref{fig:scale-all-time}(a), we exclude approaches that run more than 72 hours.}}
Specifically, \ouralgo consistently outperforms the baseline regarding running time while ensuring the same approximation guarantee across all cases. Notably, \ouralgo is at least \eat{three orders of magnitude}$1,000\times$ faster than \naimm in \mrim and at least \eat{two orders}$100\times$ faster than \advimm in \advim.
\revise{In addition, \ouralgo completes within 66 hours for \rim on Twitter-large when $\epsilon=0.3$, whereas \rma only returns results for $\epsilon\geq 0.4$ and fails to terminate even after 7 days when $\epsilon=0.3$.}
The reason for the efficiency of \ouralgo is twofold. First, \ouralgo introduces the early termination and the tightened bound \edit{$\Lambda^u_{\R_1}(S^*)$}, which allows us to generate an appropriate number of RR sets as soon as a $(1-1/e-\epsilon)$-approximation is reached. Furthermore, \cseedselect provides superior result qualities during iterations, \LL{enabling \ouralgo to reach the desired approximation faster.} In contrast, \rma, \naimm, and \advimm generate a large number of RR sets and rely on \mgreedy or \lgreedy for \edit{element} selection, leading to inferior result qualities. \shiqi{To summarize, the proposed \ouralgo is significantly faster than all competitors while providing the same result quality.}

% We also report the running time of \ouralgo with $\epsilon=0.1$ on larger datasets in Table~\ref{tab:time-scalable-others}, which demonstrates the scalability of our proposal. Note that \naimm and \advimm fail to terminate within 24 hours on larger datasets.

% \begin{table}[!t]
% \centering
% \renewcommand{\arraystretch}{1.2}
% \caption{Running time of \ouralgo with $\epsilon=0.1$ on larger datasets, in seconds.}\vspace{-2mm}
% \begin{small}
% % \resizebox{\columnwidth}{!}{
% \begin{tabular}{|c|c|c|c|c|}
% \hline
% \textbf{Problem}   & Twitter     & Google+  & Pokec  & Orkut  \\ \hline
% \mrim   & \hl{running} & \hl{running} & \hl{running} & \hl{running} \\ \hline
% \advim   & 70.605     & 2813.481  & 1351.196  & 7689.379  \\ \hline
% \end{tabular}
% % }
% \end{small}\label{tab:time-scalable-others}
% \vspace{-2mm}
% \end{table}
\section{Deployment}\label{sec:deploy}
\LL{We have deployed our proposed \ouralgo in a real industry setting for GameX \yiqiann{of Tencent}.\eat{\W{(name of game suppressed following anonymity guidelines)}.}} The details are \LL{described} below.

\stitle{Application scenario}
GameX is a \yiqiann{Tencent} multiplayer online game with hundreds of millions of users and a massive number of user-generated maps. A map in the game is a vibrant and dynamic environment featuring several obstacles and puzzle-solving settings, allowing multiple users to compete with each other simultaneously. Besides, the social network in GameX can be constructed based on friendships formed in the game, with an average of $\sim$16 friends per user, making it sufficiently dense~\cite{Backstrom2012separation}. As such, users can interact with their friends on the social network, such as \W{playing or inviting to play the game in some maps}, which contributes to cascades of maps in the game. In other words, a user might \W{play the game in a map} if their friends have played or shared it. To further promote the maps, the game \LL{environment} can generate $k$ maps as recommendations for each user, who can receive them \LL{during the game and click the maps they are interested in playing in.} The performance of GameX is measured by the number of \W{engaged user-map pairs}, each of which denotes that a user plays a user-generated map, reflecting the engagement of users in the game.

% \begin{table}[!t]
% \centering
% \caption{The percentage of users that have the same actual spread of a given map.}\label{tab:map-spread-histogram}
% \vspace{-2mm}
% \renewcommand{\arraystretch}{1.2}
% \begin{small}
% \begin{tabular}{ccccccc}
% \toprule
% Spread     & 1          & 2          & 3         & 4          & 5          & 6      \\
% \midrule
% Percent & 18.88\% & 19.50\% & 19.69\% & 15.60\% & 10.83\% & 6.07\% \\
% \bottomrule
% \end{tabular}
% \end{small}
% \end{table}

\stitle{Control and treatment groups}
We have deployed three approaches: (i) \mclick, (ii) \rma, and (iii) \ouralgo. \LL{Specifically, for the control group, we have deployed \mclick, the state-of-the-art solution for this scenario.} It operates by first ranking maps for each user $v_i$ based on the descending order of probability $w_{i,t}$ that $v_i$ will click map $t$ and then selecting the top-$k$ maps with the highest click probabilities for each user.
% Note that, the probability $w_{i,t}$ is learned using XGBoost \cite{Chen2016xgboost} based on the historical data on the behaviors of users playing maps, as well as the features of users and maps collected from the statistical data in the game. 
\LL{We learn the probability $w_{i,t}$ using XGBoost \cite{Chen2016xgboost}, a machine learning model to predict whether a given user would play the game in a given map.}
% The second control group, \mcascade, replies on the actual spread $s_{i,t}$, which represents the number of users (directly and indirectly) influenced by $v_i$ during the cascade of map $t$ in the latest month. This is equal to the size of the subtree of $t$'s diffusion tree~\cite{zhang2023capacity} rooted at $v_i$. The percentage of users that have the same actual spread is shown in Table~\ref{tab:map-spread-histogram}. Based on the actual spread, \mcascade ranks all $s_{i,t}$ scores for each user $v_i$ in descending order and selects $k$ maps where this user has generated the largest actual spread. 

\LL{We have deployed \rma and \ouralgo for two different treatment groups.} For treatment groups, we focus on addressing the \rim problem and returning $k$ user-map pairs for each user. Notably, we set each unit revenue to 1 and employ the \ic model as the diffusion model. In this model, the influence probability $p_{i,j}$ for each $e_{i,j}\in E$ is calculated as $\frac{c_{i,j}}{\sum_{v_i\in N^{in}_j}c_{i,j}}$, where $c_{i,j}$ denotes the number of historical interactions on $e_{i,j}$, such as co-playing, gifting, etc. To address \rim in this context, we first leverage the competing solution \rma~\cite{han2021efficient}, as mentioned in \S~\ref{sec:exp}, the core of which is selecting $k$ seeds for each map using \mgreedy. Next, we implement the proposed \ouralgo by targeting an enhanced objective function $\sigma^W(S)$ based on $\sigma(S)$. This function is defined as $$\sigma^W(S) = \sum_{S'\subseteq S}\prod_{(v_i,t)\in S'}w_{i,t}\prod_{(v_i,t)\in S\backslash S'}(1-w_{i,t})\cdot\sigma(S'),$$ 
where $W$ is a given \LL{(or learned)} set of click probabilities that contains $w_{i,t}$ for every user-map pair $(v_i,t)$. By adjusting the scale of $q_R$ in \cseedselect, \ouralgo can be adapted to this new objective while maintaining the same approximation guarantee and time complexity. For further details, readers are referred to \trref{\ref{sec:dep-detail}}.

\begin{table}[!t]
\centering
\caption{Numbers of engaged pairs %of each algorithm 
in GameX ($M=10^6$).}\label{tab:deploy-result}
\vspace{-3mm}
\renewcommand{\arraystretch}{1.2}
\begin{footnotesize}
\begin{tabular}{lrrr}
\toprule
\textbf{Algorithm}     & \mclick          & \rma          & \ouralgo    \\
\midrule
\textbf{\#engaged pairs} & 7.34M & 7.36M & 7.50M \\
\bottomrule
\end{tabular}
\end{footnotesize}
\end{table}

\stitle{Deployment setups}
This deployment is conducted on an in-house cluster \LL{in Tencent} consisting of hundreds of machines, each of which has 16GB memory and 12 Intel Xeon Processor E5-2670 CPU cores. \LL{To mitigate} the network effect, \shiqi{the phenomenon whereby user behaviors are influenced by others within the same network,} we follow~\cite{saveski2017detecting} and conduct cluster-level experimentation. Specifically, we first partition all users into several communities of almost the same size with high edge connectivity and node feature similarity, and then randomly assign the live traffic in the same community to the treatment or control groups. Furthermore, we set the number of recommended maps $k$ to 2. As a result, each approach is assigned to 0.7 million sampled users with 400 randomly selected maps to recommend, leading to 277.5 million user-map pairs for observation. \shiqi{In this scenario, the constraint is a partition matroid $\M=(U,\I)$, where $U$ comprises all pairs of users and maps (277.5 million) and $\mathcal{I}$ consists of all solutions, each containing $k=2$ maps per user.}

% \begin{table}[!t]
% \centering
% \caption{The number of engaged pairs of each algorithm in GameX ($M=10^6$).}\label{tab:deploy-result}
% \vspace{-2mm}
% \renewcommand{\arraystretch}{1.2}
% \begin{small}
% \begin{tabular}{cccc}
% \toprule
% Algorithm     & \mclick          & \mcascade          & \ouralgo    \\
% \midrule
% \# engaged pairs & 7.34M & 7.42M & \textbf{7.50M} \\
% \bottomrule
% \end{tabular}
% \end{small}
% \end{table}

\stitle{Online performance}
Table~\ref{tab:deploy-result} reports the number of engaged pairs for each algorithm. Notably, the treatment groups \eat{\rma and \ouralgo}achieve more engaged pairs compared to the control group \mclick \LL{ -- 20,000 (0.27\%) for \rma and 160,000 (2.17\%) for \ouralgo,} demonstrating the usefulness of involving the word-of-mouth effect in this scenario. In addition, \ouralgo yields 140,000 (1.9\%)\eat{and 160 thousand} more engaged pairs than \rma. \eat{and \mclick, respectively.} \shiqi{To explain, unlike \mclick which solely focuses on the probability of a seed adopting the promoted map, \ouralgo takes into account not only this inclination but also the subsequent word-of-mouth effect, as captured by the \ic model with its associated influence probability. The improvement of \ouralgo over \rma can be attributed to two factors. First, \ouralgo achieves a $(1-1/e-\epsilon)$-approximation, while the result of \rma is \LL{$(1/2-\epsilon)$-approximate.} Second, the enhanced objective $\sigma^W(S)$ comprehensively considers the adoption inclination of any subset of $S$. In contrast, even if the revenue unit $\alpha_t$ is leveraged, \rma can only involve the average inclination for each map $t$.} \LL{While the percentage improvement may appear modest, the improved numbers of engaged pairs lead to substantial additional revenue for the industry.}

% 1.1\% for \mcascade, 2.2\% for \mclick
% Table~\ref{tab:deploy-result} reports the number of engaged pairs for each algorithm. As demonstrated, \ouralgo surpasses both \mclick and \mcascade. Specifically, \ouralgo yields 80 thousand and 160 thousand more engaged pairs than \mcascade and \mclick, respectively. The superior performance of \ouralgo can be attributed to its comprehensive approach, which considers not only the probability of a seed adopting the promoted map but also the ensuing word-of-mouth effect, as modeled by the \ic model with its associated influence probability. In contrast, \mclick and \mcascade focus solely on one of these aspects.

\section{Conclusions}
In this paper, we focus on applications requiring multiple seed sets for \im, which are formulated as instances of \im subject to a matroid constraint. For effectiveness, we propose \cseedselect, \eat{the first ever \LL{scalable}}\sq{which achieves} a $(1-1/e-\epsilon)$-approximation guarantee for this problem. For enhanced efficiency, we also propose a fast implementation called \ouralgo. Our comprehensive experiments demonstrate that our proposal outperforms state-of-the-art methods in both effectiveness and efficiency. Moreover, we have successfully deployed \ouralgo in an online gaming propagation scenario in a real industry setting, yielding considerable improvements. \LL{In the future, to\eat{align even more closely with the deployment environment} exploit the infrastructure in the industry further, we will consider solutions in the distributed setting. It is also interesting to explore real-world applications of \im under other types of matroid constraints and develop scalable and effective solutions for them.}

\bibliographystyle{ACM-Reference-Format}
\bibliography{refs}
\balance

\clearpage
\appendix
\section{Correction on Liao et al.'s Assertion}~\label{sec:prm-note}

\stitle{Popularity ratio maximization (\prm)~\cite{liao2023popularity}} Given a graph $G=(V,E)$, the number of campaigns $T$, an overall capacity $k$, and three parameters $d^n_0, d^p_0, z$ which define the objective function $\sigma_{PRM}$, \prm aims to find $$S_1,\dots, S_T=\argmax{S_1,\dots, S_T\subseteq V,\sum_{t=1}^{T}|S_t|\leq k,S_i\cap S_j=\emptyset\forall i\not=j}\sigma_{PRM}(S_1,\dots, S_T).$$

Similar to \rim, we denote $U=V\times[T]$ and divide $U$ into $|V|$ disjoint paritions $U_1, U_2,\dots,U_{|V|}$, where $U_i=\{(v_i,t): t\in[T]\}$. In addition, we define a uniform matroid $\M_u=(U,\I_u)$ and a partition matroid $\M_p=(U,\I_p)$, where $\I_u=\{S\subseteq U:|S|\leq k\}$ and $\I_p=\{S\subseteq U:\forall v_i\in V, |S\cap U_i|\leq 1\}$.
Accordingly, the constraint of \prm is $\M_{PRM}=(U,\I_{PRM})$, where $\I_{PRM}=\{S:S\text{ is feasible}\}=\I_u\cap \I_p$. 

\stitle{Liao et al.'s Assertion} \citet{liao2023popularity} claim that $\M_{PRM}$ is a partition matroid. We review its definition below.

\begin{definition}[Partition Matroid]\label{def:partition-matroid}
A matroid $\M=(U,\I)$ is a partition matroid iff there exist $h\geq 1$ sets $U_1, \dots, U_h$, and $h$ positive integers $k_1, \dots, k_h$, such that $\I=\{S \subseteq U : |S \cap U_i| \leq k_i, \forall i \in[h]\}$.
\end{definition}

However, this claim is incorrect, as we demonstrate in the following theorem.

\begin{theorem}
$\M_{PRM}$ is not a partition matroid.
\end{theorem}

\begin{proof}
    We prove it by constructing a counterexample of $\M_{PRM}$ that is not a partition matroid. Before introducing the counterexample, we first present Lemmas~\ref{lem:judge1}-\ref{lem:judge3} which will be frequently utilized for showing that a matroid is not a partition matroid. \textit{It is easy to verify their correctness.}

\begin{figure}[t]
    \centering
    \includegraphics[width=0.4\linewidth]{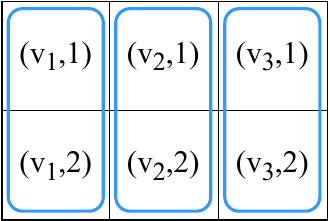}
    \caption{The illustration of $U$ in the counterexample. The squares represent the elements in $U$ and the blue boxes represent the partitions in $\M_p$.}
    \label{fig:prm-example}
\end{figure}

\begin{lemma}\label{lem:judge1}
A matroid $\M=(U,\I)$ is not a partition matroid iff $\M$ is not identical to any possible partition matroid with ground set $U$, \ie the set of their independent sets are different.
\end{lemma}

\begin{lemma}\label{lem:judge2}
For two matroids $\M$ and $\M'$, if their ranks are not equal, then $\M$ and $\M'$ are not identical.
\end{lemma}

\begin{lemma}\label{lem:judge3}
For two matroids $\M$ and $\M'$, denote the set of all bases of them as $\BPoly$ and $\BPoly'$, respectively. If $|\BPoly|\not=|\BPoly'|$, then $\M$ and $\M'$ are not identical.
\end{lemma}

\noindent\textit{The counterexample.}
Consider a special case of $\M_u,\M_p$, and $\M_{PRM}$ where $V=\{v_1,v_2,v_3\},T=2,k=2$. Figure~\ref{fig:prm-example} illustrates the ground set $U$ in this case, where $\M_p$ constrains that at most 1 square can be selected per column (the blue boxes), $\M_u$ constrains that a maximum of $k=2$ squares can be selected in total, and $\M_{PRM}$ incorporates these constraints. Notably, $\M_{PRM}$ is a general matroid with the rank $=2$, and it has $C^3_2\times 2^2=12$ different bases\footnote{We can obtain a base in Figure~\ref{fig:prm-example} by first selecting 2 columns out of 3 ($C^3_2$), then selecting 1 square out of 2 in each column ($2^2$).}. In light of Lemma~\ref{lem:judge1}, we next show that $\M_{PRM}$ is not identical to any partition matroid.

Given an integer $x$, denote a partition matroid with ground set $U$ and the number of partitions $h=x$ as $\M_{h=x}$. For any $\M_{h=3}$, since each $k_i$ (the cardinality constraint in partition $U_i$) of $\M_{h=3}$ is at least $1$ (see Definition~\ref{def:partition-matroid}), the rank of $\M_{h=3}$ must exceed 2. By lemma~\ref{lem:judge2}, $\M_{PRM}$ is not identical to any possible $\M_{h=3}$. Anagolously, $\M_{PRM}$ is not identical to any possible $\M_{h=4},\M_{h=5},$ or $\M_{h=6}$. Among all $\M_{h=1}$, there is only one with the rank equal to $\M_{PRM}$, which is $(U,\{S\subseteq U:|S|\leq 2\})=\M_u\not=\M_{PRM}.$ Therefore, $\M_{PRM}$ is also not identical to any possible $\M_{h=1}$.

Regarding the case of $h=2$, if $k_1>1$ or $k_2>1$ in one $\M_{h=2}$, the rank of $\M_{h=2}$ will also exceed 2, hence $\M_{PRM}$ is not identical to $\M_{h=2}$ due to Lemma~\ref{lem:judge2}. Therefore, we only consider any $\M_{h=2}$ with $k_1=k_2=1$, denoted as $\M^*_{h=2}$. We further divide all possible $\M^*_{h=2}$ into three categories based on the size of their partitions $|U_1|$, $|U_2|$, and denote the number of bases in each $\M^*_{h=2}$ as $|\BPoly^*_{h=2}|$: (i) if $|U_1|=1,|U_2|=5$, then $|\BPoly^*_{h=2}|=1\times 5=5$; (ii) if $|U_1|=2,|U_2|=4$, then $|\BPoly^*_{h=2}|=2\times 4=8$; (iii) if $|U_1|=3,|U_2|=3$, then $|\BPoly^*_{h=2}|=3\times 3=9$. As per Lemma~\ref{lem:judge3}, none of them is identical to $\M_{PRM}$ as $\M_{PRM}$ has 12 different bases.

Consequently, $\M_{PRM}$ is not identical to any $\M_{h=1},\dots,\M_{h=6}$. Given that $\M_{h=1},\dots,\M_{h=6}$ contains all possible partition matroids on $U$, by Lemma~\ref{lem:judge1}, $\M_{PRM}$ is not a partition matroid.
\end{proof}
\section{Proofs}\label{sec:proof-all}

\subsection{Proofs for Results in Section~\ref{sec:greedy}}\label{sec:proof-amp}

This section provides the proofs for Lemma~\ref{lem:marginal-reflect}, Lemma~\ref{lem:greedy-approx}, and Lemma~\ref{the:cround-approx}.
We first introduce some basic properties of the multilinear extension, which will be frequently utilized in the following proofs.
Specifically, given the multilinear extension $F$, we have the following properties.
\begin{proposition}~\label{prop:linear-multilinear}
For any $i\in [n]$ and $\epsilon\geq0$ (assuming that $\x[i]+\epsilon\leq 1$), $F(\x+\epsilon\cdot\mathbf{1}_i)=F(\x)+\epsilon\frac{\partial F(\x)}{\partial \x[i]}$.
\end{proposition}

\begin{proposition}~\label{prop:mono-multilinear}
If $f$ is non-decreasing, then $\frac{\partial F(\x)}{\partial \x[i]}\geq 0$ for any $i$. If $f$ is submodular, then $\frac{\partial^2 F(\x)}{\partial x[i] \partial x[j]}\leq 0$ for any $i,j$, which implies that $\frac{\partial F(\x)}{\partial \x[i]}$ is non-increasing.
\end{proposition}

\stitle{\bf Proof of Lemma~\ref{lem:marginal-reflect}}
As per Proposition~\ref{prop:linear-multilinear} and the definition of $F$ in Eq.\eqref{eq:multilinear-full-format}, we can derive that
\begin{align*}
    &\frac{\partial F(\y)}{\partial \y[i]}=\frac{1}{\epsilon}\left(F(\y+\epsilon\cdot\mathbf{1}_i)-F(\y)\right)\\
    % =&\sum_{S\subseteq U\backslash\{u_i\}}\left(\prod_{u_j\in S}\y[j]\prod_{u_j\in U\backslash\{u_i\}\backslash S}(1-\y[j])\right)\cdot \left(\Lambda_\R(S\cup \{u_i\})-\Lambda_\R(S)\right)\\
    =&\sum_{S\subseteq U\backslash\{u_i\}}\left(\prod_{u_j\in S}\y[j]\prod_{u_j\in U\backslash\{u_i\}\backslash S}(1-\y[j])\right)\cdot \Lambda_\R(u_i|S).
\end{align*}
On the other hand,
\begin{align*}
    &\sum_{S\subseteq U}\left(\prod_{u_j\in S}\y[j]\prod_{u_j\in U\backslash S}(1-\y[j])\right)\cdot \Lambda_\R(u_i|S)\\
    =&\sum_{S\subseteq U:u_i\notin S}\left((1-\y[i])\prod_{u_j\in S}\y[j]\prod_{u_j\in U\backslash\{u_i\}\backslash S}(1-\y[j])\right) \Lambda_\R(u_i|S)\\
    +&\sum_{S\subseteq U:u_i\in S}\left(\prod_{u_j\in S}\y[j]\prod_{u_j\in U\backslash S}(1-\y[j])\right) \cdot0\\
    =&\sum_{S\subseteq U\backslash\{u_i\}}\left((1-\y[i])\prod_{u_j\in S}\y[j]\prod_{u_j\in U\backslash\{u_i\}\backslash S}(1-\y[j])\right)\Lambda_\R(u_i|S).
\end{align*}
Hence, the above-said partial derivative turns to
\begin{align}
    &\frac{1}{1-\y[i]}\left(\sum_{S\subseteq U}\left(\prod_{u_j\in S}\y[j]\prod_{u_j\in U\backslash S}(1-\y[j])\right)\cdot \Lambda_\R(u_i|S)\right)\notag\\
    =&\frac{1}{1-\y[i]}\mathbb{E}\left[\Lambda_\R(u_i|\Omega(\y))\right]\label{eq:derive-expectation}\\
    =&\frac{1}{1-\y[i]}\sum_{R\in \R: u_i \in R}\mathbb{E}\left[\mathbb{I}(R\cap \Omega(\y)=\emptyset)\right].\notag
\end{align}
In addition, for any $R\in\R$, $\mathbb{E}\left[\mathbb{I}(R\cap \Omega(\y)=\emptyset)\right]=\prod_{u_i\in R}(1-\y[i])$. By the definition of $q_R$ in Eq.\eqref{eq:qr}, we can derive that 
$$\frac{\partial F(\y)}{\partial \y[i]} = \sum_{R\in \R: u_i\in R}\frac{\prod_{u_j\in R}(1-\y[j])}{1-\y[i]}=\sum\limits_{R\in\R:u_i\in R}\frac{q_R}{1-\y[i]}.$$
\qed

\stitle{\bf Proof of Lemma~\ref{lem:greedy-approx}}
We denote $B^{(i)}=\{u_{b(1)},\dots,u_{b(i)}\}$ as the independent set after selecting element $u_{b(i)}$ in iteration $i$ with $B=B^{(r)}$, and $S^o=\{u_{s(1)},\dots,u_{s(r)}\}$ as the optimal result on $\Lambda_{\R}$.
To complete this proof, we provide the following lemma.
\begin{lemma}\label{lem:feasible-s}
In \cgreedy, a permutation of $S^o$ exists such that $\forall i\in[r], u_{s(i)} \in U\backslash B^{(i-1)}$ and $B^{(i-1)}\cup \{u_{s(i)}\}\in\I$. In \cgreedyp, a permutation of $S^o$ exists such that $\forall i\in[r], u_{s(i)} \in U_l\backslash B^{(i-1)}$, where $U_l$ is the selected partition in Line 3 of Algorithm~\ref{alg:cgreedy-partition}.
\end{lemma}
Lemma~\ref{lem:feasible-s} guarantees that there exists a permutation of $S^o$ such that $u_{s(i)}$ is among the candidates when choosing $u_{b(i)}$ in both algorithms. This implies that the selection of $u_{b(i)}$, which has the largest partial derivative, ensures
\begin{equation}\label{equ:marginal-gain-exceed-s}
\frac{\partial F(\x+\epsilon\cdot\mathbf{1}_{B^{(i-1)}})}{\partial \x[b(i)]}\geq \frac{\partial F(\x+\epsilon\cdot\mathbf{1}_{B^{(i-1)}})}{\partial \x[s(i)]}.
\end{equation}
For brevity, below we let $\x'=\x+\epsilon\cdot\mathbf{1}_{B}$. Therefore, we have
\begin{align*}
    &F(\x')-F(\x)\\    =&\sum_{i=1}^{r}\left(F(\x+\epsilon\cdot\mathbf{1}_{B^{(i)}})-F(\x+\epsilon\cdot\mathbf{1}_{B^{(i-1)}})\right)\\
    =&\epsilon\cdot\sum_{i=1}^{r}\frac{\partial F(\x+\epsilon\cdot\mathbf{1}_{B^{(i-1)}})}{\partial \x[b(i)]}\tag{Proposition~\ref{prop:linear-multilinear}}\\
    \geq&\epsilon\cdot\sum_{i=1}^{r}\frac{\partial F(\x+\epsilon\cdot\mathbf{1}_{B^{(i-1)}})}{\partial \x[s(i)]}\tag{Eq.\eqref{equ:marginal-gain-exceed-s}}
\end{align*}
\begin{align*}
    \geq&\epsilon\cdot\sum_{i=1}^{r}\frac{\partial F(\x')}{\partial \x[s(i)]}\tag{Proposition~\ref{prop:mono-multilinear}}\\
    \geq&\epsilon\cdot\sum_{i=1}^{r}\mathbb{E}\left[\Lambda_\R(u_{s(i)}|\Omega(\x'))\right]\tag{Eq.\eqref{eq:derive-expectation}}\\
    \geq&\epsilon\cdot\mathbb{E}\left[\Lambda_\R(S^o\cup \Omega(\x'))-\Lambda_\R(\Omega(\x'))\right]\tag{submodularity}\\
    \geq&\epsilon\cdot\left(\Lambda_\R(S^o)-F(\x')\right)\tag{monotonicity}.
\end{align*}

\qed

\stitle{\bf Proof of Lemma~\ref{lem:feasible-s}}
We mainly utilize the following lemma.
\begin{lemma}[\cite{brualdi_1969}]\label{lem:bijection}
Given two bases $B_1, B_2$ of $\M$, there is a bijection $\phi:B_1\rightarrow B_2$ such that $\forall u_i\in B_1$, $B_1\backslash\{u_i\}\cup\{\phi(u_i)\}$ is also a base.
\end{lemma}
We let $u_{s(i)}=\phi\left(u_{b(i)}\right)$ and denote $B'=B\backslash\{u_{b(i)}\}\cup\{u_{s(i)}\}$, which is also a base in $\I$ by Lemma~\ref{lem:bijection}.
For \cgreedy in Algorithm~\ref{alg:cgreedy}, we prove $\forall i \in [r], B^{(i-1)}\cup \{u_{s(i)}\}\in\I$ and $u_{s(i)} \in U\backslash B^{(i-1)}$ by contradiction. Specifically, if $B^{(i-1)}\cup \{u_{s(i)}\}\not\in\I$, then $B'=B^{(i-1)}\cup\{u_{s(i)}, u_{b(i+1)},\dots,u_{b(r)}\}\not\in\I,$ which contradicts the fact that $B'\in \I$.
If $u_{s(i)} \not\in U\backslash B^{(i-1)}$, it implies that $\exists j \in [i-1], u_{s(i)}=u_{b(j)}$ and $B'=B\backslash\{u_{b(i)}\}$, which is not a base since $|B\backslash\{u_{b(i)}\}|<|B|=r$ and leads to a contradiction.
    
For \cgreedyp in Algorithm~\ref{alg:cgreedy-partition}, we prove $\forall i \in [r], u_{s(i)}\in U_l\backslash B^{(i-1)}$ in a similar way. In particular, if $u_{s(i)} \in U_q, q\not=l$, then $|B'\cap U_l|=k_l-1$ and $|B'\cap U_q|=k_q+1$, contradicting the definition of base.
Akin to the second case mentioned above, if $u_{s(i)} \in B^{(i-1)}$, then $B'=B\backslash\{u_{b(i)}\}$ and $B'<r$, resulting in a contradiction. \qed

\stitle{\bf Proof of Lemma~\ref{the:cround-approx}}
To complete the proof, we provide the following corollary.
\begin{corollary}[\cite{calinescu2011maximizing}]\label{cor:Ft}
Given $F$ be the multilinear extension of a monotone and submodular function $f$. For any $\y\in[0,1]^{U}$ and $i,j\in [n]$, the function $F^{\y}_{i,j}(t)=F(\y+t(\mathbf{1}_{i}-\mathbf{1}_{j}))$ is convex.
\end{corollary}
In other words, the derivative $\frac{\mathrm{d} F^{\y}_{i,j}(t)}{\mathrm{d} t}$ is non-decreasing as $t$ increases.
Based on this, we focus on each unit step in \crounding, where it selects a conflicting element pair $(u_i, u_j)$ between $B_t$ and $B_{t+1}$, and decides on a swap direction.
In Line 6 of Algorithm~\ref{alg:crounding}, \crounding computes the derivative at $t=0$, 
\begin{equation}\label{equ:derivate}
    \frac{\mathrm{d} F^{\y}_{i,j}(0)}{\mathrm{d} t}=\frac{\partial F(\y)}{\partial \y[i]}-\frac{\partial F(\y)}{\partial \y[j]}.
\end{equation}
As per Corollary~\ref{cor:Ft}, if Eq.\eqref{equ:derivate} is non-negative, then $\forall t>0, \frac{\mathrm{d} F^{\y}_{i,j}(t)}{\mathrm{d} t}\geq 0$. This indicates that updating $\y$ to $\y+\epsilon(\mathbf{1}_{i}-\mathbf{1}_{j})$ results in an non-decreasing update in $F(\y)$ from $F^{\y}_{i,j}(0)$ to $F^{\y}_{i,j}(\epsilon)$.
Conversely, if Eq.\eqref{equ:derivate} is negative, an analogous deduction shows that $F^{\y}_{i,j}(t)$ is non-increasing for $t<0$. 
Accordingly, \crounding updates $F(\y)$ to $F^{\y}_{i,j}(-t\epsilon)$, ensuring it is non-decreasing. \qed

\subsection{Proofs for Results in Section~\ref{sec:opimc}}\label{sec:proof-ampim}

This section provides the proofs for Theorem~\ref{the:correctness-opim} and Lemma~\ref{lem:rr-set-num}.
We mainly utilize two martingale-based concentration bounds as shown in the following lemma, which can be derived from Corollary 1 and Corollary 2 of~\cite{tang2015influence}.

\begin{lemma}\label{lem:martingale-imgm}
Given a set $\Rset$ of random RR sets with $|\Rset|=\theta$ and a set $S\subseteq U$ independent of $\Rset$, for any $\lambda>0$,
\begin{align}
\Pr&\left[\Lambda_{\Rset}(S)-\frac{\theta}{\kappa}\cdot\sigma(S)\geq \lambda\right]\leq \exp{\left(-\frac{\lambda^2}{2\frac{\theta}{\kappa}\sigma(S)+\frac{2}{3}\lambda}\right)},\label{eq:martingale-lb} \\
\Pr&\left[\Lambda_{\Rset}(S)-\frac{\theta}{\kappa}\cdot\sigma(S)\leq -\lambda\right]\leq \exp{\left(-\frac{\lambda^2}{2\frac{\theta}{\kappa}\sigma(S)}\right)}.\label{eq:martingale-ub}
\end{align}
\end{lemma}

\stitle{Proof of Theorem~\ref{the:correctness-opim}}
To complete the proof, we first establish the following lemma to ensure the correctness of $\theta_{max}$.

\begin{lemma}\label{lem:theta-max}
   Given constants $\epsilon>0$, $\beta>0$, $\delta >0$ and a set $S\subseteq U$ satisfying $\Lambda_\R(S)\geq \beta\cdot\Lambda_\R(S^o)$, if $$|\R|\geq\frac{2\kappa\Big(\beta\sqrt{\ln\frac{2}{\delta}}+\sqrt{\beta(\ln{|\mathcal{B}|}+\ln\frac{2}{\delta})}\Big)^2}{\epsilon^2\sigma(S^{*})},$$ then $\Pr[\sigma(S)<(\beta-\epsilon)\sigma(S^{*})]\leq\delta$.
\end{lemma}
By replacing $\epsilon$ with $\frac{\epsilon}{2}$, $\beta$ with $1-\frac{1}{e}-\frac{\epsilon}{2}$, and $\delta$ with $\frac{\delta}{3}$, we obtain the value of $\theta_{max}$ in Eq.\eqref{equ:theta_max}. 
Regarding $\sigma^u(S^*)$ and $\sigma^l(S)$, we utilize the following lemma.
\begin{lemma}\label{lem:ub-lb-opim}
Given a constant $p_f>0$, by setting
\begin{equation}\label{eq:ub-opim}
    \sigma^{u}(S^*) = \left(\sqrt{\Lambda^u_{\R_1}(S^*)-\frac{\ln{p_f}}{2}}+\sqrt{\frac{-\ln{p_f}}{2}}\right)^2\cdot \frac{\kappa}{|\Rset_1|},
\end{equation}
\begin{equation}\label{eq:lb-opim}
\sigma^{l}(S) = \left(\left(\sqrt{\Lambda_{\Rset_2}(S)-\frac{2\ln{p_f}}{9}}-\sqrt{\frac{-\ln{p_f}}{2}}\right)^2+\frac{\ln{p_f}}{18}\right)\cdot \frac{\kappa}{|\Rset_2|},
\end{equation}
we have $\Pr[\sigma(S^*)>\sigma^{u}(S^*)]\leq p_f$ and $\Pr[\sigma(S)<\sigma^{l}(S)]\leq p_f$.
\end{lemma}
Based on Lemma~\ref{lem:ub-lb-opim}, $\sigma^u(S^*)$ in Eq.\eqref{eq:ub-opim1} and $\sigma^l(S)$ in Eq.\eqref{eq:lb-opim1} are obtained using Eq.\eqref{eq:ub-opim} and Eq.\eqref{eq:lb-opim}, respectively, with $p_f=\delta/(3\cdot i_{max})$.

As guaranteed above, each computation of $\sigma^u(S^*)$ or $\sigma^l(S)$ is correct w.p. at least $1-\delta/(3\cdot i_{max})$. In addition, when $|\R_1|=|\R_2|=\theta_{max}$, \ouralgo achieves a $(1-1/e-\epsilon)$ approximation w.p. at least $1-\delta/3$. By the union bound, \ouralgo achieves a $1-1/e-\epsilon$ approximation w.p. at least $1-\delta$. \qed

\stitle{Proof of Lemma~\ref{lem:theta-max}}
To complete the proof, we first establish the following lemma, which illustrates that when the size $|\R|=\theta$ is sufficiently large, $\frac{\kappa}{\theta}\Lambda_{\Rset}(S^*)$ is close to $\sigma(S^*)$.

\begin{lemma}\label{lem:spread-coverage}
Given constants $\epsilon_1>0$ and $\delta_1>0$, if $\theta\geq \theta'_1 = \frac{2\kappa\ln{(1/\delta_1)}}{\sigma(S^*)\cdot\epsilon^2_1}$, then
$$\Pr\left[\Lambda_{\Rset}(S^*)\geq (1-\epsilon_1)\frac{\theta}{\kappa}\cdot \sigma(S^*)\right] \geq 1-\delta_1.$$
\end{lemma}
Combining it with the fact that $\Lambda_\R(S)\geq \beta\cdot\Lambda_\R(S^o)$, we can derive that Eq.\eqref{eq:lem9} holds w.p. at least $1-\delta_1$ if $\theta\geq \theta'_1$:
\begin{equation}\label{eq:lem9}
\Lambda_{\Rset}(S)\geq \beta\cdot\Lambda_\R(S^o)\geq \beta\cdot\Lambda_\R(S^*)\geq\beta\cdot(1-\epsilon_1)\frac{\theta}{\kappa}\sigma(S^*).
\end{equation}

We now use the following lemma to connect $\sigma(S)$ with $\sigma(S^*)$.

\begin{lemma}\label{lem:spread-spread}
Given constants $\epsilon_2=\epsilon-\beta\cdot{\epsilon_1}\geq 0$ and $\delta_2>0$, if Eq.\eqref{eq:lem9} holds and 
$$\theta\geq \theta'_2 = \frac{2\beta \kappa\cdot\left(\ln {|\mathcal{B}|} +\ln{\frac{1}{\delta_2}}\right)}{\sigma(S^*)\cdot\epsilon^2_2},$$
then
$$\Pr\left[\sigma(S)\geq (\beta-\epsilon)\cdot \sigma(S^*)\right] \geq 1-\delta_2.$$
\end{lemma}
Combining Lemmas~\ref{lem:spread-coverage}-\ref{lem:spread-spread} and the union bound, we derive that if $\theta\geq \max(\theta'_1,\theta'_2)$, we have $\sigma(S)\geq (\beta-\epsilon)\cdot \sigma(S^*)$ w.p. of at least $1-\delta_1-\delta_2$.
By setting $\delta_1=\delta_2=\delta/2$ and $$\epsilon_1 = \frac{\epsilon}{\left(\beta+\sqrt{\frac{\ln{|\mathcal{B}|}+\ln{\frac{1}{\delta_2}}}{\frac{1}{\beta}\ln{\frac{1}{\delta_1}}}}\right)},$$ we have
$$\theta'_1=\theta'_2=\frac{2\kappa\Big(\beta\sqrt{\ln\frac{2}{\delta}}+\sqrt{\beta(\ln{|\mathcal{B}|}+\ln\frac{2}{\delta})}\Big)^2}{\epsilon^2\sigma(S^{*})}.$$ \qed

\stitle{Proof of Lemma~\ref{lem:spread-coverage}}
As per Eq.\eqref{eq:martingale-ub}, we have
\begin{align*}
    &\Pr\left[\Lambda_{\Rset}(S^*)\leq (1-\epsilon_1)\frac{\theta}{\kappa}\cdot \sigma(S^*)\right]\\
    \leq&\Pr\left[\Lambda_{\Rset}(S^*)-\frac{\theta}{\kappa}\cdot \sigma(S^*)\leq -\epsilon_1\frac{\theta}{\kappa}\cdot \sigma(S^*)\right]\\
    \leq &\exp\left(-\frac{\epsilon^2_1 \sigma(S^*) {\theta}}{2\kappa}\right) \leq \exp\left(-\frac{\epsilon^2_1 \sigma(S^*) {\theta'_1}}{2\kappa}\right) \leq \delta_1.
\end{align*}
\qed

\stitle{Proof of Lemma~\ref{lem:spread-spread}}
For a set $S'\in \BPoly$, we say $S'$ is bad if $\sigma(S')<(\beta-\epsilon)\sigma(S^*)$.
Based on Eq.\eqref{eq:martingale-lb}, for a bad $S'$, we have
\begin{equation*}
    \begin{split}
        &\Pr\left[\frac{\kappa}{\theta}\cdot\Lambda_{\Rset}(S')-\sigma(S')\geq \epsilon_2\cdot\sigma(S^*)\right]\\
        % =&\textstyle\pr\left[\sum_{i=1}^{\theta}x_i-\theta\cdot\frac{\sigma(\Sset')}{n_p}\geq \frac{\theta}{n_p}\cdot\epsilon_2\cdot\sigma(\Sset^*)\right]\\
        \leq&\exp{\left(-\frac{\epsilon_2^2\cdot\sigma(S^*)^2}{2 \sigma(S')+\frac{2}{3}\epsilon_2 \sigma(S^*)}\cdot \frac{\theta}{\kappa} \right)}\\
        \leq&\exp{\left(-\frac{\epsilon_2^2\cdot\sigma(S^*)}{2\cdot (\beta-\epsilon)+\frac{2}{3}\cdot \epsilon_2}\cdot \frac{\theta}{\kappa} \right)}\\
        \leq&\exp{\left(-\frac{\epsilon_2^2\cdot\sigma(S^*)\theta'_2}{2\beta\cdot \kappa}\right)}\\
        \leq&{\delta_2}\big/|\mathcal{B}|.\\
    \end{split}
\end{equation*}
Since there exist at most $|\mathcal{B}|$ possible $S'$, by the union bound, we obtain that for a $S$ that satisfies Eq.\eqref{eq:lem9},
$$\sigma(S)\leq \frac{\kappa}{\theta}\cdot\Lambda_{\Rset}(S')-\epsilon_2\cdot\sigma(S^*)\leq (\beta-\epsilon)\cdot\sigma(S^*)$$
w.p. at most $\delta_2$. \qed

\stitle{Proof of Lemma~\ref{lem:ub-lb-opim}}
We denote $\theta^{(1)}=|\R_1|$ and $\theta^{(2)}=|\R_2|$.
Regarding the upper bound $\sigma^{u}(S^*)$, based on Eq.\eqref{eq:martingale-ub} in Lemma~\ref{lem:martingale-imgm}, we have
\begin{equation*}
    \begin{split}
        & \Pr\left[\sigma(S^*)> \left(\sqrt{\Lambda^u_{\Rset_1}(S^*)-\frac{\ln{p_f}}{2}}+\sqrt{\frac{-\ln{p_f}}{2}}\right)^2\cdot \frac{\kappa}{\theta^{(1)}}\right] \\
        \leq & \Pr\left[\sigma(S^*)> \left(\sqrt{\Lambda_{\Rset_1}(S^*)-\frac{\ln{p_f}}{2}}+\sqrt{\frac{-\ln{p_f}}{2}}\right)^2\cdot \frac{\kappa}{\theta^{(1)}}\right] \\
        \leq & \Pr\left[\left(\sqrt{\frac{\theta^{(1)}\cdot\sigma(S^*)}{\kappa}}-\sqrt{\frac{-\ln{p_f}}{2}}\right)^2>\Lambda_{\Rset_1}(S^*)-\frac{\ln{p_f}}{2}\right] \\
        = & \Pr\left[-\sqrt{-2\cdot\ln{p_f}\cdot\frac{\theta^{(1)}\cdot\sigma(S^*)}{\kappa}}>\Lambda_{\Rset_1}(S^*)-\frac{\theta^{(1)}\cdot\sigma(S^*)}{\kappa}\right] \\
        \leq & \exp{\left(\frac{2\cdot\ln{p_f}\cdot\frac{\theta^{(1)}\cdot\sigma(S^*)}{\kappa}}{2\cdot\frac{\theta^{(1)}\cdot\sigma(S^*)}{\kappa}}\right)} \\
        = & p_f. \\
    \end{split}
\end{equation*}
Regarding the lower bound $\sigma^{l}(S)$, by Eq.\eqref{eq:martingale-lb} in Lemma~\ref{lem:martingale-imgm}, we have
\begin{equation*}
    \begin{split}
        & \Pr{\textstyle \left[\sigma(S)<\left(\left(\sqrt{\Lambda_{\Rset_2}(S)-\frac{2\ln{p_f}}{9}}-\sqrt{\frac{-\ln{p_f}}{2}}\right)^2+\frac{\ln{p_f}}{18}\right)\cdot \frac{\kappa}{\theta^{(2)}}\right]} \\
        = & \Pr{\textstyle \left[\sqrt{\frac{\sigma(S)\cdot\theta^{(2)}}{\kappa}-\frac{\ln{p_f}}{18}}<\sqrt{\Lambda_{\Rset_2}(S)-\frac{2\ln{p_f}}{9}}-\sqrt{\frac{-\ln{p_f}}{2}}\right]} \\
        = & \Pr{\textstyle\left[\sqrt{-\frac{2\ln{p_f}\cdot\sigma(S)\theta^{(2)}}{\kappa}+\frac{\ln^2{p_f}}{9}}-\frac{\ln{p_f}}{3}<\Lambda_{\Rset_2}(S)-\frac{\sigma(S)\theta^{(2)}}{\kappa}\right]} \\
        \leq & \exp{\left(-\frac{\left(\sqrt{-\frac{2\ln{p_f}\cdot\sigma(S)\theta^{(2)}}{\kappa}+\frac{\ln^2{p_f}}{9}}-\frac{\ln{p_f}}{3}\right)^2}{2\cdot \frac{\sigma(S)\theta^{(2)}}{\kappa}+\frac{2}{3}\cdot\left(\sqrt{-\frac{2\ln{p_f}\cdot\sigma(S)\theta^{(2)}}{\kappa}+\frac{\ln^2{p_f}}{9}}-\frac{\ln{p_f}}{3}\right)}\right)}\\
        = & p_f. \\
    \end{split}
\end{equation*}
This completes the proof. \qed

\stitle{Proof of Lemma~\ref{lem:rr-set-num}}
    We denote
    \begin{align*}
        &\theta^{(1)}=|\R_1|, \theta^{(2)}=|\R_2|,\\
        &\beta=1-\frac{1}{e}-\frac{\epsilon}{2},~~\epsilon_1=\epsilon/2,~~\widetilde{\epsilon}_1=(1-\beta)\epsilon_1,~~\widehat{\epsilon}_1=\sqrt{\frac{2a\kappa}{\sigma(S^*)\theta^{(1)}}},\\
        &\epsilon_2=\sqrt{\frac{2a\kappa}{\sigma(S)\theta^{(2)}}},~~\widetilde{\epsilon}_2=\Big(\sqrt{\frac{2a\sigma(S)\theta^{(2)}}{\kappa}+\frac{a^2}{9}}+\frac{a}{3}\Big)\cdot\frac{\kappa}{\sigma(S)\theta^{(2)}},
    \end{align*}
    where $a=c\ln{\frac{3i_{max}}{\delta}}$ for any $c\geq 1$. Furthermore, we denote $\theta'=\max\{\theta'_1, \theta'_2\}$, where
    \begin{equation*}
\theta'_1=\frac{(2+2\widetilde{\epsilon}_1/3)\kappa\ln{\frac{6|\mathcal{B}|}{\delta}}}{\widetilde{\epsilon}^2_1\sigma(S^*)},~~\theta'_2=\frac{27\kappa\ln{\frac{3i_{max}}{\delta}}}{(\beta-\epsilon_1)\epsilon^2_1\sigma(S^*)}.
    \end{equation*}
    One can verify that $\theta'=O((\ln{|\mathcal{B}|}+\ln{(1/\delta)})\kappa\epsilon^{-2}_t/\sigma(S^*))$. According to Lemma~\ref{lem:martingale-imgm}, for any $c\geq 1$, if $\theta^{(1)}=\theta^{(2)}=c\theta'$, we can derive the following results.
    \begin{align}
        &\Pr\left[\frac{\kappa}{\theta^{(1)}}\cdot\Lambda_{\R_1}(S^*)<(1-\epsilon_1)\sigma(S^*)\right]\leq \left(\frac{\delta}{6}\right)^c,\label{eq:event-a}\\
        &\Pr\left[\frac{\kappa}{\theta^{(1)}}\cdot\Lambda_{\R_1}(S)>\sigma(S)+\widetilde{\epsilon}_1\cdot\sigma(S^*)\right]\leq \left(\frac{\delta}{6|\mathcal{B}|}\right)^c,\label{eq:event-b}\\
        &\Pr\left[\frac{\kappa}{\theta^{(1)}}\cdot\Lambda_{\R_1}(S^*)<(1-\widehat{\epsilon}_1)\sigma(S^*)\right]\leq\left(\frac{\delta}{3i_{max}}\right)^c,\label{eq:event-c}\\
        &\Pr\left[\frac{\kappa}{\theta^{(2)}}\cdot\Lambda_{\R_2}(S)<(1-\epsilon_2)\sigma(S)\right]\leq \left(\frac{\delta}{3i_{max}}\right)^c,\label{eq:event-d}\\
        &\Pr\left[\frac{\kappa}{\theta^{(2)}}\cdot\Lambda_{\R_2}(S)>(1+\widetilde{\epsilon}_2)\sigma(S)\right]\leq \left(\frac{\delta}{3i_{max}}\right)^c.\label{eq:event-e}
    \end{align}
    Note that by definition, $\widetilde{\epsilon}_2=\sqrt{\frac{a(2+2\widetilde{\epsilon}_2/3)\kappa}{\sigma(S)\theta^{(2)}}}$, so Eq.\eqref{eq:event-e} is obatined by
    \begin{align*}
    l.h.s.\leq\exp\left(-\frac{\frac{a(2+2\widetilde{\epsilon}_2/3)\kappa}{\sigma(S)\theta^{(2)}}\sigma(S)}{2+2\widetilde{\epsilon}_2/3}\cdot\frac{\theta^{(2)}}{\kappa}\right)=\left(\frac{\delta}{3i_{max}}\right)^c.
    \end{align*}
    Recall in \ouralgo that $S$ is derived from $\R_1$. To satisfy the requirement of Lemma~\ref{lem:martingale-imgm}, we consider any of the $|\mathcal{B}|$ possible values of $S$ in Eq.\eqref{eq:event-b}. By the union bound, the probability that at least one of the events in Eq.\eqref{eq:event-a}-Eq.\eqref{eq:event-e} happens is at most
    \begin{equation*}
        \left(\frac{\delta}{6}\right)^c+\left(\frac{\delta}{6|\mathcal{B}|}\right)^c\cdot|\mathcal{B}|+3\cdot\left(\frac{\delta}{3i_{max}}\right)^c\leq \delta^c.
    \end{equation*}
    
    In the following, we elaborate on the connections implied by these events. If the events in Eq.\eqref{eq:event-a} and Eq.\eqref{eq:event-b} do not happen, we have
    \begin{align*}
        \sigma(S)&\geq\Lambda_{\R_1}(S)-\widetilde{\epsilon}_1\sigma(S^*)\\
        &\geq \beta\cdot\Lambda_{\R_1}(S^*)-\widetilde{\epsilon}_1\sigma(S^*)\\
        &\geq \beta\cdot(1-\epsilon_1)\sigma(S^*)-\widetilde{\epsilon}_1\sigma(S^*)\\
        &=(\beta-\epsilon_1)\sigma(S^*).
    \end{align*}
    Therefore, by the definition of $\widehat{\epsilon}_1, \epsilon_2, \widetilde{\epsilon}_2$ and the fact that $c\theta'\geq \theta'_2$, we have
    \begin{align*}
        \widehat{\epsilon}_1&\leq\sqrt{\frac{2(\beta-\epsilon_1)\epsilon^2_1}{27}}<\epsilon_1/3,\\
        \epsilon_2&\leq\sqrt{\frac{2(\beta-\epsilon)\sigma(S^*)\epsilon^2_1}{27\sigma(S)}}<\epsilon_1/3,\\
        \widetilde{\epsilon}_2&=\sqrt{\frac{a(2+2\widetilde{\epsilon}_2/3)\kappa}{\sigma(S)\theta^{(2)}}}\leq\sqrt{\frac{(2+2\widetilde{\epsilon}_2/3)\epsilon^2_1}{27}}<\epsilon_1/3.
    \end{align*}
    In addition, if the event in Eq.\eqref{eq:event-c} does not happen, we have
    \begin{align*}
        &\frac{\kappa}{\theta^{(1)}}\Lambda_{\R_1}(S^*)\geq\left(1-\sqrt{\frac{2a\kappa}{\sigma(S^*)\theta^{(1)}}}\right)\sigma(S^*)\\
        \Leftrightarrow&\frac{\kappa}{\theta^{(1)}\sigma(S^*)}\geq\frac{1-\sqrt{\frac{2a\kappa}{\sigma(S^*)\theta^{(1)}}}}{\Lambda_{\R_1}(S^*)}\\
        \Leftrightarrow&\frac{\kappa}{\theta^{(1)}\sigma(S^*)}\geq\left(\frac{1}{\sqrt{\Lambda_{\R_1}(S^*)+\frac{a}{2}}+\sqrt{\frac{a}{2}}}\right)^2\\
        \Leftrightarrow&\left(\sqrt{\Lambda_{\R_1}(S^*)+\frac{a}{2}}+\sqrt{\frac{a}{2}}\right)^2\cdot\frac{\kappa}{\theta^{(1)}}\geq\sigma(S^*).
    \end{align*}
    Therefore, we can infer that
    \begin{align*}
        1-\widehat{\epsilon}_1&=1-\sqrt{\frac{2a\kappa}{\sigma(S^*)\theta^{(1)}}}\\
        &\leq 1-\frac{\sqrt{2a}}{\sqrt{\Lambda_{\R_1}(S^*)+\frac{a}{2}}+\sqrt{\frac{a}{2}}}\\
        &\leq\frac{\Lambda^u_{\R_1}(S^*)}{\left(\sqrt{\Lambda^u_{\R_1}(S^*)+\frac{a}{2}}+\sqrt{\frac{a}{2}}\right)^2}.
    \end{align*}
    By setting $p_f=\frac{\delta}{3i_{max}}$ in Eq.\eqref{eq:ub-opim} such that $\ln{\frac{1}{p_f}}\leq a$, we have
    \begin{equation}\label{equ:ub-ub}
    \sigma^u(S^*)\leq \left(\sqrt{\Lambda^u_{\R_1}(S^*)+\frac{a}{2}}+\sqrt{\frac{a}{2}}\right)^2\cdot\frac{\kappa}{\theta^{(1)}}\leq\frac{\Lambda^u_{\R_1}(S^*)}{1-\widehat{\epsilon}_1}\cdot\frac{\kappa}{\theta^{(1)}}.
    \end{equation}
    Similarly, when the event in Eq.\eqref{eq:event-e} does not happen, we have
    \begin{equation*}
        \left(\sqrt{\Lambda_{\R_2}(S)+\frac{2a}{9}}-\sqrt{\frac{a}{2}}\right)^2-\frac{a}{18}\geq\sigma(S)\cdot\frac{\theta^{(2)}}{\kappa}.
    \end{equation*}
    Therefore, we can infer that
    \begin{align*}
        &\Lambda_{\R_2}(S)-\widetilde{\epsilon}_2\sigma(S)\cdot\frac{\theta^{(2)}}{\kappa}\\
        &=\Lambda_{\R_2}(S)-\Big(\sqrt{\frac{2a\sigma(S)\theta^{(2)}}{\kappa}+\frac{a^2}{9}}+\frac{a}{3}\Big)\\
        &\leq\Lambda_{\R_2}(S)-\Big(\sqrt{2a\Lambda_{\R_2}(S)+\frac{4a^2}{9}}-a+\frac{a}{3}\Big)\\
        &=\left(\sqrt{\Lambda_{\R_2}(S)+\frac{2a}{9}}-\sqrt{\frac{a}{2}}\right)^2-\frac{a}{18}
    \end{align*}
    By setting $p_f=\frac{\delta}{3i_{max}}$ in Eq.\eqref{eq:lb-opim} such that $\ln{\frac{1}{p_f}}\leq a$, we have
    \begin{equation}\label{equ:lb-lb}
        \sigma^l(S)\geq\Lambda_{\R_2}(S)\cdot\frac{\kappa}{\theta^{(2)}}-\widetilde{\epsilon}_2\sigma(S)=\Lambda_{\R_2}(S)-\widetilde{\epsilon}_2\sigma(S).
    \end{equation}
    Putting \eqref{equ:ub-ub} and \eqref{equ:lb-lb} together, when none of the events in Eq.\eqref{eq:event-a}-Eq.\eqref{eq:event-e} happen, we have
    \begin{align*}
        \frac{\sigma^l(S)}{\sigma^u(S^*)}&\geq\frac{\Lambda_{\R_2}(S)-\widetilde{\epsilon}_2\sigma(S)\cdot\frac{\theta^{(2)}}{\kappa}}{\Lambda^u_{\R_1}(S^*)/(1-\widehat{\epsilon}_1)}\\
        &\geq\frac{(1-\widehat{\epsilon}_1)(1-\epsilon_2-\widetilde{\epsilon}_2)\sigma(S)\cdot\frac{\theta^{(2)}}{\kappa}}{\Lambda^u_{\R_1}(S^*)}
    \end{align*}
    \begin{align*}
        &>(1-\widehat{\epsilon}_1-\epsilon_2-\widetilde{\epsilon}_2)\frac{\sigma(S)\cdot\frac{\theta^{(2)}}{\kappa}}{\Lambda_{\R_1}(S)}\frac{\Lambda_{\R_1}(S)}{\Lambda^u_{\R_1}(S^*)}\\
        &
        >(1-\epsilon_1)\frac{\Lambda_{\R_1}(S)-\widetilde{\epsilon}_1\cdot\sigma(S^*)\cdot\frac{\theta^{(2)}}{\kappa}}{\Lambda_{\R_1}(S)}\cdot \beta\\
        &\geq (1-\epsilon_1) \frac{\Lambda_{\R_1}(S)-\widetilde{\epsilon}_1 \Lambda_{\R_1}(S^*)/(1-\epsilon_1)}{\Lambda_{\R_1}(S)}\cdot \beta\\
        &\geq (1-\epsilon_1)\left(1-\frac{\widetilde{\epsilon}_1/(1-\epsilon_1)}{\beta}\right)\cdot \beta\\
        &=\beta-\epsilon_1=1-1/e-\epsilon.
    \end{align*}
    This implies that when we generate $\theta^{(1)}=\theta^{(2)}=c\theta'$ RR sets, \ouralgo does not terminate only if at least one of the events in Eq.\eqref{eq:event-a}-Eq.\eqref{eq:event-e} happens, whose probability is at most $ \delta^c_t$.
    Let $j$ be the first iteration that $\theta^{(1)}=\theta^{(2)}\geq\theta'$. From this iteration, the expected number of RR sets generated thereafter is at most
    \begin{align*}
    &2\cdot\sum_{z=j}^{i_{max}}\theta_1\cdot2^z\cdot\delta^{2^{z-j}}\\&=2\cdot2^j\cdot\theta_1\sum_{z=0}^{i_{max}-j}2^z\cdot\delta^{2^z}\\
    &\leq 4\theta'\sum_{z=0}^{i_{max}-j}2^{-2^z+z} \tag{$2^{j-1}\theta_1<\theta'$ and $\delta\leq 1/2$}\\
    &\leq 4\theta'\sum_{z=0}^{i_{max}-j}2^{-z} \tag{$-2^z+z\leq -z$}\\
    &\leq 8\theta'.
    \end{align*}
    Therefore, the expected number of RR sets generated is less than $10\theta'$, which is also $O((\ln{|\mathcal{B}|}+\ln{(1/\delta)})\kappa\epsilon^{-2}/\sigma(S^*))$.
\qed
\section{Tightened Upper Bound}\label{sec:tight-bound}

In Section~\ref{sec:opim-main-idea}, we set the tightened upper bound $\Lambda^u_{\R_1}(S^*)$ to
\begin{equation}\label{eq:ub-tight}
    \textstyle
    \Lambda^u_{\R_1}(S^*)=\min\left\{\min\limits_{0\leq t \leq 1\backslash \epsilon_s}\left\{\widehat{\Lambda}_{\R_1}(\x_t)\right\}, \frac{F(\x_{1/\epsilon_s})}{1-\frac{1}{(1+\epsilon_s)^{\frac{1}{\epsilon_s}}}}, |\R_1|\right\},
\end{equation}
where $\x_0=\mathbf{0}$, $\x_t=\sum_{l=1}^{t}\epsilon_s\cdot\mathbf{1}_{B_l}$, and $$\widehat{\Lambda}_{\R}(\x)=F(\x) + \mathop{\max}_{S\in\I}\left\{\sum_{u_i\in S}\sum_{R\in\R:u_i\in R}q_R\right\}.$$ 

In particular, $\frac{F(\x_{1/\epsilon_s})}{1-\frac{1}{(1+\epsilon_s)^{\frac{1}{\epsilon_s}}}}\geq \Lambda_{\R_1}(S^*)$ is derived from Theorem~\ref{thm:gans-correct}, and $|\R_1|\geq \Lambda_{\R_1}(S^*)$ holds clearly. The following lemma ensures that $\min\limits_{0\leq t \leq 1\backslash \epsilon_s}\left\{\widehat{\Lambda}_{\R_1}(\x_t)\right\}\geq \Lambda_{\R_1}(S^*)$.

\begin{lemma}
    For any $\x\in[0,1]^n$ and a collection $\R$ of RR sets, we have $\widehat{\Lambda}_{\R}(\x)\geq\Lambda_\R(S^o)\geq \Lambda_\R(S^*).$
\end{lemma}

\begin{proof}
By the monotonicity and submodularity of $\Lambda_\R(\cdot)$, we have
\begin{align*}
        \Lambda_\R(S^o)&\leq \mathbb{E}[\Lambda_\R(S^o\cup\Omega(\x))]\\
        &\leq F(\x)+\sum_{u_i\in S^o}\mathbb{E}[\Lambda_\R(u_i|\Omega(\x))]\\
        &\leq F(\x) +  \mathop{\max}_{S\in\I}\left\{\sum_{u_i\in S}\mathbb{E}[\Lambda_\R(u_i|\Omega(\x)]\right\}.
\end{align*}
Derived from Eq.\eqref{eq:derive-expectation}, we have $\mathbb{E}[\Lambda_\R(u_i|\Omega(\x)]=\sum\limits_{R\in\R:u_i\in R}q_R$. Therefore, the above fromula is equal to $$F(\x) +  \mathop{\max}_{S\in\I}\left\{\sum_{u_i\in S}\left(\sum\limits_{R\in\R:u_i\in R}q_R\right)\right\},$$ which completes the proof.
\end{proof}

To compute Eq.\eqref{eq:ub-tight}, we utilize a greedy algorithm~\cite{edmonds1971matroids} to obtain the term $\mathop{\max}_{S\in\I}\{\sum\limits_{u_i\in S}\sum\limits_{R\in\R:u_i\in R}q_R\}$. Its pseudocode is shown in Algorithm~\ref{alg:max-weight}. We invoke \mwgreedy after each execution of \cgreedy or \cgreedyp in Line 9 of Algorithm~\ref{alg:greedy-framework}. In addition, $F(\x_{t})$ can be obtained by accumulating the partial derivatives of the selected $u_j$ at all $r$ iterations in Line 3 (resp. Line 4) of \cgreedy (resp. \cgreedyp).

\begin{algorithm}[!t]
$\forall u_i\in U, w_i=\sum_{R\in\R:u_i\in R}q_R$;\\
$S\gets\emptyset$;\\
\For{$i=1,2,\dots,n$}{
    $u_j\gets$ the element with the $i$-th largest $w_j$;\\
    \lIf{$S\cup\{u_j\}\in\I$}{$S\gets S\cup\{u_j\}$}
}
\Return{$\sum_{u_j\in S}w_j$;}
\caption{\mwgreedy~$(\R,\x,q_R)$}
\label{alg:max-weight}
\end{algorithm}
\section{Implementation Details}\label{sec:exp-detail}

In each \ourproblem instance, we adopt the same approach that constructs one RR set $R$ as in its original work. Apart from \mrim, the details for other \ourproblem instances are described below.

\begin{itemize}[topsep=2pt,itemsep=1pt,parsep=0pt,partopsep=0pt,leftmargin=11pt]
    \item \textit{\im}: We first select a node $v_i\in V$ uniformly at random, and then simulate a reverse diffusion process in $G$ starting from $v_i$. We let $R$ be the set of all examined nodes.
    % \item \textit{\mrim}: We first randomly select a node $v_i\in V$ uniformly, and then independently simulate $T$ reverse diffusion processes starting from $v_i$, denoting the set of examined nodes in each process as $R'_1, \dots, R'_T$. We finally let $R \gets R'_1\times\{1\}\cup\dots\cup R'_T\times\{T\}$.
    \item \textit{\rim}: We first randomly select a node $v_i\in V$ uniformly, and a campaign index $j\in [T]$ w.p. $\frac{\alpha_j}{\sum_{j\in [T]}\alpha_j}$. We simulate a reverse diffusion process starting from $v_i$, denoting the set of examined nodes as $R'$. Then, we let $R \gets R'\times\{j\}$.
    \item \textit{\advim}: We first select a node $v_i\in V\backslash A$ uniformly at random, and then simulate the reverse diffusion process from $v_i$ and include the examined nodes into $R$. If $R\cap A=\emptyset$, we clear $R$ and repeat the process until $R\cap A\neq\emptyset$.
\end{itemize}

\begin{algorithm}[!t]
$\theta_{max}\gets\frac{2|V|}{\epsilon^2}\left(\frac{1}{2}\sqrt{\ln\frac{16}{\delta}}+\sqrt{\frac{1}{2}(T\cdot|V|+\ln\frac{16}{\delta})}\right)^2$;\\
$i_{max}\gets \log_2{\theta_{max}}$;\\
Generate $\R_1$ and $\R_2$ with $|\R_1|=|\R_2|=1$;\\
\For{$i=1,2,\dots,i_{max}$}{
$S\gets$~\mgreedy$(\R_1, \M)$;\\
Compute $\sigma^u(\Sset^*)$ by Eq.\eqref{eq:ub-opim} on $\Rset_1$ with $p_f=\frac{\delta}{4(T+2)i_{max}}$ and $\Lambda^u_{\R_1}(S^*)= {2\cdot\Lambda_{\R_1}(S)}$\;
Compute $\sigma^l(\Sset)$ by Eq.\eqref{eq:lb-opim} on $\Rset_2$ with $p_f=\frac{\delta}{4(T+2)i_{max}}$\;
\lIf{$\frac{\sigma^l(S)}{\sigma^u(S^*)}\geq 1/2-\epsilon$~\textbf{or}~$i=i_{max}$}{\Return{$S$}}
Double the sizes of $\Rset_1$ and $\Rset_2$ with new random RR sets;\\
}
\caption{\textsf{RM-A-Simplified}$(\G, \kappa, \M, \epsilon, \delta)$}
\label{alg:rma}
\end{algorithm}

In \advim, we compute $\theta_{\max}$ in Eq.\eqref{equ:theta_max} setting $\kappa \gets |V| - |A|$. Note that this may slightly increase $\theta_{\max}$ but does not affect the correctness and overall time complexity of \ouralgo.

Regarding the lazy evaluation technique on \cseedselect and \cseedselectp, this process can be efficiently executed using a max-heap and improves the practical runtime performance. Specifically, we maintain an upper bound of $\frac{\partial F(\y)}{\partial \y[i]}$ for each $u_i\in U$, denoted as $F^u_i$, initially set to $+\infty$. During each iteration, when selecting $u_j$, we avoid traversing all elements. Instead, we repeatedly choose the element $u_j$ with the largest $F^u_j$, updating $F^u_j$ to $\frac{\partial F(\y)}{\partial \y[j]}=\sum_{R\in\R:{u_j}\in R} q_R/(1-\y[j])$. If $F^u_j$ continues to be the largest one among the candidates after this update, it confirms that $u_j$ has the largest partial derivative. 

It is worth noting that the original version of \rma~\cite{han2021efficient} includes two additional sets of input: $c_{i,j}$, the cost of selecting the node $v_i$ for campaign $j$, and $b_j$, the maximum tolerable cost for campaign $j$. By setting $c_{i,j}=1,\forall v_i\in V,j\in[T]$ and $b_j=+\infty,\forall j\in[T]$, its simplified pseudocode is presented in Algorithm~\ref{alg:rma}, with the notations aligned to our paper. Algorithm~\ref{alg:rma} is similar to \ouralgo, but it differs in (i) the parameterization of $\theta_{max}$, $i_{max}$, and $p_f$, and (ii) the utilization \mgreedy\footnote{By our settings of $c_{i,j}$ and $b_j$, the element selection algorithm \textsf{RM\_with\_Oracle}$(\tau)$ in~\cite{han2021efficient} is essentially equivalent to \mgreedy, for any given $\tau$.} to yield the set $S$.
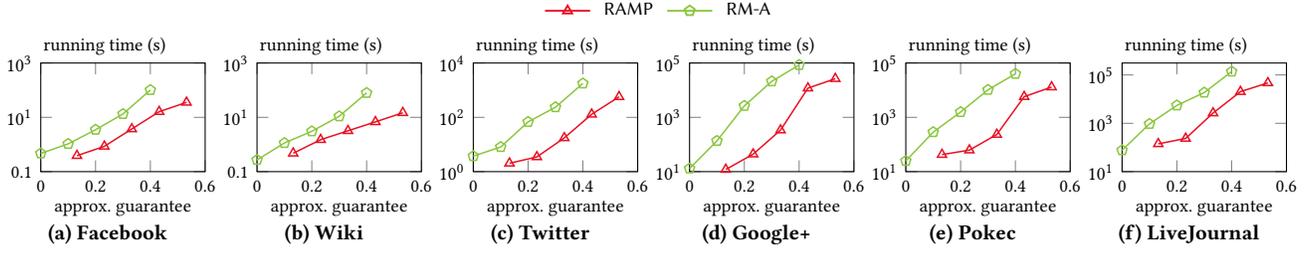
\begin{figure*}[!t]
\centering
\vspace{-1mm}
\begin{tikzpicture}
    \begin{customlegend}[legend columns=2,
        legend entries={\ouralgo, \rma
        }
        ,
        legend style={at={(0.45,1.05)},anchor=north,draw=none,font=\footnotesize,column sep=0.1cm}]
    \addlegendimage{line width=0.25mm,mark size=2pt,mark=triangle,color=Red}
    \addlegendimage{line width=0.25mm,mark size=2pt,mark=pentagon,color=Pink}
    \end{customlegend}
    %  \begin{customlegend}[legend columns=6,
    %     legend entries={},
    %     legend style={at={(0.45,0.7)},anchor=north,draw=none,font=\footnotesize,column sep=0.1cm}]
 
    % \end{customlegend}
\end{tikzpicture}
\\[-\lineskip]
\hspace{-3mm}\subfloat[Facebook]{
\begin{tikzpicture}[scale=1]
    \begin{axis}[
        height=\columnwidth/2.8,
        width=\columnwidth/2.25,
        ylabel={running time (s)},
        xlabel={approx. guarantee},
        xmin=0, xmax=0.6,
        ymin=0.1, ymax=1000,
        xtick={0,0.2, 0.4,0.6},
        xticklabel style = {font=\footnotesize},
        xticklabels={0,0.2, 0.4,0.6},
        ytick={0.1, 10, 1000},
        yticklabels={0.1, $10^1$, $10^3$},
        ymode=log,
        log basis y={10},
        every axis y label/.style={at={(current axis.north west)},right=8.5mm,above=0mm},
        label style={font=\footnotesize},
        tick label style={font=\footnotesize},
        every axis x label/.style={at={(current axis.south)},right=0mm,above=-7mm},
        label style={font=\footnotesize},
        tick label style={font=\footnotesize},
    ]

    \addplot[line width=0.2mm,mark size=2.0pt,mark=pentagon,color=Pink]
        plot coordinates { 
        (0,0.469)
        (0.1,1.051)
        (0.2,3.490)
        (0.3,13.252)
        (0.4,101.423)
    };

    \addplot[line width=0.2mm,mark size=2.0pt,mark=triangle,color=Red]
        plot coordinates { % ours
        (0.13212+0, 0.387)
        (0.13212+0.1, 0.854)
        (0.13212+0.2, 3.719)
        (0.13212+0.3, 16.120)
        (0.13212+0.4, 35.460)
    };
    
    \end{axis}
\end{tikzpicture}\vspace{1mm}\hspace{1mm}%
}%
\hspace{-3mm}\subfloat[Wiki]{
\begin{tikzpicture}[scale=1]
    \begin{axis}[
        height=\columnwidth/2.8,
        width=\columnwidth/2.25,
        ylabel={running time (s)},
        xlabel={approx. guarantee},
        xmin=0, xmax=0.6,
        ymin=0.1, ymax=1000,
        xtick={0,0.2, 0.4,0.6},
        xticklabel style = {font=\footnotesize},
        xticklabels={0,0.2, 0.4,0.6},
        ytick={0.1, 10, 1000},
        yticklabels={0.1, $10^1$, $10^3$},
        ymode=log,
        log basis y={10},
        every axis y label/.style={at={(current axis.north west)},right=8.5mm,above=0mm},
        label style={font=\footnotesize},
        tick label style={font=\footnotesize},
        every axis x label/.style={at={(current axis.south)},right=0mm,above=-7mm},
        label style={font=\footnotesize},
        tick label style={font=\footnotesize},
    ]

    \addplot[line width=0.2mm,mark size=2.0pt,mark=pentagon,color=Pink]
        plot coordinates { 
        (0, 0.270)
        (0.1, 1.134)
        (0.2, 3.059)
        (0.3, 10.911)
        (0.4, 79.288)
    };

    \addplot[line width=0.2mm,mark size=2.0pt,mark=triangle,color=Red]
        plot coordinates { % ours
        (0.13212+0, 0.475)
        (0.13212+0.1, 1.500)
        (0.13212+0.2, 3.195)
        (0.13212+0.3, 6.791)
        (0.13212+0.4, 14.730)
    };
    
    \end{axis}
\end{tikzpicture}\vspace{1mm}\hspace{1mm}%
}%
\hspace{-3mm}\subfloat[Twitter]{
\begin{tikzpicture}[scale=1]
    \begin{axis}[
        height=\columnwidth/2.8,
        width=\columnwidth/2.25,
        ylabel={running time (s)},
        xlabel={approx. guarantee},
        xmin=0, xmax=0.6,
        ymin=1, ymax=10000,
        xtick={0,0.2, 0.4,0.6},
        xticklabel style = {font=\footnotesize},
        xticklabels={0,0.2, 0.4,0.6},
        ytick={1, 100, 10000},
        yticklabels={$10^0$, $10^2$, $10^4$},
        ymode=log,
        log basis y={10},
        every axis y label/.style={at={(current axis.north west)},right=8.5mm,above=0mm},
        label style={font=\footnotesize},
        tick label style={font=\footnotesize},
        every axis x label/.style={at={(current axis.south)},right=0mm,above=-7mm},
        label style={font=\footnotesize},
        tick label style={font=\footnotesize},
    ]

    \addplot[line width=0.2mm,mark size=2.0pt,mark=pentagon,color=Pink]
        plot coordinates { 
        (0, 3.729)
        (0.1, 8.066)
        (0.2, 67.960)
        (0.3, 235.848)
        (0.4, 1767.774)
    };

    \addplot[line width=0.2mm,mark size=2.0pt,mark=triangle,color=Red]
        plot coordinates { % ours
        (0.13212+0, 2.033)
        (0.13212+0.1, 3.485)
        (0.13212+0.2, 17.456)
        (0.13212+0.3, 129.155)
        (0.13212+0.4, 570.433)
    };
    
    \end{axis}
\end{tikzpicture}\vspace{1mm}\hspace{1mm}%
}%
\hspace{-3mm}\subfloat[Google+]{
\begin{tikzpicture}[scale=1]
    \begin{axis}[
        height=\columnwidth/2.8,
        width=\columnwidth/2.25,
        ylabel={running time (s)},
        xlabel={approx. guarantee},
        xmin=0, xmax=0.6,
        ymin=10, ymax=100000,
        xtick={0,0.2, 0.4,0.6},
        xticklabel style = {font=\footnotesize},
        xticklabels={0,0.2, 0.4,0.6},
        ytick={10, 1000, 100000},
        yticklabels={$10^1$, $10^3$, $10^5$},
        ymode=log,
        log basis y={10},
        every axis y label/.style={at={(current axis.north west)},right=8.5mm,above=0mm},
        label style={font=\footnotesize},
        tick label style={font=\footnotesize},
        every axis x label/.style={at={(current axis.south)},right=0mm,above=-7mm},
        label style={font=\footnotesize},
        tick label style={font=\footnotesize},
    ]

    \addplot[line width=0.2mm,mark size=2.0pt,mark=pentagon,color=Pink]
        plot coordinates { 
        (0, 13.037)
        (0.1, 135.461)
        (0.2, 2640.824)
        (0.3, 21148.543)
        (0.4, 84756.630)
    };

    \addplot[line width=0.2mm,mark size=2.0pt,mark=triangle,color=Red]
        plot coordinates { % ours
        (0.13212+0, 11.971)
        (0.13212+0.1, 44.034)
        (0.13212+0.2, 337.722)
        (0.13212+0.3, 11890.766)
        (0.13212+0.4, 26451.165)
    };
    
    \end{axis}
\end{tikzpicture}\vspace{1mm}\hspace{1mm}%
}%
\hspace{-3mm}\subfloat[Pokec]{
\begin{tikzpicture}[scale=1]
    \begin{axis}[
        height=\columnwidth/2.8,
        width=\columnwidth/2.25,
        ylabel={running time (s)},
        xlabel={approx. guarantee},
        xmin=0, xmax=0.6,
        ymin=10, ymax=100000,
        xtick={0,0.2, 0.4,0.6},
        xticklabel style = {font=\footnotesize},
        xticklabels={0,0.2, 0.4,0.6},
        ytick={10, 1000, 100000},
        yticklabels={$10^1$, $10^3$, $10^5$},
        ymode=log,
        log basis y={10},
        every axis y label/.style={at={(current axis.north west)},right=8.5mm,above=0mm},
        label style={font=\footnotesize},
        tick label style={font=\footnotesize},
        every axis x label/.style={at={(current axis.south)},right=0mm,above=-7mm},
        label style={font=\footnotesize},
        tick label style={font=\footnotesize},
    ]

    \addplot[line width=0.2mm,mark size=2.0pt,mark=pentagon,color=Pink]
        plot coordinates { 
        (0, 24.793)
        (0.1, 285.135)
        (0.2, 1565.096)
        (0.3, 10276.820)
        (0.4, 39892.794)
    };

    \addplot[line width=0.2mm,mark size=2.0pt,mark=triangle,color=Red]
        plot coordinates { % ours
        (0.13212+0, 42.438)
        (0.13212+0.1, 61.914)
        (0.13212+0.2, 232.752)
        (0.13212+0.3, 5758.771)
        (0.13212+0.4, 13014.785)
    };
    
    \end{axis}
\end{tikzpicture}\vspace{1mm}\hspace{1mm}%
}%
\hspace{-3mm}\subfloat[LiveJournal]{
\begin{tikzpicture}[scale=1]
    \begin{axis}[
        height=\columnwidth/2.8,
        width=\columnwidth/2.25,
        ylabel={running time (s)},
        xlabel={approx. guarantee},
        xmin=0, xmax=0.6,
        ymin=10, ymax=312000,
        xtick={0,0.2, 0.4,0.6},
        xticklabel style = {font=\footnotesize},
        xticklabels={0,0.2, 0.4,0.6},
        ytick={10, 1000, 100000},
        yticklabels={$10^1$, $10^3$, $10^5$},
        ymode=log,
        log basis y={10},
        every axis y label/.style={at={(current axis.north west)},right=8.5mm,above=0mm},
        label style={font=\footnotesize},
        tick label style={font=\footnotesize},
        every axis x label/.style={at={(current axis.south)},right=0mm,above=-7mm},
        label style={font=\footnotesize},
        tick label style={font=\footnotesize},
    ]

    \addplot[line width=0.2mm,mark size=2.0pt,mark=pentagon,color=Pink]
        plot coordinates { 
        (0, 76.729)
        (0.1, 945.583)
        (0.2, 5497.649)
        (0.3, 18694.367)
        (0.4, 138222.436)
    };

    \addplot[line width=0.2mm,mark size=2.0pt,mark=triangle,color=Red]
        plot coordinates { % ours
        (0.13212+0, 140.973)
        (0.13212+0.1, 236.005)
        (0.13212+0.2, 2658.967)
        (0.13212+0.3, 20382.425)
        (0.13212+0.4, 46882.561)
    };
    
    \end{axis}
\end{tikzpicture}\vspace{1mm}\hspace{1mm}%
}%
\vspace{-3mm}
\caption{Total running time in \rim.} \label{fig:scale-rm-time}
\end{figure*}
\begin{figure*}[!t]
\centering
% \vspace{-2.5mm}
% \begin{tikzpicture}
%     \begin{customlegend}[legend columns=2,
%         legend entries={\ouralgo, \rma
%         }
%         ,
%         legend style={at={(0.45,1.05)},anchor=north,draw=none,font=\footnotesize,column sep=0.1cm}]
%     \addlegendimage{line width=0.25mm,mark size=2pt,mark=triangle,color=Red}
%     \addlegendimage{line width=0.25mm,mark size=2pt,mark=pentagon,color=Pink}
%     % \addlegendimage{line width=0.25mm,mark size=2pt,mark=o, color=LightBlue}
%     % \addlegendimage{line width=0.25mm,mark size=2pt,mark=square, color=Green}
%     \end{customlegend}
%     %  \begin{customlegend}[legend columns=6,
%     %     legend entries={},
%     %     legend style={at={(0.45,0.7)},anchor=north,draw=none,font=\footnotesize,column sep=0.1cm}]
 
%     % \end{customlegend}
% \end{tikzpicture}
% \\[-\lineskip]
\hspace{-3mm}\subfloat[\revise{Facebook}]{
\begin{tikzpicture}[scale=1]
    \begin{axis}[
        height=\columnwidth/2.8,
        width=\columnwidth/2.25,
        ylabel={memory usage (MB)},
        xlabel={approx. guarantee},
        xmin=0, xmax=0.6,
        ymin=10000, ymax=160000,
        xtick={0,0.2, 0.4,0.6},
        xticklabel style = {font=\footnotesize},
        xticklabels={0,0.2, 0.4,0.6},
        ytick={10000,100000},
        yticklabels={$10^1$, $10^2$},
        ymode=log,
        log basis y={10},
        every axis y label/.style={at={(current axis.north west)},right=10.3mm,above=0mm},
        label style={font=\footnotesize},
        tick label style={font=\footnotesize},
        every axis x label/.style={at={(current axis.south)},right=0mm,above=-7mm},
        label style={font=\footnotesize},
        tick label style={font=\footnotesize},
    ]

    \addplot[line width=0.2mm,mark size=2.0pt,mark=pentagon,color=Pink]
        plot coordinates { 
        (0,16596)
        (0.1,17132)
        (0.2,23624)
        (0.3,44636)
        (0.4,128328)
    };

    \addplot[line width=0.2mm,mark size=2.0pt,mark=triangle,color=Red]
        plot coordinates { % ours
        (0.13212+0, 16296)
        (0.13212+0.1, 16180)
        (0.13212+0.2, 19572)
        (0.13212+0.3, 33360)
        (0.13212+0.4, 54736)
    };
    
    \end{axis}
\end{tikzpicture}\vspace{1mm}\hspace{1mm}%
}%
\hspace{-3mm}\subfloat[\revise{Wiki}]{
\begin{tikzpicture}[scale=1]
    \begin{axis}[
        height=\columnwidth/2.8,
        width=\columnwidth/2.25,
        ylabel={memory usage (MB)},
        xlabel={approx. guarantee},
        xmin=0, xmax=0.6,
        ymin=10000, ymax=1000000,
        xtick={0,0.2, 0.4,0.6},
        xticklabel style = {font=\footnotesize},
        xticklabels={0,0.2, 0.4,0.6},
        ytick={10000,100000,1000000},
        yticklabels={$10^1$, $10^2$, $10^3$},
        ymode=log,
        log basis y={10},
        every axis y label/.style={at={(current axis.north west)},right=10.3mm,above=0mm},
        label style={font=\footnotesize},
        tick label style={font=\footnotesize},
        every axis x label/.style={at={(current axis.south)},right=0mm,above=-7mm},
        label style={font=\footnotesize},
        tick label style={font=\footnotesize},
    ]

    \addplot[line width=0.2mm,mark size=2.0pt,mark=pentagon,color=Pink]
        plot coordinates { 
        (0,16288)
        (0.1,19260)
        (0.2,28916)
        (0.3,111392)
        (0.4,395068)
    };

    \addplot[line width=0.2mm,mark size=2.0pt,mark=triangle,color=Red]
        plot coordinates { % ours
        (0.13212+0,16068)
        (0.13212+0.1,21208)
        (0.13212+0.2,28956)
        (0.13212+0.3,45032)
        (0.13212+0.4,81008)
    };
    
    \end{axis}
\end{tikzpicture}\vspace{1mm}\hspace{1mm}%
}%
\hspace{-3mm}\subfloat[\revise{Twitter}]{
\begin{tikzpicture}[scale=1]
    \begin{axis}[
        height=\columnwidth/2.8,
        width=\columnwidth/2.25,
        ylabel={memory usage (MB)},
        xlabel={approx. guarantee},
        xmin=0, xmax=0.6,
        ymin=100000, ymax=3120000,
        xtick={0,0.2, 0.4,0.6},
        xticklabel style = {font=\footnotesize},
        xticklabels={0,0.2, 0.4,0.6},
        ytick={100000,1000000},
        yticklabels={$10^2$, $10^3$},
        ymode=log,
        log basis y={10},
        every axis y label/.style={at={(current axis.north west)},right=10.3mm,above=0mm},
        label style={font=\footnotesize},
        tick label style={font=\footnotesize},
        every axis x label/.style={at={(current axis.south)},right=0mm,above=-7mm},
        label style={font=\footnotesize},
        tick label style={font=\footnotesize},
    ]

    \addplot[line width=0.2mm,mark size=2.0pt,mark=pentagon,color=Pink]
        plot coordinates { 
        (0,143088)
        (0.1,158984)
        (0.2,240212)
        (0.3,430168)
        (0.4,2122760)
    };

    \addplot[line width=0.2mm,mark size=2.0pt,mark=triangle,color=Red]
        plot coordinates { % ours
        (0.13212+0,154512)
        (0.13212+0.1,159160)
        (0.13212+0.2,173436)
        (0.13212+0.3,305628)
        (0.13212+0.4,790584)
    };
    
    \end{axis}
\end{tikzpicture}\vspace{1mm}\hspace{1mm}%
}%
\hspace{-3mm}\subfloat[\revise{Google+}]{
\begin{tikzpicture}[scale=1]
    \begin{axis}[
        height=\columnwidth/2.8,
        width=\columnwidth/2.25,
        ylabel={memory usage (MB)},
        xlabel={approx. guarantee},
        xmin=0, xmax=0.6,
        ymin=1000000, ymax=10000000,
        xtick={0,0.2, 0.4,0.6},
        xticklabel style = {font=\footnotesize},
        xticklabels={0,0.2, 0.4,0.6},
        ytick={1000000,10000000},
        yticklabels={$10^3$, $10^4$},
        ymode=log,
        log basis y={10},
        every axis y label/.style={at={(current axis.north west)},right=10.3mm,above=0mm},
        label style={font=\footnotesize},
        tick label style={font=\footnotesize},
        every axis x label/.style={at={(current axis.south)},right=0mm,above=-7mm},
        label style={font=\footnotesize},
        tick label style={font=\footnotesize},
    ]

    \addplot[line width=0.2mm,mark size=2.0pt,mark=pentagon,color=Pink]
        plot coordinates { 
        (0,1265028)
        (0.1,1299956)
        (0.2,1397500)
        (0.3,2246180)
        (0.4,5038128)
    };

    \addplot[line width=0.2mm,mark size=2.0pt,mark=triangle,color=Red]
        plot coordinates { % ours
        (0.13212+0,1271964)
        (0.13212+0.1,1271916)
        (0.13212+0.2,1289132)
        (0.13212+0.3,1832272)
        (0.13212+0.4,2510488)
    };
    
    \end{axis}
\end{tikzpicture}\vspace{1mm}\hspace{1mm}%
}%
\hspace{-3mm}\subfloat[\revise{Pokec}]{
\begin{tikzpicture}[scale=1]
    \begin{axis}[
        height=\columnwidth/2.8,
        width=\columnwidth/2.25,
        ylabel={memory usage (MB)},
        xlabel={approx. guarantee},
        xmin=0, xmax=0.6,
        ymin=1000000, ymax=100000000,
        xtick={0,0.2, 0.4,0.6},
        xticklabel style = {font=\footnotesize},
        xticklabels={0,0.2, 0.4,0.6},
        ytick={1000000,10000000,100000000},
        yticklabels={$10^3$, $10^4$, $10^5$},
        ymode=log,
        log basis y={10},
        every axis y label/.style={at={(current axis.north west)},right=10.3mm,above=0mm},
        label style={font=\footnotesize},
        tick label style={font=\footnotesize},
        every axis x label/.style={at={(current axis.south)},right=0mm,above=-7mm},
        label style={font=\footnotesize},
        tick label style={font=\footnotesize},
    ]

    \addplot[line width=0.2mm,mark size=2.0pt,mark=pentagon,color=Pink]
        plot coordinates { 
        (0,2549844)
        (0.1,2741108)
        (0.2,4297512)
        (0.3,14855112)
        (0.4,50378040)
    };

    \addplot[line width=0.2mm,mark size=2.0pt,mark=triangle,color=Red]
        plot coordinates { % ours
        (0.13212+0,2592460)
        (0.13212+0.1,2621324)
        (0.13212+0.2,2877532)
        (0.13212+0.3,9599256)
        (0.13212+0.4,17792136)
    };
    
    \end{axis}
\end{tikzpicture}\vspace{1mm}\hspace{1mm}%
}%
\hspace{-3mm}\subfloat[\revise{LiveJournal}]{
\begin{tikzpicture}[scale=1]
    \begin{axis}[
        height=\columnwidth/2.8,
        width=\columnwidth/2.25,
        ylabel={memory usage (MB)},
        xlabel={approx. guarantee},
        xmin=0, xmax=0.6,
        ymin=3160000, ymax=312000000,
        xtick={0,0.2, 0.4,0.6},
        xticklabel style = {font=\footnotesize},
        xticklabels={0,0.2, 0.4,0.6},
        ytick={10000000, 100000000},
        yticklabels={$10^4$, $10^5$},
        ymode=log,
        log basis y={10},
        every axis y label/.style={at={(current axis.north west)},right=10.3mm,above=0mm},
        label style={font=\footnotesize},
        tick label style={font=\footnotesize},
        every axis x label/.style={at={(current axis.south)},right=0mm,above=-7mm},
        label style={font=\footnotesize},
        tick label style={font=\footnotesize},
    ]

    \addplot[line width=0.2mm,mark size=2.0pt,mark=pentagon,color=Pink]
        plot coordinates { 
        (0,6596336)
        (0.1,7016296)
        (0.2,11440888)
        (0.3,23991124)
        (0.4,138597016)
    };

    \addplot[line width=0.2mm,mark size=2.0pt,mark=triangle,color=Red]
        plot coordinates { % ours
        (0.13212+0,6620992)
        (0.13212+0.1,6696448)
        (0.13212+0.2,9220792)
        (0.13212+0.3,26044400)
        (0.13212+0.4,48664084)
    };
    
    \end{axis}
\end{tikzpicture}\vspace{1mm}\hspace{1mm}%
}%
\vspace{-3mm}
\caption{Memory usage in \rim.} \label{fig:memory-formal-rm}
\end{figure*}
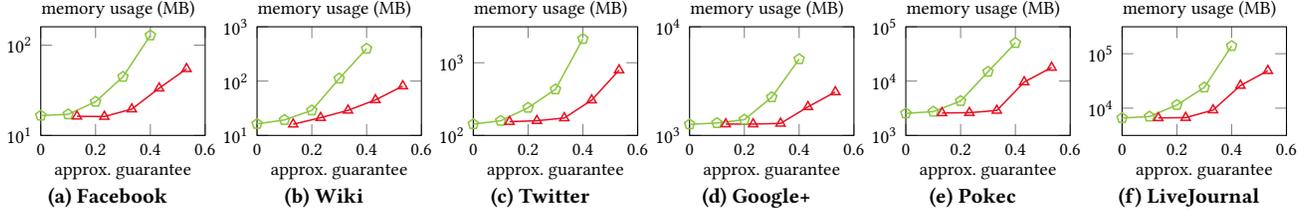

\section{More Experimental Results}\label{sec:more-exp}

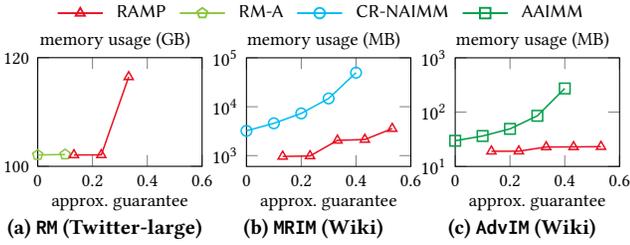
\begin{figure}[!t]
\centering

\begin{tikzpicture}
    \begin{customlegend}[legend columns=4,
        legend entries={\ouralgo, \rma, \naimm, \advimm
        }
        ,
        legend style={at={(0.45,1.05)},anchor=north,draw=none,font=\footnotesize,column sep=0.1cm}]
    \addlegendimage{line width=0.25mm,mark size=2pt,mark=triangle,color=Red}
    \addlegendimage{line width=0.25mm,mark size=2pt,mark=pentagon,color=Pink}
    \addlegendimage{line width=0.25mm,mark size=2pt,mark=o, color=LightBlue}
    \addlegendimage{line width=0.25mm,mark size=2pt,mark=square, color=Green}
    \end{customlegend}
    %  \begin{customlegend}[legend columns=6,
    %     legend entries={},
    %     legend style={at={(0.45,0.7)},anchor=north,draw=none,font=\footnotesize,column sep=0.1cm}]
 
    % \end{customlegend}
\end{tikzpicture}
\vspace{-1mm}
\\[-\lineskip]
\hspace{-3.1mm}\subfloat[\revise{\rim (Twitter-large)}]{
\begin{tikzpicture}[scale=1]
    \begin{axis}[
        height=\columnwidth/2.8,
        width=\columnwidth/2.25,
        ylabel={memory usage (GB)},
        xlabel={approx. guarantee},
        xmin=0, xmax=0.6,
        ymin=100000000, ymax=120000000,
        xtick={0,0.2, 0.4,0.6},
        xticklabel style = {font=\footnotesize},
        xticklabels={0,0.2, 0.4,0.6},
        ytick={100000000, 120000000},
        yticklabels={100,120},
        ymode=log,
        log basis y={10},
        every axis y label/.style={at={(current axis.north west)},right=10.3mm,above=0mm},
        label style={font=\footnotesize},
        tick label style={font=\footnotesize},
        every axis x label/.style={at={(current axis.south)},right=0mm,above=-7mm},
        label style={font=\footnotesize},
        tick label style={font=\footnotesize},
    ]

    \addplot[line width=0.2mm,mark size=2.0pt,mark=pentagon,color=Pink]
        plot coordinates { 
        (0,101925836)
        (0.1,102007436)
        % (0.2,155030664)
    };

    \addplot[line width=0.2mm,mark size=2.0pt,mark=triangle,color=Red]
        plot coordinates { % ours
        (0.13212+0, 101920900)
        (0.13212+0.1,101939008)
        (0.13212+0.2,116192608)
        % (0.13212+0.3,220358456)
    };
    
    \end{axis}
\end{tikzpicture}\vspace{1mm}\hspace{1mm}%
}%
\hspace{-4mm}\subfloat[\revise{\mrim (Wiki)}]{
\begin{tikzpicture}[scale=1]
    \begin{axis}[
        height=\columnwidth/2.8,
        width=\columnwidth/2.25,
        ylabel={memory usage (MB)},
        xlabel={approx. guarantee},
        xmin=0, xmax=0.6,
        ymin=60000, ymax=10000000,
        xtick={0,0.2, 0.4,0.6},
        xticklabel style = {font=\footnotesize},
        xticklabels={0,0.2, 0.4,0.6},
        ytick={100000, 1000000, 10000000},
        yticklabels={$10^3$, $10^4$, $10^5$},
        ymode=log,
        log basis y={10},
        every axis y label/.style={at={(current axis.north west)},right=10.3mm,above=0mm},
        label style={font=\footnotesize},
        tick label style={font=\footnotesize},
        every axis x label/.style={at={(current axis.south)},right=0mm,above=-7mm},
        label style={font=\footnotesize},
        tick label style={font=\footnotesize},
    ]

    \addplot[line width=0.2mm,mark size=2.0pt,mark=o, color=LightBlue]
        plot coordinates { % CR-MRIMM
        (0, 320772)
        (0.1, 457388)
        (0.2, 729404)
        (0.3, 1473896)
        (0.4, 4963396)
    };

    \addplot[line width=0.2mm,mark size=2.0pt,mark=triangle,color=Red]
        plot coordinates { % ours
        (0.13212+0, 96132)
        (0.13212+0.1, 98240)
        (0.13212+0.2, 204908)
        (0.13212+0.3, 213112)
        (0.13212+0.4, 356000)
    };
    \end{axis}
\end{tikzpicture}\vspace{1mm}\hspace{1mm}%
}%
\hspace{-4mm}\subfloat[\revise{\advim (Wiki)}]{
\begin{tikzpicture}[scale=1]
    \begin{axis}[
        height=\columnwidth/2.8,
        width=\columnwidth/2.25,
        ylabel={memory usage (MB)},
        xlabel={approx. guarantee},
        xmin=0, xmax=0.6,
        ymin=10000, ymax=1000000,
        xtick={0,0.2, 0.4,0.6},
        xticklabel style = {font=\footnotesize},
        xticklabels={0,0.2, 0.4,0.6},
        ytick={10000, 100000, 1000000},
        yticklabels={$10^1$, $10^2$, $10^3$},
        ymode=log,
        log basis y={10},
        every axis y label/.style={at={(current axis.north west)},right=10.3mm,above=0mm},
        label style={font=\footnotesize},
        tick label style={font=\footnotesize},
        every axis x label/.style={at={(current axis.south)},right=0mm,above=-7mm},
        label style={font=\footnotesize},
        tick label style={font=\footnotesize},
    ]

    \addplot[line width=0.2mm,mark size=2.0pt,mark=square, color=Green]
        plot coordinates { % AAIMM
        (0, 29712)
        (0.1, 36344)
        (0.2, 49204)
        (0.3, 85568)
        (0.4, 272916)
    };

    \addplot[line width=0.2mm,mark size=2.0pt,mark=triangle,color=Red]
        plot coordinates { % ours
        (0.13212+0, 19044)
        (0.13212+0.1, 19096)
        (0.13212+0.2, 22820)
        (0.13212+0.3, 22912)
        (0.13212+0.4, 23164)
    };
    \end{axis}
\end{tikzpicture}\vspace{1mm}\hspace{1mm}%
}%
\vspace{-3mm}
\caption{\revise{Memory usage of scalable algorithms.}} \label{fig:scale-all-memory}
\end{figure}

\noindent
\textbf{Memory usage.}
We report the memory usage of each scalable implementation for \rim, \mrim, or \advim at its peak memory consumption in Figure \ref{fig:scale-all-memory}. Similar to Figure \ref{fig:scale-all-time}, we report results for the largest dataset on which the baseline algorithm with $\epsilon=0.5$ can terminate within 72 hours.
As shown, the memory usage of each algorithm is roughly proportional to its running time reported in Figure \ref{fig:scale-all-time}. Most notably, the memory occupancy of \ouralgo is consistently lower than that of \rma, \naimm and \advimm due to the smaller number of sampled RR sets. Specifically, the memory footprint of \ouralgo in \advim is up to \eat{an order of magnitude}$10\times$ lower than \advimm on Wiki. 
Regarding \rim, \ouralgo or \rma has a memory footprint of 102GB on Twitter-large when $\epsilon \geq 0.4$. Of this, 62GB is allocated for storing the dataset (with half for the original graph and half for the reversed graph, on which the RR sets are generated), while the remaining 40 GB ($\sim$39\%) is used for storing the sampled RR sets. In conclusion, the resource utilization of the proposed \ouralgo is superior to all competitors.

\noindent
\textbf{More experiments for \rim.} Figures \ref{fig:scale-rm-time}-\ref{fig:memory-formal-rm} report the running times and memory footprints of scalable algorithms for \rim on smaller datasets, respectively. As shown in Figure \ref{fig:scale-rm-time}, \ouralgo is more efficient than \rma while ensuring the same approximation guarantee across all cases. In particular, \ouralgo is an order of
magnitude faster than \rma on Google+, Pokec, and LiveJournal. Additionally, Figure \ref{fig:memory-formal-rm} shows that \ouralgo occupies significantly lower memory than \rma. For example, the memory usage of \ouralgo is $8.7\times$ lower than \rma on Wiki when ensuring a $0.4$-approximation.

\begin{figure}[!t]
\centering

\begin{tikzpicture}
    \begin{customlegend}[legend columns=2,
        legend entries={\ouralgo, \naimm
        }
        ,
        legend style={at={(0.45,1.05)},anchor=north,draw=none,font=\footnotesize,column sep=0.1cm}]
    \addlegendimage{line width=0.25mm,mark size=2pt,mark=triangle,color=Red}
    \addlegendimage{line width=0.25mm,mark size=2pt,mark=o, color=LightBlue}
    \end{customlegend}
    %  \begin{customlegend}[legend columns=6,
    %     legend entries={},
    %     legend style={at={(0.45,0.7)},anchor=north,draw=none,font=\footnotesize,column sep=0.1cm}]
 
    % \end{customlegend}
\end{tikzpicture}
\\[-\lineskip]
\subfloat[Running time]{
\begin{tikzpicture}[scale=1]
    \begin{axis}[
        height=\columnwidth/2.8,
        width=\columnwidth/2.25,
        ylabel={running time (s)},
        xlabel={approx. guarantee},
        xmin=0, xmax=0.6,
        ymin=1, ymax=100000,
        xtick={0,0.2, 0.4,0.6},
        xticklabel style = {font=\footnotesize},
        xticklabels={0,0.2, 0.4,0.6},
        ytick={1, 100, 10000},
        yticklabels={$10^0$, $10^2$, $10^4$},
        ymode=log,
        log basis y={10},
        every axis y label/.style={at={(current axis.north west)},right=8.5mm,above=0mm},
        label style={font=\footnotesize},
        tick label style={font=\footnotesize},
        every axis x label/.style={at={(current axis.south)},right=0mm,above=-7mm},
        label style={font=\footnotesize},
        tick label style={font=\footnotesize},
    ]

    \addplot[line width=0.2mm,mark size=2.0pt,mark=o, color=LightBlue]
        plot coordinates { % CR-MRIMM
        (0, 5134.514)
        (0.1, 7382.662)
        (0.2, 12030.421)
        (0.3, 24722.275)
        (0.4, 89950.386)
    };

    \addplot[line width=0.2mm,mark size=2.0pt,mark=triangle,color=Red]
        plot coordinates { % ours
        (0.13212+0, 1.293)
        (0.13212+0.1, 1.437)
        (0.13212+0.2, 1.574)
        (0.13212+0.3, 1.710)
        (0.13212+0.4, 4.310)
    };
    \end{axis}
\end{tikzpicture}\vspace{1mm}\hspace{1mm}%
}%
\subfloat[Memory usage]{
\begin{tikzpicture}[scale=1]
    \begin{axis}[
        height=\columnwidth/2.8,
        width=\columnwidth/2.25,
        ylabel={memory usage (MB)},
        xlabel={approx. guarantee},
        xmin=0, xmax=0.6,
        ymin=10000, ymax=10000000,
        xtick={0,0.2, 0.4,0.6},
        xticklabel style = {font=\footnotesize},
        xticklabels={0,0.2, 0.4,0.6},
        ytick={10000, 100000, 1000000, 10000000},
        yticklabels={$10^2$, $10^3$, $10^4$, $10^5$},
        ymode=log,
        log basis y={10},
        every axis y label/.style={at={(current axis.north west)},right=10.3mm,above=0mm},
        label style={font=\footnotesize},
        tick label style={font=\footnotesize},
        every axis x label/.style={at={(current axis.south)},right=0mm,above=-7mm},
        label style={font=\footnotesize},
        tick label style={font=\footnotesize},
    ]

    \addplot[line width=0.2mm,mark size=2.0pt,mark=o, color=LightBlue]
        plot coordinates { % CR-MRIMM
        (0, 226640)
        (0.1, 321736)
        (0.2, 503772)
        (0.3, 987056)
        (0.4, 3317692)
    };

    \addplot[line width=0.2mm,mark size=2.0pt,mark=triangle,color=Red]
        plot coordinates { % ours
        (0.13212+0, 48692)
        (0.13212+0.1, 64484)
        (0.13212+0.2, 65808)
        (0.13212+0.3, 66952)
        (0.13212+0.4, 79532)
    };
    \end{axis}
\end{tikzpicture}\vspace{1mm}\hspace{1mm}%
}%
\vspace{-3mm}
\caption{Running time and memory usage in \mrim on Wiki.} \label{fig:more-mrim}
\end{figure}
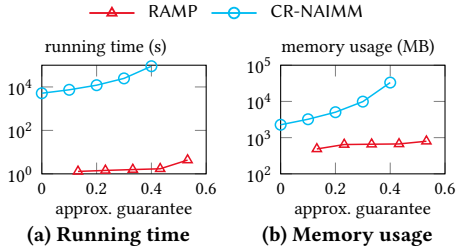

\noindent
\textbf{More experiments for \mrim.}
Figure \ref{fig:more-mrim} reports the running time and memory usage of each algorithm for \mrim on Wiki. Note that we exclude results on other datasets as \naimm and \advimm fail to terminate within 72 hours even when $\epsilon = 0.5$. Here, \ouralgo outperforms \naimm at least three orders of magnitude in terms of running time and at least an order of magnitude in terms of memory usage, respectively. To summarize, Figures \ref{fig:scale-rm-time}-\ref{fig:more-mrim} present similar outcomes to those in Figures \ref{fig:scale-all-time}-\ref{fig:scale-all-memory}, where \ouralgo is more efficient regarding running time and resource utilization.
\section{Adjusted Algorithm for Deployment}\label{sec:dep-detail}
Note that the enhanced objective function $\sigma^W(S)$ is also monotone and submodular. Towards the new objective $\sigma^W(S)$, we modify \cseedselect for maximizing the following function: $$\Lambda^W_\R(S)=\sum_{S'\subseteq S}\prod_{u_j\in S'}w_j\prod_{u_j\in S\backslash S'}(1-w_j)\Lambda_\R(S'),$$ which is an unbiased estimate of $\sigma^W(S)$ satisfying $\sigma^W(S)=\frac{\kappa}{|\R|}\cdot\Lambda^W_\R(S)$. Specifically, at the beginning of \cseedselect (Algorithm~\ref{alg:greedy-framework}), we maintain a variable 
\begin{equation*}
    q'_R=\prod_{u_i\in R}(1-\x[i]\cdot w_i)
\end{equation*}
instead of $q_R$ for each $R\in\R$ in $\mathcal{Q}$. It is also initialized to $1$. Throughout Algorithms~\ref{alg:greedy-framework}-\ref{alg:crounding}, when calculating the partial derivative $\frac{\partial F(\y)}{\partial \y[i]}$ w.r.t. $u_i$, we replace $\sum\limits_{R\in\R:~u_i\in R}\frac{q_R}{1-\y[i]}$ with $\frac{w_i}{1-\y[i]w_i}\sum\limits_{R\in\R:~u_i\in R}{q'_R}$. Furthermore, when updating each $q_R$, we replace $q_R\gets q_R\cdot\frac{1-\y[j]}{1-\y[j]+z}$ with $q'_R\gets q'_R\cdot\frac{1-w_j\y[j]}{1-w_j(\y[j]+z)}$, where $z\in\{\epsilon, -\epsilon, t\epsilon, -t\epsilon\}$. The correctness of these adjustments is assured by the following lemma.

\begin{lemma}\label{lem:extend-objective}
Let $F^W$ denote the multilinear extension of $\Lambda^W_\R(S)$. For any $\x\in[0,1]^U$ we have \begin{equation*}
    \frac{\partial F^W(\x)}{\partial \x[i]}=\frac{w_i}{1-\x[i]\cdot w_i}\sum_{R\in\R:u_i\in R}\prod_{u_j\in R}(1-\x[j]\cdot w_j).
\end{equation*}
\end{lemma}
\begin{proof}
    Similar to Eq.\eqref{eq:derive-expectation}, we have
\begin{equation}\label{eq:derived-multilinear}
    \frac{\partial F^W(\x)}{\partial \x[i]}=\frac{1}{1-\x[i]}\mathbb{E}\left[\Lambda^W_\R(u_i|\Omega(\x))\right].
\end{equation}
In addition, for any $S\subseteq U$ and $u_i\in U\backslash S$, we have
\begin{align*}
    \Lambda^W_\R(u_i|S)&=\sum_{S'\subseteq S\cup\{u_i\}}\prod_{u_j\in S'}w_j\prod_{u_j\in S\backslash S'}(1-w_j)\Lambda_\R(S')\\
    &-\sum_{S'\subseteq S}\prod_{u_j\in S'}w_j\prod_{u_j\in S\backslash S'}(1-w_j)\Lambda_\R(S')\\
    &=\Bigg(w_i\cdot\sum_{S'\subseteq S}\prod_{u_j\in S'}w_j\prod_{u_j\in S\backslash S'}(1-w_j)\Lambda_\R(S'\cup\{u_i\})\\
    &+(1-w_i)\cdot\sum_{S'\subseteq S}\prod_{u_j\in S'}w_j\prod_{u_j\in S\backslash S'}(1-w_j)\Lambda_\R(S')\Bigg)\\
    &-\sum_{S'\subseteq S}\prod_{u_j\in S'}w_j\prod_{u_j\in S\backslash S'}(1-w_j)\Lambda_\R(S')\\
    &=w_i\cdot\sum_{S'\subseteq S}\prod_{u_j\in S'}w_j\prod_{u_j\in S\backslash S'}(1-w_j)\Lambda_\R(u_i|S').
\end{align*}
Let $\Phi_S$ denote a random set that independently includes each element $u_j\in S$ w.p. $w_j$. Hence, for any $S\subseteq U$ and $u_i\in U$, $\Lambda^W_\R(u_i|S)$ can be rewritten to
\begin{align*}
w_i\cdot\sum_{R\in\R:u_i\in R}\mathbb{E}\left[\mathbb{I}(\Phi_S \cap R = \emptyset)\right]\cdot\mathbb{I}(u_i \not\in S).
\end{align*}
By plugging the above result into Eq.\eqref{eq:derived-multilinear}, we have
\begin{align*}
    \frac{\partial F^W(\x)}{\partial \x[i]}&=\frac{w_i}{1-\x[i]}\sum_{R\in\R:u_i\in R}\mathbb{E}\left[\mathbb{I}(\Phi_{\Omega(\x)} \cap R = \emptyset)\cdot\mathbb{I}(u_i \not\in \Omega(\x))\right]\\
    &=w_i\cdot \sum_{R\in\R:u_i\in R}\mathbb{E}\left[\mathbb{I}(\Phi_{\Omega(\x)\backslash \{u_i\}} \cap R = \emptyset)\right]
\end{align*}
Equivalently, $\Phi_{\Omega(\x)\backslash \{u_i\}}$ is a random set that independently includes each $u_j\in U\backslash \{u_i\}$ with probability $\x[j]\cdot w_j$. Hence, we can further derive that
\begin{align*}
    \frac{\partial F^W(\x)}{\partial \x[i]}&={w_i}\cdot\sum_{R\in\R:u_i\in R}\prod_{u_j\in R\backslash\{u_i\}}(1-\x[j]\cdot w_j)\\
    &=\frac{w_i}{1-\x[i]\cdot w_i}\sum_{R\in\R:u_i\in R}\prod_{u_j\in R}(1-\x[j]\cdot w_j).
\end{align*}
This completes the proof.
\end{proof}

\end{document}